\documentclass[11pt]{article} % default is 10pt
\usepackage{microtype}
\usepackage{amsmath,amssymb,verbatim}
\usepackage{bbm} % for mathematical symbols
\usepackage[sans]{dsfont}
\usepackage{slashed,mathtools} % for Dirac slash
\usepackage[normalem]{ulem}
\usepackage{graphicx,enumerate,bm} % for \includegraphics command and options
\pdfoutput=1

\usepackage[style=ext-alphabetic,backend=bibtex8,maxnames=8,maxalphanames=4,
minalphanames=4,articlein=false,giveninits=true,isbn=false]{biblatex}
\usepackage{makerobust}

\usepackage{geometry} % to change the page dimensions
\usepackage{fancyhdr} % This should be set AFTER setting up the page geometry
\usepackage[titles]{tocloft}
\usepackage{parskip} % needed to not screw up 'other' spacings!
\usepackage[unicode]{hyperref} % for bookmarks in Acrobat Reader and using hyperlinks
\hypersetup{bookmarksnumbered=true, bookmarksopen=true,
  bookmarksopenlevel=2, breaklinks=true, citecolor=blue,
  colorlinks=true, linkcolor=blue, linktoc=page, pdfborder={0 0 0},
  pdfkeywords={Projective superspace}, pdfstartview=FitH,
  pdfauthor={NSP}, pdfsubject={Projective superspace},
  pdftitle={(0,4) projective superspace}, plainpages=false, unicode=true, urlcolor=blue}

\makeatletter
\let\save@mathaccent\mathaccent
\newcommand*\if@single[3]{%
  \setbox0\hbox{${\mathaccent"0362{#1}}^H$}%
  \setbox2\hbox{${\mathaccent"0362{\kern0pt#1}}^H$}%
  \ifdim\ht0=\ht2 #3\else #2\fi
  }
\newcommand*\rel@kern[1]{\kern#1\dimexpr\macc@kerna}
\newcommand*\widebar[1]{\@ifnextchar^{{\wide@bar{#1}{0}}}{\wide@bar{#1}{1}}}
\newcommand*\wide@bar[2]{\if@single{#1}{\wide@bar@{#1}{#2}{1}}{\wide@bar@{#1}{#2}{2}}}
\newcommand*\wide@bar@[3]{%
  \begingroup
  \def\mathaccent##1##2{%
    \let\mathaccent\save@mathaccent
    \if#32 \let\macc@nucleus\first@char \fi
    \setbox\z@\hbox{$\macc@style{\macc@nucleus}_{}$}%
    \setbox\tw@\hbox{$\macc@style{\macc@nucleus}{}_{}$}%
    \dimen@\wd\tw@
    \advance\dimen@-\wd\z@
    \divide\dimen@ 3
    \@tempdima\wd\tw@
    \advance\@tempdima-\scriptspace
    \divide\@tempdima 10
    \advance\dimen@-\@tempdima
    \ifdim\dimen@>\z@ \dimen@0pt\fi
    \rel@kern{0.6}\kern-\dimen@
    \if#31
      \overline{\rel@kern{-0.6}\kern\dimen@\macc@nucleus\rel@kern{0.4}\kern\dimen@}%
      \advance\dimen@0.4\dimexpr\macc@kerna
      \let\final@kern#2%
      \ifdim\dimen@<\z@ \let\final@kern1\fi
      \if\final@kern1 \kern-\dimen@\fi
    \else
      \overline{\rel@kern{-0.6}\kern\dimen@#1}%
    \fi
  }%
  \macc@depth\@ne
  \let\math@bgroup\@empty \let\math@egroup\macc@set@skewchar
  \mathsurround\z@ \frozen@everymath{\mathgroup\macc@group\relax}%
  \macc@set@skewchar\relax
  \let\mathaccentV\macc@nested@a
  \if#31
    \macc@nested@a\relax111{#1}%
  \else
    \def\gobble@till@marker##1\endmarker{}%
    \futurelet\first@char\gobble@till@marker#1\endmarker
    \ifcat\noexpand\first@char A\else
      \def\first@char{}%
    \fi
    \macc@nested@a\relax111{\first@char}%
  \fi
  \endgroup
}
\makeatother

\makeatletter
\g@addto@macro\bfseries{\boldmath}
\makeatother

\newcommand\ol[1] {\overline{#1}}

\newcommand\pmat[1] {\begin{pmatrix}#1\end{pmatrix}}

\newcommand{\bvert}{{\pmb{\rvert}}}
\newcommand{\diag}{\text{diag}}
\newcommand{\wt}[1]{\widetilde{#1}}
\newcommand{\wh}[1]{\widehat{#1}}

\newcommand{\msf}[1]{\mathsf{#1}}
\newcommand{\mc}[1]{\mathcal{#1}}

\newcommand{\ud}{\text{d}}
\newcommand{\uD}{\mathrm{D}}
\newcommand{\uQ}{\mathrm{Q}}
\newcommand{\uz}{\underline{z}}
\newcommand{\mbb}[1]{\mathbf{#1}}

\newcommand{\mbbm}[1]{\mathbbm{#1}}

\newcommand{\w}[1]{\widebar{#1}}

\newcommand{\pD}{\mathbf{D}}
\newcommand{\mC}{\bm{C}}

\newcommand{\mF}{{\bm{F}}}
\newcommand{\mH}{\bm{H}}

\newcommand{\mX}{{\bm{X}}}
\newcommand{\mY}{{\bm{Y}}}
\newcommand{\mZ}{{\bm{Z}}}
\newcommand{\mK}{{\bm{K}}}

\newcommand{\mS}{{\bm{S}}}

\newcommand{\bUpsilon}{{\mbb\Upsilon}}
\newcommand{\tp}[2]{\texorpdfstring{#1}{#2}}
\renewcommand{\i}{\text{i}}

\newcommand{\iUpsilon}{\mathnormal{\Upsilon}}
\newcommand{\Hom}{\text{Hom}}

\DeclareMathOperator{\im}{im}

\MakeRobustCommand\widebar
\MakeRobustCommand\bvert

\makeatletter
\let\@fnsymbol\@arabic
\makeatother

\numberwithin{equation}{section} % Equations numbered as <section>.#
\title{\textsc{$(0,4)$ Projective Superspaces I: \\ {\large Interacting Linear Sigma Models}}}
\author{Naveen S.~Prabhakar \footnote{International
    Centre for Theoretical Sciences-TIFR, Shivakote, Bengaluru -
    560089, India. \newline email: naveen.s.prabhakar@gmail.com} \and
  Martin Ro\v{c}ek \footnote{C.~N.~Yang Institute of Theoretical
    Physics, Stony Brook University, Stony Brook, NY 11794-3840,
    USA. \newline email: martin.rocek@stonybrook.edu}} \date{}

\begin{document}
\maketitle

\begin{abstract}
  We describe the projective superspace approach to supersymmetric
  models with off-shell $(0,4)$ supersymmetry in two dimensions. In
  addition to the usual superspace coordinates, projective superspace
  has extra bosonic variables -- one doublet for each $\text{SU}(2)$
  in the R-symmetry $\text{SU}(2) \times \text{SU}(2)$ which are
  interpreted as homogeneous coordinates on
  $\mbb{CP}^1 \times \mbb{CP}^1$. The superfields are analytic in the
  $\mbb{CP}^1$ coordinates and this analyticity plays an important
  role in our description. For instance, it leads to stringent
  constraints on the interactions one can write down for a given
  superfield content of the model. As an example, we describe in
  projective superspace Witten's ADHM sigma model -- a linear sigma
  model with non-derivative interactions whose target is $\mbb{R}^4$
  with a Yang-Mills instanton solution. The hyperk\"ahler nature of
  target space and the twistor description of instantons by Ward, and
  Atiyah, Hitchin, Drinfeld and Manin are natural outputs of our
  construction.
\end{abstract}

\newpage
\tableofcontents
\newpage
\section{Introduction}\label{intro}

Two dimensional quantum field theories with chiral supersymmetry have
appeared in a variety of physical and mathematical contexts. The most
familiar example is the construction of heterotic string models which
have $(0,1)$ supersymmetry on the worldsheet
\cite{Gross:1984dd}. Conformal theories with $(0,2)$ supersymmetry
were explored \cite{Candelas:1985en,Hull:1985jv} in the context of
compactifications of the type $\mbb{R}^4 \times K$ where $K$ is a
compact Calabi-Yau threefold. $(0,2)$ Landau-Ginzburg models were also
found to furnish a large class of $(0,2)$ heterotic sigma models
\cite{Distler:1993mk}. $(0,4)$ worldsheet conformal theories are also
interesting: they describe compactifications to six dimensions
\cite{Banks:1988yz,Eguchi:1988vra,Seiberg:1988pf} and are useful in
worldsheet descriptions of five-brane instantons
\cite{Callan:1991dj,Callan:1991ky}.

% On the mathematics side, $(0,1)$ models have been used to derive index
% theorems \cite{}. $(0,2)$ theories are related to the theory of
% \emph{chiral differential operators} \cite{}. (To be expanded)

Since the brane revolution, many two dimensional spacetime models with
chiral supersymmetry have been constructed -- these appear as
low-energy effective theories on two dimensional intersections of
D-branes or on D1-branes probing manifolds with special
holonomy. Depending on the brane setup, the models on the intersection
may have $(0,1)$, $(0,2)$, $(0,4)$ or even $(0,8)$ supersymmetry
\cite{Green:1996dd,Berkooz:1996km}. Typically, D-branes have gauge
fields as part of their low-energy dynamics and the chiral
supersymmetric theory is a gauged linear sigma model.

For example, a D1-brane probing a $\text{Spin}(7)$ manifold has
$(0,1)$ supersymmetry on its worldvolume whereas it has $(0,2)$
supersymmetry when probing a Calabi-Yau fourfold. The intersection of
two stacks of D5-branes on a two-dimensional plane has $(0,8)$
supersymmetry on the common intersection \cite{Itzhaki:2005tu};
including a probe D1-brane on the common intersection gives $(0,4)$
susy on the intersection
\cite{Gukov:2004ym,Tong:2014yna,Nekrasov:2015wsu,Nekrasov:2016gud}. Another
system of D-branes which has $(0,4)$ susy is the
$\text{D}1 \subset \text{D}5 \subset \text{D}9$ system which is a
D1-brane probe of a gauge theory instanton on $\mbb{R}^4$ realized by
the $\text{D}5 \subset \text{D}9$ system \cite{Douglas:1996u}, or
instantons on an ALE space realized by taking the four transverse
directions of the D9-brane relative to the D5-brane
\cite{Douglas:1996sw}. Other brane realizations include the
worldvolume theory on M5-branes wrapped on a coassociative submanifold
of a $G_2$-manifold which has $(0,2)$ supersymmetry
\cite{Gadde:2013sca} and M5-branes wrapped on a four dimensional
submanifold of a Calabi-Yau threefold which has $(0,4)$ supersymmetry
\cite{Maldacena:1997de,Gadde:2014wma,Putrov:2015jpa}.

Superspace has proven to be powerful in understanding supersymmetric
theories primarily because it realizes the supersymmetry algebra
off-shell. The advantage of an off-shell realization is that, as long
as the constraints on superfields do not themselves introduce
interactions, we have a clean separation of kinematics and dynamics
and the sum of two supersymmetric actions is automatically
supersymmetric. This has been useful in uncovering the geometric
structures hidden in supersymmetric theories and also understanding
dualities between very different-looking models
\cite{Lindstrom:1983rt,Hitchin:1986ea}. However, the presence of
so-called $E$-terms can mix dynamics with kinematics, and then
supersymmetry restricts the structure of the action even in
superspace; we shall see that this plays a crucial role in our
construction of interacting models.

Superspace descriptions of $(0,1)$, $(1,1)$, $(0,2)$, $(1,2)$ and
$(2,2)$ theories exist
\cite{Sakamoto:1984zk,Siegel:1983es,Fairlie:1973jw,Brink:1975qk,Zumino:1975icn,Dine:1986by,Ademollo:1975an,Brooks:1986uh}
and are well-understood. For theories with a higher amount of
supersymmetry, for instance $(4,4)$ in two dimensions (more generally,
theories with eight supercharges in other dimensions), it is well
known that ordinary superspace is not sufficient to describe off-shell
charged hypermultiplets since the superspace constraints for the
hypermultiplet put it on-shell (see \cite[Section 4.6]{Gates:1983nr}).

There are at least two approaches that address these issues, harmonic
superspace \cite{Galperin:1984av,Galperin:1984mln} and the closely
related isotwistor superspace
\cite{rosly1983super,Roslyi:1985hn,Rosly_1985}, and projective
superspace \cite{Gates:1984nk,Karlhede:1984vr}. All approaches
introduce a new set of bosonic coordinates $u$ which are coordinates
on an $S^2$. In the harmonic approach the $u$ are viewed as harmonic
coordinates on $S^2 \simeq \text{SU(2)} / \text{U(1)}$ where
$\text{SU}(2)$ is the R-symmetry group or a subgroup thereof, and one
considers superfields which are harmonic functions on $S^2$. In the
projective approach, the $S^2$ is viewed as
$\mbb{CP}^1 \simeq \{\mbb{C}^2\smallsetminus 0\} / \mbb{C}^\star$ and
the $u$ are homogeneous coordinates on the $\mbb{CP}^1$ and the
superfields are analytic functions on $\mbb{CP}^1$. These two
approaches are in fact related \cite{Kuzenko:1998xm,Jain:2009aj}.

Projective superspace has been successful in describing many
supersymmetric models with eight supercharges
\cite{Karlhede:1984vr,Karlhede:1986mg,Ivanov:1995cy,Lindstrom:1987ks,Lindstrom:1989ne,Lindstrom:2009afn,Arai:2006gg,Grundberg:1984xr,Gates:1998si}.
In projective superspace, one can write down new kinds of superfields
and superspace constraints which depend on the coordinates $u$. More
precisely, they are fibred over the coset space
$\mbb{CP}^1$. Superfields over projective superspace typically contain
an infinite number of ordinary superfields (the coefficients in a
Taylor expansion in $u$) and these turn out to be crucial in realizing
the off-shell version of the hypermultiplet. Dynamically, most of
these superfields turn out to be auxiliary and thus do not change the
on-shell content of the hypermultiplet.

$(0,4)$ projective superspace has been introduced in
\cite{Hull:2017hfa,Hull:2016khc} and has been used to give off-shell
formulations of nonlinear sigma models involving hypermultiplets. In
this paper, we describe linear sigma models with manifest off-shell
$(0,4)$ supersymmetry in projective superspace.

The R-symmetry of the $(0,4)$ supersymmetry algebra is
$\text{SU}(2) \times \text{SU}(2)'$ and thus one has two projective
superspaces with the $\mbb{CP}^1$s corresponding to the two
$\text{SU}(2)$ subgroups. The hypermultiplets are also of two kinds,
transforming as a doublet under either $\text{SU}(2)$ or
$\text{SU}(2)'$. We call them standard hypermultiplets and, following
\cite{Witten:1994tz}, twisted hypermultiplets respectively. We
describe these in detail in Section \ref{hypermult}. We shall see that
a hyper can be realized either as a linear polynomial in the
homogeneous coordinates (the $\mc{O}(1)$ superfield) or as a power
series in a local coordinate on the $\mbb{CP}^1$ (the `polar'
superfield). $\mc{O}(1)$ superfields are treated in some detail in
\cite{Hull:2017hfa,Hull:2016khc}. $(0,p)$ supersymmetry allows
independent fermionic multiplets with chirality opposite to that of
the supercharges. These are the fermi multiplets; we realize them in
projective superspace in Section \ref{fermimult}.

In $(0,2)$ models, we have interactions of the nonlinear sigma model
type or the non-derivative type. Non-derivative interactions between
chiral multiplets, gauge multiplets and fermi multiplets are described
by modifying their superspace constraints with the so-called
$E$-terms, or by including superpotential-like $J$-terms in the
Lagrangian (see Appendix \ref{02superspace} of this paper). In Section
\ref{interaction}, we describe the $E$-term type non-derivative
interactions for $(0,4)$ models containing standard hypers, twisted
hypers and fermis (it turns out that $(0,4)$ $J$-terms are not
possible). In a companion paper \cite{PRcomp}, we describe gauge
multiplets and their interactions with hypers and fermis in projective
superspace.

In Section \ref{ADHM}, we describe in projective superspace a
prominent $(0,4)$ supersymmetric model: a linear sigma model which
flows down to a sigma model with target being an instanton solution in
four dimensions. The couplings of the linear sigma model and the
constraints they satisfy as a consequence of $(0,4)$ supersymmetry
encode the data that enters ADHM construction of instantons
\cite{Atiyah:1978ri}. This was demonstrated in $(0,1)$ superspace by
Witten \cite{Witten:1994tz}, and it was given a D-brane interpretation
by Douglas \cite{Douglas:1996u}. A manifest $(0,4)$ construction was
given in harmonic superspace in \cite{Galperin:1994qn,Galperin:1995pq}
(see \cite{Gates:1994bu} for some partial results in ordinary $(0,4)$
superspace). In our construction in $(0,4)$ projective superspace, the
hyperk\"ahler nature of the target space is manifest, and the monads
which describe holomorphic bundles on twistor space $\mbb{CP}^3$
\cite{Atiyah:1978ri} appear explicitly. We also extend this
construction to self-dual solutions on $\mbb{R}^{4k'}$ with $k' > 1$,
i.e., of dimension greater than 4 \cite{Corrigan:1984si,Witten:1994tz}.

The appendix includes a quick review of $(0,1)$ and $(0,2)$
superspaces (Appendix \ref{0102app}), a realization of the $(4,4)$
hypermultiplet in $(4,4)$ projective superspace and its reduction to
$(0,4)$ projective superspace (Appendix \ref{44app}), and finally a
detailed derivation of the ordinary space component actions for the
general $(0,4)$ supersymmetric interacting linear sigma model
(Appendix \ref{components}).

\section{\tp{$(0,4)$}{(0,4)} projective superspace}\label{sspace}

\subsection{Introduction}
The $(0,4)$ supersymmetric algebra consists of four real supercharges
$\mc{Q}_{\mu+}$, $\mu = 1,\ldots,4$, of right-handed chirality. It is
useful to write these real supercharges in terms of a $2 \times 2$
matrix $\mc{Q}_{aa'+}$ that satisfies the reality conditions
\begin{equation}\label{Qreal}
  \mc{Q}_{aa'+} = \w{\mc{Q}}{}^{bb'}_+\,\varepsilon_{ba}\varepsilon_{b'a'}\ ,
\end{equation}
where $\w{\mc{Q}}{}^{bb'}_+ = \w{(\mc{Q}_{\emph{bb}'+})}$. Here,
$a = 1,2$ and $a'=1',2'$ are $\text{SU}(2)$-doublet indices. The
R-symmetry group is then
$\text{SO}(4) = (\text{SU}(2) \times \text{SU}(2))/\mbb{Z}_2$. We will
be interested in the representations of the supersymmetry algebra
which are charged under just one of the $\text{SU}(2)$s and hence it
is useful to consider the double cover
$\text{Spin}(4) \approx \text{SU}(2) \times \text{SU}(2) := F \times
F'$. The $a$ and $a'$ indices are lowered using the invariant tensors
$\varepsilon_{ab}$ and $\varepsilon_{a'b'}$ which satisfy
$\varepsilon^{ab}\varepsilon_{bc} = -\delta^a{}_c$,
$\varepsilon^{a'b'}\varepsilon_{b'c'} = -\delta^{a'}{}_{c'}$ and
$\varepsilon_{12} = \varepsilon_{1'2'} = +1$.

The supersymmetry algebra is
\begin{equation}\label{susyalg}
  \{\mc{Q}_{aa'+}\,, \mc{Q}_{bb'+}\} = -2\i \varepsilon_{ab}\varepsilon_{a'b'} \partial_{++}\ .
\end{equation}

$(0,4)$ superspace $\mbb{R}^{1,1|0,4}$ is described by the
supercoordinates $\uz = (x^{\pm\pm}, \theta^{aa'+})$ where
$x^{\pm\pm} = \tfrac{1}{2}(x^0 \pm x^1)$. The corresponding
supercovariant derivatives are
$\partial_{\pm\pm} = \partial_0 \pm \partial_1$ and $\uD_{aa'+}$ with
the algebra
\begin{equation}\label{04basic}
  \{\uD_{aa'+}\, ,\uD_{bb'+}\} = 2\i\varepsilon_{ab}\varepsilon_{a'b'} \partial_{++}\ ,\quad [\uD_{aa'+}\, ,\partial_{\pm\pm}] = 0\ .
\end{equation}
The derivatives $\uD_{aa'+}$ also satisfy the same reality condition
as for the supersymmetry generators \eqref{Qreal}. We loosely refer to
\eqref{04basic} as the supersymmetry algebra though it differs from
\eqref{susyalg} by a sign. The supersymmetry generators
$\mc{Q}_{aa'+}$ and the derivatives $\uD_{aa'+}$ mutually anticommute:
$\{\mc{Q}_{aa'+}\, , \uD_{bb'+}\} = 0$.

In this paper, we work
exclusively with the derivatives $\uD_{aa'+}$ rather than the
supersymmetry generators $\mc{Q}_{aa'+}$. Supersymmetry
transformations of some component of a superfield $\Phi$ can be
expressed in terms of $\uD_{aa'+}$ because of the following fact which
can be easily verified by using the explicit superspace expressions
for $\mc{Q}_{aa'+}$ and $\uD_{aa'+}$:
\begin{equation}\label{susytr}
  \delta\Phi_\bvert = \Big(\epsilon^{aa'+} \mc{Q}_{aa'+}\Phi\Big)_\bvert = \Big(\epsilon^{aa'+} \uD_{aa'+}\Phi\Big)_\bvert\ ,
\end{equation}
where $\epsilon^{aa'+}$ are constant Grassmann parameters, and
$(X)_\bvert$ stands for the operation of setting the Grassmann
coordinates $\theta^{aa'+}$ to zero in the expression $X$. The
$\ud\theta^{aa'+}$ that appear in the superspace measure can also
replaced by the corresponding $\uD_{aa'+}$ up to total
derivatives\footnote{This is standard procedure, see
  e.g.~\cite{Gates:1983nr}.}.

It is convenient to define
\begin{equation}\label{ordderdef}
  \uD_+ := \uD_{11'+}\ ,\quad \w\uD_+ := \uD_{22'+}\ ,\quad \uQ_+ := \uD_{21'+}\ ,\quad \w\uQ_+ := -\uD_{12'+}\ .
\end{equation}
These derivatives span two (anti)commuting $(0,2)$ subalgebras:
\begin{equation}\label{02subalg}
  \{\uD_+, \w\uD_+\} = 2\i\partial_{++}\ ,\quad \{\uQ_+, \w\uQ_+\} = 2\i\partial_{++}\ ,\ \text{with other anticommutators equal to zero.}
\end{equation}

\subsection{Algebras, superfields and actions}\label{algsupact}

Consider two sets of commuting coordinates $u^a$ and $v^{a'}$ which
are doublets under the R-symmetry $\text{SU}(2)$ subgroups $F$ and
$F'$ respectively. These are most usefully interpreted in our context
as homogeneous coordinates on $\mbb{CP}^1 \times \mbb{CP}^1{}'$ (we
label the second $\mbb{CP}^1$ as $\mbb{CP}^{1}{}'$ to indicate its
relation to $F'$). The superspace with the coordinates
$(x^{\pm\pm}, \theta^{aa'+}, u^a, v^{a'})$ is
$\mbb{R}^{1,1|0,4} \times \mbb{CP}^1 \times \mbb{CP}^1{}'$ which we
refer to as \emph{projective superspace}. The subspaces
$\mbb{R}^{1,1|0,4} \times \mbb{CP}^1$ and
$\mbb{R}^{1,1|0,4} \times \mbb{CP}^1{}'$ are important for us.

We also introduce \emph{conjugate} doublets $\wt{u}^a$ and
$\wt{v}{}^{a'}$ which satisfy
\begin{equation}\label{utleq}
  \varepsilon_{ab}\wt{u}^a {u}^b = 1\ ,\quad \varepsilon_{a'b'}\wt{v}^{a'} {v}^{b'} = 1\ .
\end{equation}
\paragraph{A shift symmetry} Note that there is more than one solution
to the equation $\varepsilon_{ab}\wt{u}^a {u}^b = 1$. If $\wt{u}_0^b$
is one solution, then so is $\wt{u}_0^b + \omega u^b$ for any
$\omega \in \mbb{C}$. Thus there is a shift symmetry on the
$\wt{u}^a$:
\begin{equation}\label{shiftutl}
  \wt{u}^a \to \wt{u}^a + \omega u^a\ ,\quad\text{for}\quad \omega \in \mbb{C}\ .
\end{equation}
There is a similar shift symmetry for the conjugate doublet
$\wt{v}^{a'}$.

\paragraph{Derivatives on projective superspace} Consider the
derivatives
\begin{equation}\label{projder}
  \pD_{a'+} := u^a \uD_{aa'+}\ ,\quad  \widetilde{\pD}_{a'+} := \wt{u}^a \uD_{aa'+}\ ,\quad \pD_{a+} := v^{a'} \uD_{aa'+}\ ,\quad  \widetilde{\pD}_{a+} := \wt{v}^{a'} \uD_{aa'+}\ ,
\end{equation}
where $\wt{u}^a$ and $\wt{v}^{a'}$ are any solutions to the equations
\eqref{utleq}. The algebra of the derivatives \eqref{projder} is
obtained from \eqref{04basic}:
\begin{align}\label{projalg}
  &\{\pD_{a'+}\, , \pD_{b'+}\} = 0\ ,\quad \{\widetilde{\pD}_{a'+}\, , \widetilde{\pD}_{b'+}\} = 0\ ,\quad \{\pD_{a'+}\, ,\widetilde{\pD}_{b'+}\} = -2\i\varepsilon_{a'b'}\partial_{++}\ ,\nonumber\\
    &\{\pD_{a+}\, , \pD_{b+}\} = 0\ ,\quad \{\widetilde{\pD}_{a+}\, , \widetilde{\pD}_{b+}\} = 0\ ,\quad \{\pD_{a+}\, ,\widetilde{\pD}_{b+}\} = -2\i\varepsilon_{ab}\partial_{++}\ .
\end{align}
Note that the shift symmetry \eqref{shiftutl} shifts the derivatives
$\wt\pD_{a'+}$ by $\omega \pD_{a'+}$ but it leaves the algebra
\eqref{projalg} unchanged. We shall see below that the action is also
invariant under the shift symmetry up to total derivative terms.

We also introduce the fully contracted derivative
\begin{equation}\label{dsqzero}
  \pD_+ = u^a v^{a'} \uD_{aa'+} = u^a \pD_{a+} = v^{a'} \pD_{a'+}\quad\text{which satisfies}\quad   \pD_+^2 = 0\ ,
\end{equation}
due to the anticommutation relations \eqref{04basic}. We can recover
the algebra in \eqref{projalg} by writing
$\pD_+^2 = u^a u^b \{\pD_{a+}, \pD_{b+}\}$ or
$\pD_+^2 = v^{a'} v^{b'}\{\pD_{a'+}, \pD_{b'+}\}$.

\paragraph{Projective superfields} An $F$-\emph{projective} superfield
$\bm{\Phi}(\uz,u)$ is a function of the superspace coordinates
$\uz = (x_{\pm\pm}, \theta^{aa'+})$ and the $\mbb{CP}^1$ coordinates
$u^a$ which satisfy the following:
\begin{enumerate}
\item[(1)] $\bm\Phi$ is holomorphic in a domain in $\mbb{CP}^1$,

\item[(2)] $\bm\Phi$ satisfies the \emph{projective} constraints
  $\pD_{a'+}\bm\Phi(\uz,u) = 0$,

\item[(3)] $\bm\Phi$ may be in non-trivial
  representations of the R-symmetry group
  $\text{SU}(2) \times \text{SU}(2)'$ and the Lorentz group
  $\text{SO}(1,1)$.
\end{enumerate}
An $F'$-projective superfield is analogously a function of the
superspace coordinates $\uz$ and the $\mbb{CP}^1{}'$ coordinate
$v^{a'}$ and is annihilated by $\pD_{a+}$. We discuss the different
types of projective superfields in Section \ref{projsfield}.

The $F$-projective constraints $\pD_{a'+}\bm{\Phi}(\uz,u) = 0$ can be
encoded more economically in terms of the fully contracted derivative
\eqref{dsqzero} $\pD_+ = v^{a'}\pD_{a'+}$:
\begin{equation}\label{econconst}
  \pD_{+} \bm{\Phi} = 0\ .
\end{equation}
Since $\bm{\Phi}$ depends only on $u$ and not on $v$,
$\pD_+\bm\Phi = v^{a'} \pD_{a'+}\bm\Phi$ implies
$\pD_{a'+}\bm{\Phi} = 0$. The advantage of \eqref{econconst} is that
it takes the same form for $F'$-projective superfields
$\bm{\Phi}(\uz, v)$ as well, since we can now recover
$\pD_{a+}\bm{\Phi} = 0$ using $\pD_+ = u^a \pD_{a+}$. We frequently
use the derivative $\pD_+$ in the paper.

\paragraph{Actions} The constraints $\pD_{a'+} \bm{\Phi} = 0$ on a
projective superfield $\bm{\Phi}$ imply that $\bm{\Phi}$ depends on
only half of the Grassmann coordinates. The appropriate superspace
measure which ensures $(0,4)$ invariance of an action composed of
projective superfields is then quadratic in the derivatives
$\wt\pD_{a'+}$, i.e., $\widetilde{\pD}_{1'+}\widetilde{\pD}_{2'+}$.
The $(0,4)$ supersymmetric action is then given by
\begin{equation}\label{actdef}
  \mc{S}[\bm{\Phi}] = \int\ud^2x \left(\frac{1}{2\pi\i}\oint_\gamma \varepsilon_{ab} u^a \ud u^b\ \wt\pD_{1'+}\wt\pD_{2'+}\ \mK_{--}(\bm{\Phi})\right)_\bvert\ ,
\end{equation}
where 
\begin{enumerate}
\item $\bvert$ sets all the Grassmann coordinates to zero (we
  frequently omit the $\bvert$ from our expressions).

\item $\mK_{--}$ is the superspace Lagrangian which satisfies
  $\pD_{a'+}\mK_{--} = 0$. It must carry the $--$ Lorentz
  representation (left-moving part of a vector) in order to compensate
  the $++$ in the projective superspace measure.

\item The contour $\gamma \in \mbb{CP}^1$ is chosen to avoid possible
  singularities in $\wt\pD_{1'+}\wt\pD_{2'+} \mK_{--}$.
\end{enumerate}
The action is invariant (up to total spacetime derivatives) under the
shift symmetry \eqref{shiftutl}
$\wt\pD_{a'+} \to \wt\pD_{a'+} + \omega \pD_{a'+}$ since the
Lagrangian $\mK_{--}$ satisfies $\pD_{a'+} \mK_{--} = 0$. Since the
superspace measure
$\varepsilon_{ab} u^a \ud u^b \wt\pD_{1'+} \wt\pD_{2'+}$ is invariant
under $F$ and $F'$, the action \eqref{actdef} is manifestly invariant
under $F$ and $F'$ if the superspace Lagrangian is invariant.

\paragraph{Non-derivative interactions} Suppose a projective
superfield $\Phi_s$ is in the spin $s$ representation of the Lorentz
group $\text{SO}(1,1)$. The requirement that $\pD_{a'+}\bm\Phi_s = 0$
can be relaxed to have a non-zero right hand side:
\begin{equation}\label{modconst}
  \pD_{a'+} \bm\Phi_s = \mS_{a',s+1}\ ,
\end{equation}
where $\mS_{a',s+1}$ is a function of other superfields in the model
and is in the spin $s+1$ representation of $\text{SO}(1,1)$. This
allows us to introduce interactions (the so-called $E$-terms) as we
will see later in Section \ref{interaction}:

The modified constraints \eqref{modconst} are consistent with the
algebra $\{\pD_{a'+}, \pD_{b'+}\} = 0$ only if the function satisfies
\begin{equation}
  \pD_{a'+} \mS_{b',s+1} + \pD_{b'+} \mS_{a',s+1} = 0\ .
\end{equation}
To ensure $(0,4)$ invariance of the action, we require that the
superspace Lagrangian $\mK_{--}(\bm\Phi)$ satisfies
$\pD_{a'+}\mK_{--} = 0$ even if $\pD_{a'+}\bm\Phi$ is not zero. This
further constrains the $\mS_{a',s+1}$.

Thus, any $(0,4)$ supersymmetric model must satisfy the following
constraints:
\begin{enumerate}
\item The $(0,4)$ algebra $\pD_+^2 = 0$ must be satisfied on every superfield in the model,
\item The superspace Lagrangian $\mK_{--}$ must satisfy
  $\pD_{a'+} \mK_{--} = 0$ to ensure $(0,4)$ supersymmetry of the action.
\end{enumerate}
These criteria place stringent constraints on the superfield
content and the interactions in a model.

\subsection{Projective superspace in inhomogeneous coordinates}

\paragraph{A primer on $\mbb{CP}^1$}
The projective space $\mbb{CP}^1$ is constructed as the quotient space
$\{\mbb{C}^2 \smallsetminus 0\} / \sim$, where $\sim$ is the following
equivalence relation on the coordinates of $\mbb{C}^2$:
$(u^1, u^2) \sim (\lambda u^1, \lambda u^2)$,
$\lambda \in \mbb{C}^\star$. We describe $\mbb{CP}^1$ in terms of two
charts $\mathsf{U}_1$ and $\mathsf{U}_2$:
\begin{equation}\label{chart}
  \msf{U}_a := \{(u^1, u^2) \in \mbb{C}^2\ |\ u^a \neq 0\}\ .
\end{equation}
The map $\mc{S} \in \text{SU}(2)$ which acts on the homogeneous
coordinates as
\begin{equation}\label{Smap}
\mc{S}:\  \pmat{u^1 \\ u^2}  \longmapsto \pmat{0 & 1 \\ -1 & 0}  \pmat{u^1 \\ u^2} =   \pmat{-u^2 \\ u^1}\ ,
\end{equation}
interchanges the two charts. Using the equivalence
$(u^1, u^2) \sim (\lambda u^1, \lambda u^2)$,
$\lambda \in \mbb{C}^\star$, we can scale out the non-zero coordinate
in each of the charts and obtain a description in terms of
\emph{inhomogeneous} coordinates:
\begin{equation}\label{charts}
  \msf{U}_1 = \{(1, -\rho)\ |\ \rho \in \mbb{C}\}\ ,\quad \msf{U}_2 = \{(\zeta, 1)\ |\ \zeta \in \mbb{C}\}\ ,
\end{equation}
with $\rho = - u^2 / u^1$ and $\zeta = u^1 / u^2$. On the intersection
$\msf{U}_{12} := \msf{U}_1 \cap \msf{U}_2$, the local coordinates
$\zeta$ and $\rho$ are related by the $\mc{S}$ map \eqref{Smap}
\begin{equation}
  \mc{S}:\ \zeta \longmapsto -1/\zeta = \rho\ .
\end{equation}
We can thus express all our results exclusively in terms of one of the
inhomogeneous coordinates, say $\zeta$, by appending the point
$\zeta = \infty$ to the chart $\msf{U}_2$. We frequently adopt this
usage to avoid cluttering of notation.

\paragraph{The derivatives $\pD_{a'+}$, $\wt\pD_{a'+}$} We next
express the derivatives $\pD_{a'+}$ and $\wt\pD_{a'+}$ in terms of the
local coordinates $\zeta$ and $\rho$ in the charts $\msf{U}_2$ and
$\msf{U}_1$ respectively. In the chart $\msf{U}_2$, we have
$u^2 \neq 0$ and $u^a = (u^2) (\zeta, 1)$. Thus, we can choose
$\wt{u}^a = (u^2)^{-1} (1,0)$ which indeed satisfies
$\wt{u}{}^a u^b \varepsilon_{ab} = 1$. Using the scale invariance
$u^a \to \lambda u^a$, we can set $u^2 = 1$ as discussed above
\eqref{charts}. The derivatives $\pD_{a'+} = u^a \uD_{aa'+}$ and
$\wt\pD_{a'+} = \wt{u}{}^{a} \uD_{aa'+}$ are then given by
\begin{align}\label{derdef}
\text{In $\msf{U}_2$}:\quad  &\pD_{1'+} = \zeta \uD_+ + \uQ_+\ ,\quad \pD_{2'+} = -\zeta \w\uQ_+ + \w\uD_+\ ,\quad \wt\pD_{1'+} =  \uD_+\ ,\quad \wt\pD_{2'+} = -\w\uQ_+\ ,
\end{align}
where we have used the expressions \eqref{ordderdef} for
$\uD_{aa'+}$. A similar description can be obtained in the chart
$\msf{U}_1$ in which $u^1 \neq 0$. Writing $u^a = u^1 (1, -\rho)$,
choosing $\wt{u}^a = (u^1)^{-1}(0,-1)$ and setting $u^1 = 1$ by scale
invariance, we have
\begin{equation}\label{derdefu1}
  \text{In $\msf{U}_1$}:\quad \pD_{1'+} = \uD_+ - \rho \uQ_+\ ,\quad \pD_{2'+} = -\w\uQ_+ - \rho\w\uD_+\ ,\quad \wt\pD_{1'+} = -\uQ_+\ ,\quad \wt\pD_{2'+} = -\w\uD_+\ .
\end{equation}
Observe that, in the intersection $\msf{U}_{12}$, the derivatives
$\pD_{a'+}(\rho)$ defined in $\msf{U}_1$ are related to the
$\pD_{a'+}(\zeta)$ defined in $\msf{U}_2$ as
\begin{equation}\label{O1pd}
  \pD_{a'+}(\rho) = (-\rho) \pD_{a'+}(\zeta(\rho))\ ,
\end{equation}
which is the gluing rule for a global section of the line bundle
$\mc{O}(1) \to \mbb{CP}^1$ (we have used that $\zeta(\rho) = -\rho^{-1}$
on the overlap).

Similarly, the derivatives $\wt\pD_{a'+}(\rho)$ in $\msf{U}_1$ and $\wt\pD_{a'+}(\zeta)$ in $\msf{U}_2$ are related on the overlap as
\begin{equation}\label{affinetr}
  \wt\pD_{a'+}(\rho) = (-\rho)^{-1} \wt\pD_{a'+}(\zeta(\rho)) + \rho^{-1} \pD_{a'+}(\rho)  = (-\rho)^{-1} \wt\pD_{a'+}(\zeta(\rho))  - \pD_{a'+}(\zeta(\rho)) \ ,
\end{equation}
where we have used \eqref{O1pd} in going to the last expression. The
transformation \eqref{affinetr} can be viewed as the usual
transformation of a section of $\mc{O}(-1)$ plus a shift term
proportional to $\pD_{a'+}$ generated by the shift symmetry
\eqref{shiftutl}. This allows us to define $\wt\pD_{a'+}$ globally on
$\mbb{CP}^1$, not as a section of $\mc{O}(-1)$ but as a section of the
affine bundle modelled on $\mc{O}(-1)$.

\textbf{Note:} There is an alternate way of writing the $(0,4)$
algebra using the derivatives $\pD_{a'+}$ and
$\frac{\partial}{\partial\zeta}$:
\begin{equation}\label{altzetader}
  \{ \pD_{a'+}\, , [\partial_\zeta\, , \pD_{b'+}]\} = -2\i \varepsilon_{a'b'} \partial_{++}\ .
\end{equation}
Thus, one may use the derivatives $\pD_{a'+}$ and
$\partial / \partial \zeta$ instead of $\pD_{a'+}$ and $\wt\pD_{a'+}$
in describing projective superspace. Observe that
$\partial_\zeta \pD_{b'+}$ coincides with $\wt\pD_{b'+}$ in
$\msf{U}_1$ and $\partial_\rho \pD_{b'+}$ coincides with $\wt\pD_{b'+}$
in $\msf{U}_2$. Further, $\partial_\zeta\pD_{a'+}$ also satisfies the
rule \eqref{affinetr}. However, this is expected since the derivative
of a section of $\mc{O}(1)$ transforms as a section of the affine bundle
modelled on $\mc{O}(-1)$.

\paragraph{A $(0,2)$ action which is $(0,4)$ supersymmetric} Plugging
in the derivatives \eqref{derdef} in the action \eqref{actdef}, we get
\begin{equation}
  \mc{S}[\bm{\Phi}] = \int\ud^2x \oint_\gamma \frac{\ud\zeta}{2\pi\i} \uD_+\w\uQ_+ \mK_{--}(\bm{\Phi})\ .
\end{equation}
We can also rewrite the above action in $(0,2)$ superspace. Using
$-\w\uD_+ = -\zeta^{-1}\w{\uD}_+ + \zeta^{-1}\pD_{2'+}$ and
$\pD_{a'+} \mK_{--} = 0$, we get
\begin{equation}\label{04action}
  \mc{S}[\bm{\Phi}] = \int\ud^2x\ \oint_\gamma \frac{\ud\zeta}{2\pi \i\zeta} \uD_{+}\w{\uD}_+\mK_{--}(\bm{\Phi}) =  \int\ud^2x\ \uD_{+}\w{\uD}_+\oint_\gamma \frac{\ud\zeta}{2\pi \i\zeta}\mK_{--}(\bm{\Phi})\ .
\end{equation}

\paragraph{$F'$-projective superspace} For completeness, we explicitly
describe some analogous aspects of $F'$-projective superspace. We have
the inhomogeneous coordinate $\zeta'$ for the $\mbb{CP}^1$
corresponding to the $F'$ doublet $v^{a'}$. We then choose
$v^{a'} = (\zeta', 1)$ and $\wt{v}{}^{a'} = (1 ,0)$ using the scale
invariance $v^{a'} \to \lambda' v^{a'}$, $\lambda' \in
\mbb{C}^\star$. The $F'$-projective derivatives $\pD_{a+}$ and
$\wt\pD_{a+}$ are then
\begin{align}\label{prprojder}
 \pD_{1+} = \zeta' \uD_+ - \w\uQ_+\ ,\quad \pD_{2+} = \zeta'\uQ_+ + \w\uD_+\ ,\quad \wt\pD_{1+} =  \uD_+  \ ,\quad \wt\pD_{2+} = \uQ_+\ .
\end{align}
A $(0,4)$ supersymmetric action in $(0,2)$ superspace for
$F'$-projective superfields $\bm{\Phi}'$ is given by
\begin{equation}\label{F'actdef}
  \mc{S}[\bm{\Phi}'] = -\int\ud^2x\ \uD_{+}\w{\uD}_+\oint_{\gamma'} \frac{\ud\zeta'}{2\pi \i\zeta'}\mK'_{--}(\bm{\Phi}')\ .
\end{equation}
The actions we consider in this paper will only have a single contour
integral over either $\zeta$ or $\zeta'$.

The fully contracted derivative $\pD_+ = u^a v^{a'} \uD_{aa'+}$
\eqref{dsqzero} in terms of $\zeta$ and $\zeta'$ is
\begin{equation}\label{derfullcont}
  \pD_{+} = \zeta \zeta' \uD_{11'+} + \zeta \uD_{12'+} + \zeta' \uD_{21'+} + \uD_{22'+} = \zeta\zeta' \uD_+ - \zeta \w\uQ_+ + \zeta' \uQ_+ + \w\uD_+\ .
\end{equation}

\subsection{Analytic structure of projective
  superfields}\label{projsfield}
Recall that $F$-projective superfields are holomorphic in a connected
open subset of $\mbb{CP}^1$ and that they are annihilated by the
derivatives $\pD_{a'+}$. We now describe the different types of
projective superfields which differ in their analytic structure on the
$\mbb{CP}^1$. $F'$-projective superfields are defined analogously.

\paragraph{$\mc{O}(p)$ superfields} The superfield is a homogeneous
polynomial in the $u^a$ of degree $p > 0$:
\begin{equation}
  \bm{\eta}(\uz,u) = {\eta}_{a_1\cdots a_p}(\uz) u^{a_1} \cdots u^{a_p} = \sum_{i=0}^p {\eta}_i(\uz) (u^1)^{i} (u^2)^{p-i}\ .
\end{equation}
The components $\eta_{a_1\cdots a_p}(\uz)$ are ordinary $(0,4)$
superfields, i.e., functions on $\mbb{R}^{1,1|0,4}$. Note that
$\bm{\eta}$ is a global section of the line bundle
$\mc{O}(p) \to \mbb{CP}^1$. We thus call such superfields
\textbf{$\mc{O}(p)$ superfields}. In the chart $\msf{U}_2$ where
$u^2 \neq 0$ we can write $\bm{\eta}$ as
\begin{equation}\label{etapzeta}
  \bm{\eta}(\uz,u) = (u^2)^p \bm{\eta}(\uz,\zeta) = (u^2)^p \sum_{j=0}^p \eta_j(\uz)\, \zeta^j\ ,
\end{equation}
which becomes a polynomial in the inhomogeneous coordinate
$\zeta = u^1 / u^2$ when we set $u^2 = 1$.

\paragraph{Meromorphic $\mc{O}(n)$ superfields} The $\mc{O}(n)$
superfields discussed above are global holomorphic sections of
$\mc{O}(n) \to \mbb{CP}^1$. We can consider more general superfields
which are only local sections of $\mc{O}(n)$ and cannot be extended to
all of $\mbb{CP}^1$. A familiar class of examples are the meromorphic
sections of $\mc{O}(n)$ which are rational functions of $u^a$:
\begin{equation}
  \bm{\eta}(\uz, u) = \frac{\bm{P}(\uz, u)}{\bm{Q}(\uz, u)} = \frac{ P_{i_1\cdots i_p}(\uz) u^{i_1}\cdots u^{i_p}}{Q_{i_1\cdots i_q}(\uz) u^{i_1}\cdots u^{i_q}}\ ,
\end{equation}
where $\bm{P}$ and $\bm{Q}$ are homogeneous polynomials of degree $p$
and $q$ respectively. The domain of definition $\mc{D}_{\bm{\eta}}$ of
$\bm{\eta}$ on $\mbb{CP}^1$ is restricted to the open set where
$\bm{Q}(\uz,u) \neq 0$. The degree of homogeneity of $\bm{\eta}$ is
then $n = p-q$ and thus $\bm{\eta}$ is a local section of
$\mc{O}(p-q) \to \mbb{CP}^1$ defined on $\mc{D}_{\bm{\eta}}$. In terms
of the inhomogeneous coordinate $\zeta$, we have
\begin{equation}
  \bm{\eta}(\uz, \zeta) = \frac{a_0(\uz) + a_1(\uz) \zeta + \cdots + a_p(\uz) \zeta^p}{b_0(\uz) + b_1(\uz) \zeta + \cdots + b_q(\uz) \zeta^q}\ ,
\end{equation}
where the $a_i(\uz)$ are appropriate combinations of
$P_{i_1\cdots i_p}(\uz)$ and similarly, $b_i(\uz)$ are combinations of
the $Q_{i_1\cdots i_q}$.

\paragraph{Local superfields} Consider superfields which are formal
power series in $\zeta$ or $\zeta^{-1}$ or both. These appear as
series expansions of local holomorphic sections in the neighbourhoods
of $\zeta = 0$, $\zeta = \infty$ or in the annulus
$\mbb{CP}^1 \smallsetminus \{0, \infty\}$. Consider a power series in
$\zeta$:
\begin{equation}
  \bUpsilon(\uz,\zeta) = \sum_{j = 0}^\infty \iUpsilon_j(\uz) \zeta^j\ .
\end{equation}
Such superfields shall be termed \textbf{arctic} since they are
well-defined at the north pole $\zeta = 0$ of $\mbb{CP}^1$ (and
possibly in a neighbourhood of $\zeta = 0$ as well). Similarly, a
superfield which is a power series in $\zeta^{-1}$ is designated
\textbf{antarctic}. 

Finally, a superfield which is defined in the annulus and is real
under the extended complex conjugation given below in Section
\ref{extconjdef} is called \textbf{equatorial}.

\subsection{R-symmetry in projective superspace}\label{Rsym}
We consider the R-symmetry transformation of the various objects in
projective superspace for the subgroup $F = \text{SU}(2)$ in this
subsection \cite{Karlhede:1984vr} (the discussion for
$F' = \text{SU}(2)'$ proceeds analogously).  The homogeneous
coordinates $u^a = (u^1, u^2)$ on $\mbb{CP}^1$ transforms as a doublet
under $F$:
\begin{equation}\label{gmat}
  u^c \to (g\cdot u)^c = g^c{}_d u^d\ ,\quad g = \pmat{a & b \\ -\w{b} & \w{a}}\quad\text{with}\quad a\w{a} + b\w{b} = 1\ .
\end{equation}
Accordingly, the inhomogeneous coordinate $\zeta = u^1 / u^2$
transforms fractional-linearly:
\begin{equation}
 \zeta \to g\cdot\zeta = \frac{a\zeta + b}{-\w{b}\zeta + \w{a}}\ .
\end{equation}
Also, a doublet $u_a = \varepsilon_{ab} u^b$ with a lower index $a$ transforms as
\begin{equation}\label{lowerind}
  u_a \to (g\cdot u)_a := u_b (g^{-1})^b{}_a\ .
\end{equation}

\paragraph{Factor of automorphy} We define a \emph{factor of
  automorphy} $j: F \times \mbb{CP}^1 \to \mbb{C}$ for the action of
$F$ on $\mbb{CP}^1$ as follows. Let
$g = \pmat{a & b \\ -\w{b} & \w{a}} \in F$ and $\zeta \in
\mbb{CP}^1$. Then we have
\begin{equation}\label{factauto}
j(g,\zeta) := (\w{a} - \w{b}\zeta)\ .
\end{equation}
It is easy to check that $j(g,\zeta)$ satisfies
$j(g_1g_2, \zeta) = j(g_1, g_2\cdot\zeta)\, j(g_2,\zeta)$. Suppose we
have an object $\bm\Phi(\zeta)$ that depends holomorphically on
$\zeta$. The transformation of $\bm\Phi$ by a $F$-transformation $g$
is denoted by $g\cdot\bm\Phi$. An object $\bm\Phi(\zeta)$ is said to
have $F$-weight $n$ if it satisfies
\begin{equation}\label{weightdef}
  \bm\Phi(\zeta) = j(g,\zeta)^n \times [g\cdot\bm\Phi](g\cdot \zeta)\ ,\quad g\in F\ ,
\end{equation}
That is, $\bm\Phi$ is a local section of the line bundle
$\mc{O}(n) \to \mbb{CP}^1$. Note that weight $0$ objects are simply
local functions on $\mbb{CP}^1$. 

Next, we describe the $R$-symmetry of $\mc{O}(n)$ superfields and
arctic superfields.

\paragraph{$\mc{O}(n)$ superfields}
Consider an $\mc{O}(n)$ superfield $\bm{\eta}$ given by
$\bm{\eta}(u) = \eta_{a_1\ldots a_{n}} u^{a_1} \cdots
u^{a_{n}}$. Since all $F$-doublet indices are contracted in
$\bm{\eta}(u)$, it is invariant under $F$. That is,
\begin{equation}
  \bm{\eta}(u) = [g\cdot\bm{\eta}](g\cdot u)\ ,\quad g\in F\ ,
\end{equation}
where $[g\cdot\bm{\eta}](g\cdot u)$ on the right hand side is a new
$\mc{O}(n)$ superfield $[g\cdot\bm\eta]$ obtained by transforming the
components $\eta_{a_1\cdots a_n}$, and evaluated at the transformed
coordinates $g\cdot u$. In terms of the inhomogeneous coordinate
$\zeta$, we have
\begin{equation}
  \bm{\eta}(u) := (u^2)^{n} \bm{\eta}(\zeta)\ ,\quad\text{with}\quad \bm{\eta}(\zeta) := \sum_{j=0}^{n} \eta_j \zeta^j\ ,
\end{equation}
where $\eta_j$ are appropriate combinations of the
$\eta_{a_1\cdots a_n}$. Similarly,
\begin{equation}
  [g\cdot\bm{\eta}](g\cdot u) = (\w{a} u^2 -\w{b} u^1)^n \times {}^g\bm{\eta}(g\cdot\zeta) = (u^2)^n j(g,\zeta)^n \times [g\cdot\bm{\eta}](g\cdot\zeta)\ .
\end{equation}
This leads to
\begin{equation}\label{OnRsym}
  \bm{\eta}(\zeta) = j(g, \zeta)^n [g\cdot\bm{\eta}](g\cdot\zeta)\ .
\end{equation}
We define the transformation of a $\mc{O}(n)$ superfield
$\bm\eta(\zeta)$ by an element $g \in F$ as
\begin{equation}\label{Oninhomtr}
  \bm\eta(\zeta) \to \ [g^{-1}\cdot\bm\eta](\zeta) := j(g,\zeta)^n\, \bm\eta(g\cdot\zeta)\ ,
\end{equation}
where the right hand side must be expanded about $\zeta = 0$ so that
it is a function of $\zeta$ rather than $g\cdot\zeta$. Thus, an
$\mc{O}(n)$ superfield has weight $n$ (note that this is also the
degree of the line bundle $\mc{O}(n) \to \mbb{CP}^1$). Meromorphic
sections of $\mc{O}(n)$ also transform similarly under R-symmetry.

\paragraph{An example} We are primarily interested in describing
hypermultiplets which correspond to $n = 1$. In this case the
components $\eta_a$ of $\bm{\eta}(\uz, u) = \eta_{a}(\uz) u^a$
transform as an $F$-doublet. We check that $\bm\eta$ satisfies
\eqref{OnRsym} for $n =1$:
\begin{equation}
  j(g,\zeta) \times\,  [g\cdot\bm{\eta}](g\cdot \zeta) = (\w{a}\eta_1 + \w{b} \eta_2) (a\zeta + b) + (-b \eta_1 + a \eta_2) (\w{a} - \w{b} \zeta) = \eta_1 \zeta +  \eta_2 = \bm{\eta}(\zeta)\ ,
\end{equation}
where we have used the $\text{SU}(2)$ transformation of a doublet
$\eta_a$ with a lower index as described in eq.~\eqref{lowerind}. It can be easily checked that the
conjugate $\w{\bm{\eta}} = \w\eta{}^{1} - \zeta\w\eta{}^{2}$
(cf.~\eqref{extendedconj}) also transforms as an $\mc{O}(1)$
multiplet.

\paragraph{Arctic superfields}
Arctic superfields are typically defined only in a neighbourhood of
$\zeta = 0$ and not globally on $\mbb{CP}^1$. As a result, we may only
consider infinitesimal R-symmetry transformations of arctic
superfields since they retain $\zeta$ in a neighbourhood of
$\zeta=0$. These we obtain by setting $a = 1 + \i\alpha$ and
$b = \beta$, with $\alpha$ and $\beta$ infinitesimal, in the formula
for the $F$-transformation $g$ in \eqref{gmat}. The determinant
condition $a\w{a} + b \w{b} = 1$ then gives
$\i(\alpha - \w\alpha) = 0$ to first order in the infinitesimals,
i.e., $\alpha$ is real. The infinitesimal $F$-transformation of
$\zeta$ is then (cf.~\cite{Lindstrom:2009afn})
\begin{equation}\label{zetainf}
  \delta\zeta = \beta + 2\i \alpha \zeta + \w\beta\zeta^2\ .
\end{equation}
The $F$-transformation of an arctic superfield
$\bUpsilon(\zeta) = \sum_{0}^\infty \iUpsilon_j\,\zeta^j$ of
\emph{weight} $k$ is then given by the infinitesimal version of
$[g^{-1}\cdot\bUpsilon](\zeta) = j(g,\zeta)^k\times
\bUpsilon(g\cdot\zeta)$:
\begin{equation}\label{Rarctic}
  \delta\bUpsilon(\zeta) = -k(\i\alpha + \w\beta\zeta) \bUpsilon(\zeta) + \frac{\partial\bUpsilon}{\partial\zeta}\,\delta\zeta\ ,\quad k \in \mbb{Z}\ .
\end{equation}
It is important to note that arctic superfields can be assigned
any integral weight $k$ \emph{a priori} since arctics go to arctics
under infinitesimal transformations for any $k$ in
eq.~\eqref{Rarctic}\footnote{Explicitly, we have
  \begin{equation*}
  \delta\bUpsilon = [g^{-1}\cdot\bUpsilon](\zeta) - \bUpsilon(\zeta)
  = (1 - k \i \alpha - k\w\beta \zeta) \left(\bUpsilon(\zeta) + \frac{\partial\bUpsilon}{\partial\zeta} \delta\zeta\right) - \bUpsilon(\zeta) = - k(\i\alpha + \w\beta\zeta) \bUpsilon(\zeta) + \frac{\partial\bUpsilon}{\partial\zeta}\delta\zeta\ .
\end{equation*}
Clearly, the right hand side is also an arctic
superfield.}. Further, it is easy to check that
$\zeta^k \w\bUpsilon(-\zeta^{-1})$ also transforms as an object of
weight $k$ but is no longer an antarctic superfield.

The components $\iUpsilon_j$ transform under \eqref{Rarctic} as
\begin{equation}\label{arctictr}
  \delta \iUpsilon_j = (j+1)\beta \iUpsilon_{j+1} + (2j-k)\alpha \iUpsilon_j + (j - 1 - k)\w\beta \iUpsilon_{j-1}\ .
\end{equation}
% For the first few components, we have 
% \begin{align}\label{arctictr1}
%   -\delta \iUpsilon_0 &= \beta \iUpsilon_1 - k\alpha \iUpsilon_0\ ,\nonumber\\
%   -\delta \iUpsilon_1 &= 2\beta \iUpsilon_2 + (2-k)\alpha \iUpsilon_1 - k\w\beta \iUpsilon_0\ ,\nonumber\\ 
%   -\delta \iUpsilon_2 &= 3\beta \iUpsilon_3 + (4-k)\alpha \iUpsilon_2 - (k - 1)\w\beta \iUpsilon_1\ ,\nonumber\\ \vdots \nonumber\\
%   -\delta \iUpsilon_k &= (k+1)\beta \iUpsilon_{k+1} + k \alpha \iUpsilon_k - \w\beta \iUpsilon_{k-1}\ ,\nonumber\\
%   \vdots
% \end{align}
Let us look at $k = 1$ which will be required in our study of
hypermultiplets. We shall show below that, with our choice of action
for the arctic superfield, the components $\iUpsilon_{j \geq 2}$ will
turn out to be auxiliary and will be set to zero by their equations of
motion. The arctic superfield then truncates to an $\mc{O}(1)$
superfield after substituting $\iUpsilon_{j\geq 2} = 0$. It is then
clear that the components $\iUpsilon_0$ and $\iUpsilon_1$ decouple
from the $\iUpsilon_{j \geq 2}$ components in \eqref{arctictr} and
$\iUpsilon_0$ and $\iUpsilon_1$ transform as
\begin{equation}
  \delta \iUpsilon_0 = -\alpha \iUpsilon_0 + \beta \iUpsilon_1\ ,\quad \delta \iUpsilon_1 = \alpha \iUpsilon_1 - \w\beta \iUpsilon_0\ .
\end{equation}
These are the transformation rules for an $F$-doublet
$(\iUpsilon_1, \iUpsilon_0)$ and this is the standard transformation
of a hypermultiplet under $\text{SU}(2)$ R-symmetry.

\paragraph{The derivatives $\pD_{a'+}$,$\wt\pD_{a'+}$}
Since $\pD_{a'+} = u^a \uD_{aa'+}$, the same manipulations we did for
$\mc{O}(n)$ superfields works here and it follows from \eqref{OnRsym}
that $\pD_{a'+}$ has $F$-weight $+1$. Let us next discuss the
$F$-weight of $\wt\pD_{a'+}$. Recall from the discussion above
equation \eqref{derdef} that our chosen solution for the equation
$\varepsilon_{ab} \wt{u}^a u^b = 1$ is
\begin{equation}\label{utlsol}
  \wt{u}^a = (u^2)^{-1} \pmat{1 \\ 0}\ ,\quad\text{given}\quad u^a =
  \pmat{u^1 \\ u^2} = u^2 \pmat{\zeta \\ 1}\ .
\end{equation}
Under $F$-transformations, since $u^2$ transforms as
$u^2 \to j(g,\zeta) u^2$, $u^a$ and $\wt{u}^a$ transform as
\begin{equation}\label{utltr}
  \wt{u}^a \to j(g,\zeta)^{-1} (u^2)^{-1} \pmat{1 \\ 0}\ ,\quad u^a \to j(g,\zeta) u^2 \pmat{\frac{a\zeta+b}{\w{a}-\w{b}\zeta} \\ 1}\ .
\end{equation}
From this, it is clear that $\wt\pD_{a'+}$ has $F$-weight $-1$. This
is consistent with the algebra
$\{\pD_{a'+}\,, \wt\pD_{b'+}\} = -2\i\varepsilon_{a'b'} \partial_{++}$
since the right hand side is independent of $\zeta$ and hence, has
weight $0$.

However, the transformation \eqref{utltr} of $\wt{u}^a$ does not look
like that of an $F$-doublet. The latter looks like
\begin{equation}\label{Futltr}
  \wt{u}^a \to \pmat{a (u^2)^{-1} \\ -\w{b} (u^2)^{-1}}\ .
\end{equation}
How do we reconcile \eqref{utltr} and \eqref{Futltr}? Recall that we
had a shift symmetry \eqref{shiftutl} $\delta\wt{u}^a =\omega u^a$ in
the space of $\wt{u}^a$ that satisfy
$\varepsilon_{ab}\wt{u}^a u^b = 1$. We could add a shift in one of the
transformations, say \eqref{utltr} and see if that can be matched with
\eqref{Futltr} for a particular value of the shift parameter. Indeed, writing
\begin{equation}
  j(g,\zeta)^{-1} (u^2)^{-1} \pmat{1 \\ 0} + \omega j(g,\zeta) u_2 \pmat{\frac{a\zeta + b}{\w{a} - \w{b}\zeta} \\ 1} =  (u^2)^{-1} \pmat{a \\ -\w{b}} \ ,
\end{equation}
we get a solution for $\omega$
\begin{equation}\label{omegasol}
  \omega = -\w{b} (u^2)^{-2} j(g,\zeta)^{-1}\ .
\end{equation}
In analogy with \eqref{Oninhomtr}, we define the transformations
of the $\pD_{a'+}$ and $\wt\pD_{a'+}$ expressed in inhomogeneous coordinates as
\begin{equation}\label{inhomtr}
  \pD_{a'+}(\zeta) \to j(g,\zeta) \pD_{a'+}(g\cdot\zeta)\ ,\quad \wt\pD_{a'+}(\zeta) \to j(g,\zeta)^{-1} \wt\pD_{a'+}(g\cdot\zeta) -\w{b}\, \pD_{a'+}(g\cdot\zeta)\ .
\end{equation}

\paragraph{The projective superspace measure}

Recall that the $(0,4)$ projective superspace action \eqref{actdef}
is
\begin{equation}\label{actdef1}
  \mc{S}[\bm{\Phi}] = \int\ud^2x \frac{1}{2\pi\i}\oint_\gamma \varepsilon_{ab} u^a \ud u^b\ \wt\pD_{1'+}\wt\pD_{2'+}\, \mK_{--}(\bm{\Phi})\ .
\end{equation}
As discussed after \eqref{actdef}, the action is manifestly $F$ and
$F'$ invariant provided the superspace Lagrangian $\mK_{--}$ is
invariant. In terms of $F$-weight, it has weight $0$ since the measure
$u^a \ud u^b$ has two factors of $u^a$ and $\wt\pD_{1'+}\wt\pD_{2'+}$
has two factors of $\wt{u}^a$. Let us elaborate in terms of
inhomogeneous coordinates. The action takes the form
\begin{align}\label{localact}
  \mc{S}[\bm\Phi] &= \int \ud^2x \oint_\gamma \frac{\ud\zeta}{2\pi\i} \wt\pD_{1'+} \wt\pD_{2'+}\, \mK_{--}(\bm\Phi)\ .
\end{align} 
The measure $\ud\zeta$ transforms with $F$-weight $2$ under an
$F$-transformation $\zeta \to g\cdot\zeta$:
\begin{equation}
  \ud\zeta = j(g,\zeta)^2 \, \ud(g\cdot\zeta)\ .
\end{equation}
The superderivatives $\wt\pD_{a'+}$ effectively transform with
$F$-weight $-1$ (cf.~the first term in the transformation of
$\wt\pD_{a'+}$ in \eqref{inhomtr}; the second term in \eqref{inhomtr}
is proportional to $\pD_{a'+}$ which annihilates $\mK_{--}$). As a
result, the combination $\ud\zeta\, \wt\pD_{1'+}\wt\pD_{2'+}$ has
weight $0$, i.e., the superspace measure is invariant (up to total
derivatives). Since integrating a weight $0$ object with the invariant
measure yields an $F$-invariant answer, the action is R-symmetric if
the superspace Lagrangian $\mK_{--}$ has weight $0$.

\subsection{Extended complex conjugation}\label{extconjdef}
Recall the $\mc{S}$ map \eqref{Smap} which takes $\zeta \to
-1/\zeta$. The \emph{antipodal} map $\mc{I}$ that takes a point on
$\mbb{CP}^1$ to its antipode is the composition of the $\mc{S}$ map
and complex conjugation:
\begin{equation}\label{antip}
\mc{I}:\  \pmat{u^1 \\ u^2} \longmapsto \pmat{-\w{(u^2)} \\ \w{(u^1)}}\ ,\quad \text{that is}\quad \mc{I}:\ \zeta \longmapsto -\frac{1}{\w\zeta}\ .
\end{equation}
The antipodal map can be used to define a new real structure
\cite{Lindstrom:1987ks} on the (sheaf of) sections of a line bundle as
the action of the antipodal map on a section followed by ordinary
complex conjugation of the resulting section.

For instance, the antipodal map acts on an arctic superfield
$\bUpsilon(\zeta) = \sum_{j\geq 0} \iUpsilon_j \zeta^j$ (which is a
local section of some line bundle on $\mbb{CP}^1$) as
\begin{equation}
  \sum_{j\geq 0} \iUpsilon_j \zeta^j \to   \sum_{j\geq 0} \iUpsilon_j (-\w\zeta)^{-j}\ .
\end{equation}
Ordinary complex conjugation of the resulting local section is
\begin{equation}
  \sum_{j\geq 0} \iUpsilon_j (-\w\zeta)^{-j} \to  \sum_{j\geq 0} \w\iUpsilon_j (-\zeta)^{-j}\ .
\end{equation}
Thus, the extended complex conjugate of an arctic superfield
$\bUpsilon(\zeta)$ is
\begin{equation}\label{arcticcc}
  \w\bUpsilon(-1/\zeta) := \sum_{j \geq 0} \w\iUpsilon_j (-1/\zeta)^j\ .
\end{equation}
Let us compute the extended complex conjugate of an $\mc{O}(p)$
superfield $\bm\eta$. Since $\bm\eta$ is globally defined on
$\mbb{CP}^1$, and the antipodal map contains the $\mc{S}$ map, we can
use the $F$-transformation rule \eqref{Oninhomtr} for $\mc{O}(p)$
superfields to obtain the extended complex conjugate:
\begin{equation}
\bm\eta(\uz,\zeta) =  \sum_{j=0}^p \eta_j(\uz)  \zeta^{j}\quad \substack{\mc{I} \\ \displaystyle \longmapsto}  \quad (-\w\zeta)^p \sum_{j=0}^p \eta_j(\uz)  (-\w\zeta)^{-j}\quad \substack{\text{c.c.} \\ \displaystyle \longmapsto}\quad  \sum_{j=0}^p\w\eta_j(\uz)  (-\zeta)^{p-j}\ .
\end{equation}
The difference between the above and \eqref{arcticcc} is that there is
an additional factor of $(-\w\zeta)^p$ in the antipodal map step. This
factor makes the new section also a global section of
$\mc{O}(p)$. Thus, the extended complex conjugate of an $\mc{O}(p)$
superfield $\bm\eta$ is
\begin{equation}\label{extendedconj}
  \w{\bm{\eta}}(\uz, \zeta) :=  \sum_{j=0}^p\w\eta_j(\uz)  (-\zeta)^{p-j}\ .
\end{equation}
\paragraph{A reality condition} As is obvious from
\eqref{extendedconj}, the extended conjugate of an $\mc{O}(p)$
superfield is also an $\mc{O}(p)$ superfield. Notice that applying the
extended complex conjugate twice on $\bm{\eta}$ gives
\begin{equation}\label{conj2}
  \w{\w{\bm{\eta}}} = (-1)^p \bm{\eta}\ .
\end{equation}
Thus, we can impose a \emph{reality} condition on an $\mc{O}(p)$
superfield only when $p$ is even:
\begin{equation}\label{conj1}
  \bm{\eta}(\zeta) =  \w{\bm{\eta}}(\zeta)\ ,\quad\text{that is,}\quad \sum_{j=0}^p \eta_j(\uz) \zeta^j = \sum_{j=0}^p \w\eta_j(\uz) (-\zeta)^{p-j}\ .
\end{equation}
\paragraph{Extended complex conjugates of $\pD_{a'+}$ and
  $\wt\pD_{a'+}$} Next, consider the derivatives $\pD_{a'+}$ and
$\wt\pD_{a'+}$. Since they are globally defined (see the equations
\eqref{O1pd}, \eqref{affinetr} and the discussion around them), we use
the global $F$-transformation rules in \eqref{inhomtr} to get the
conjugates. The factor of automorphy for the $\mc{S}$-map is
$j(\mc{S},\zeta) = -\zeta$. The complex conjugate of $\pD_{a'+}$ is
then
\begin{multline}\label{antipd}
  \pD_{a'+}\quad \substack{{\mc{I}} \\ \displaystyle \longmapsto}\quad -\w\zeta (-\w\zeta{}^{-1} \uD_{1a'+} + \uD_{2a'+})\quad \substack{ {\rm c.c.} \\ \displaystyle \longmapsto}\quad -\zeta (-\zeta^{-1} \w\uD_{1a'+} + \w\uD_{2a'+}) \\ = \varepsilon^{a'b'} (\uD_{2b'+} + \zeta \uD_{1b'+}) = \varepsilon^{a'b'} \pD_{b'+}\ .
\end{multline}
The complex conjugate of $\wt\pD_{a'+}$ is obtained as follows. First,
we apply the antipodal map:
\begin{equation}\label{antipdtl}
  \wt\pD_{a'+}\quad \substack{{\mc{I}} \\ \displaystyle \longmapsto} \quad (-\w\zeta)^{-1} \wt\pD_{a'+}(-\w\zeta{}^{-1}) - \pD_{a'+}(-\w\zeta{}^{-1}) = (-\w\zeta)^{-1} \uD_{1a'+} - (-\w\zeta{}^{-1} \uD_{1a'+} + \uD_{2a'+}) = -\uD_{2a'+}\ . %\quad \substack{ {\rm c.c.} \\ \displaystyle \longmapsto}  \quad \varepsilon^{a'b'} \uD_{1b'+} = \varepsilon^{a'b'}  \wt\pD_{b'+}\ .
\end{equation}
where we have used the fact that since the $\wt\pD_{a'+}$ are
independent of $\zeta$, the expressions for
$\wt\pD_{a'+}(-\w\zeta{}^{-1})$ are the same as in \eqref{derdef}, 
i.e., $\wt\pD_{a'+}(-\w\zeta{}^{-1}) = \uD_{1a'+}$. Next, doing
ordinary complex conjugation, we get
\begin{equation}\label{antipdtl1}
- \uD_{2a'+} \quad \substack{ {\rm c.c.} \\ \displaystyle \longmapsto}  \quad \varepsilon^{a'b'} \uD_{1b'+} = \varepsilon^{a'b'} \wt\pD_{b'+}\ .%  \wt\pD_{a'+}\quad \substack{{\mc{I}} \\ \displaystyle \longmapsto} \quad 
\end{equation}
Thus, we have
\begin{equation}\label{derconj}
  \w{\pD}{}^{a'}_+ = \varepsilon^{a'b'}\pD_{b'+}\ ,\quad   \w{\wt\pD}{}^{a'}_+ = \varepsilon^{a'b'}\wt\pD_{b'+}\ .
\end{equation}
We may need to consider a slightly different version of the complex
conjugates of the derivatives when they act on arctic superfields for
the following reason. Under extended complex conjugation, an arctic
superfield goes to an antarctic superfield (see \eqref{arcticcc}). We
would like this to be true for the derivative of an arctic as
well. However, applying \eqref{derconj} on $\pD_{a'+}\bUpsilon$ gives
$\varepsilon^{a'b'} \pD_{b'+}\w\bUpsilon$ which is not antarctic due
to a term proportional to $\zeta$ in $\pD_{b'+}$. On the other hand,
treating $\pD_{a'+}\bUpsilon$ as a new arctic superfield with
components
\begin{equation}
  \pD_{a'+}\bUpsilon(\zeta) = \sum_{j\geq 0} (\zeta \uD_{1a'+} + \uD_{2a'+})\iUpsilon_j \zeta^j = \sum_{j \geq 0} \zeta^j (\uD_{2a'+}\iUpsilon_j + \uD_{1a'+} \iUpsilon_{j-1})\ ,
\end{equation}
we can apply the conjugation rule \eqref{arcticcc} to the above and
obtain
\begin{equation}\label{Darcticconj1}
 \sum_{j \geq 0} (-1/\zeta)^j \varepsilon^{a'b'} (-\uD_{1b'+}\w\iUpsilon_j + \uD_{2a'+} \w\iUpsilon_{j-1})\ ,
\end{equation}
as the conjugate antarctic superfield corresponding to
$\pD_{a'+}\bUpsilon$. Clearly, \eqref{Darcticconj1} can be written as
\begin{equation}
    \varepsilon^{a'b'}(-\uD_{1b'+} - \zeta^{-1} \uD_{2b'+})\sum_{j \geq 0} (-1/\zeta)^j \w\iUpsilon_j =  -\zeta^{-1}\varepsilon^{a'b'} \pD_{b'+} \w\bUpsilon(-1/\zeta)\ ,
\end{equation}
which suggests that we modify the conjugate of the derivative
$\pD_{a'+}$ when acting on arctic superfields to
\begin{equation}
  \pD_{a'+} \to \breve\pD{}^{a'+} = -\zeta^{-1} \varepsilon^{a'b'} \pD_{b'+}\ .
\end{equation}
Similarly, we have
\begin{equation}
  \wt\pD_{a'+}\bUpsilon(\zeta) = \sum_{j\geq 0} \zeta^j \uD_{1a'+} \iUpsilon_j\ .
\end{equation}
Applying \eqref{arcticcc} to the above, we get
\begin{equation}\label{Dtlarcticconj2}
  \sum_{j\geq 0} (-1/\zeta)^j \varepsilon^{a'b'}\uD_{2b'+} \w\iUpsilon_j\ .
\end{equation}
Note the identity
\begin{equation}
  \uD_{2b'+} = -\zeta\uD_{1b'+} + \pD_{b'+} = -\zeta \wt\pD_{b'+} + \pD_{b'+}\ .
\end{equation}
This allows us to write \eqref{Dtlarcticconj2} as
\begin{equation}\label{Dtlarcticconj}
  \sum_{j\geq 0} (-1/\zeta)^j \varepsilon^{a'b'}\uD_{2b'+} \w\iUpsilon_j = \varepsilon^{a'b'}(-\zeta\wt\pD_{b'+} +\pD_{b'+}) \w\bUpsilon(-1/\zeta)\ ,
\end{equation}
which suggests the modification
\begin{equation}
  \wt\pD_{a'+} \to \breve{\wt\pD}{}^{a'+} = \varepsilon^{a'b'}(-\zeta \wt\pD_{b'+} + \pD_{b'+})\ .
\end{equation}
Thus, on arctic superfields, we can postulate the following modified extended
complex conjugates of the derivatives:
\begin{equation}\label{derconjarctic}
  \breve{\pD}{}^{a'}_+ = -\zeta^{-1} \varepsilon^{a'b'} \pD_{b'+}\ ,\quad \breve{\wt\pD}{}^{a'}_+ =  \varepsilon^{a'b'} (-\zeta \wt\pD_{b'+} + \pD_{b'+})\ .
\end{equation}
The notation $\breve{\ }$ for the above notion of the extended complex
conjugate of a derivative has been used earlier in
\cite{Kuzenko:1998xm} and has been called `smile conjugation'; we
continue to use the same notation in this paper. Note that the smile
conjugation simply treats $\pD_{a'+}(\zeta)$ and $\wt\pD_{a'+}(\zeta)$
as local sections and applies the conjugation rule \eqref{arcticcc}.

\section{Hypermultiplets} \label{hypermult}

The dynamical degrees of freedom of a $(0,4)$ hypermultiplet consists
of two $(0,2)$ chiral superfields $\phi$ and $\chi$ such that
$(\phi,\w\chi)$ form an $\text{SU}(2)$ doublet. The $\text{SU}(2)$ in
question can be either $F$ or $F'$ and the corresponding hypers are
called standard and twisted hypermultiplets respectively. A standard
hypermultiplet\footnote{See also \cite{Gates:1994bu} for a discussion
  in ordinary superspace.} can be described in $(0,4)$ projective
superspace either by an $\mc{O}(1)$ superfield \cite{Hull:2017hfa} or
by a pair of $F$-arctic superfields $(\bUpsilon, \bUpsilon_{--})$. The
analogous notation for the twisted hypers is $\mc{O}(1)'$ and
$F'$-arctic respectively. We describe free hypermultiplets in this
section and study interactions in Section \ref{interaction}.

\subsection{Standard hypermultiplets}\label{stdhyp}

%\paragraph{$\mc{O}(1)$ superfield}
\subsubsection{\tp{$\mc{O}(1)$}{O(1)} superfield}
We start with a complex $\mc{O}(1)$ superfield $\bm\eta = \eta_a
u^a$. In terms of the inhomogeneous coordinate $\zeta$, we have
$u^a = (\zeta, 1)$ and
\begin{equation}\label{O1zeta}
  \bm{\eta}(\zeta) =  \eta_{2} + \zeta\eta_{1}\ ,\quad \w{\bm{\eta}}(\zeta) = \w\eta{}^1 - \zeta \w\eta{}^2\ .
\end{equation}
The projective constraints $\pD_{a'+} \bm{\eta} = 0$ give the
following constraints on $\eta_1$ and $\eta_2$:
\begin{align}\label{etaprojconst}
  &\w\uQ_{+} \eta_{1} = 0\ ,\quad \uD_{+}\eta_{1} = 0\ ,\quad \w{\uD}_{+}\eta_{2} = 0\ ,\quad\uQ_{+} \eta_{2} = 0\ ,\quad \w\uQ_{+} \eta_{2} = \w\uD_{+} \eta_{1}\ ,\quad \uQ_+ \eta_{1} = -\uD_+ \eta_{2}\ .
\end{align}
We see that $\w\eta{}^{1}$ and $\eta_{2}$ are $(0,2)$ chiral
superfields since $\w{\uD}_+$ annihilates them. See Appendix
\ref{02superspace} for a review of $(0,2)$ superspace.

The superpartner fermions are defined as\footnote{The conjugate
  fermions are obtained as follows. $\wt\pD_{a'+}\bm\eta$ is best thought
  of as $[\wt\pD_{a'+}\, , \bm\eta]$ which, under conjugation, goes to
  $[\w{\bm\eta}\, , \w{(\wt\pD_{{\emph a}'+})}] = -\varepsilon^{a'c'} \wt\pD_{c'+} \w{\bm\eta}$.}
\begin{equation}\label{supfer}
  \sqrt{2}\xi_{a'+} := \wt\pD_{a'+}\bm\eta\ ,\quad \sqrt{2}\w\xi{}^{a'}_+ := - \varepsilon^{a'b'} \wt\pD_{b'+} \w{\bm\eta}\ .
\end{equation}
The superpartners $\xi_{a'+}$ are in the doublet of $F'$; they are
also independent of $\zeta$ since the above combinations are globally
defined weight $0$ superfields, i.e., global holomorphic functions on
$\mbb{CP}^1$ which are indeed constants in $\zeta$. Using the
expressions $\wt\pD_{a'+} = \uD_{1a'+}$ and that
$\pD_{a'+}\bm\eta = 0$, we can arrive at the following $(0,2)$
superspace definitions for the $\xi_{a'+}$:
\begin{equation}\label{supfer02}
 \sqrt{2} \xi_{1'+} = \uD_+ \eta_2 \ ,\quad \sqrt{2} \w\xi{}^{1'}_+ = -\w\uD_+\w\eta{}^2\ ,\quad  -\sqrt{2} \xi_{2'+} = \w\uD_+ \eta_1\ ,\quad \sqrt{2} \w\xi{}^{2'}_+ = \uD_+ \w\eta{}^1\ .
\end{equation}
The next superfield in the multiplet would be
$\wt\pD_{a'+}\wt\pD_{b'+}\bm\eta$ which (1) is globally defined on
$\mbb{CP}^1$, (2) has $F$-weight $-1$, (3) is antisymmetric in $a'b'$,
and (4) is a Lorentz vector. The only superfield which satisfies all these
properties is $\varepsilon_{a'b'}\partial_{++}\wt{\bm\eta}$, where
$\wt{\bm\eta} = \wt{u}^a \eta_a$. Thus, we have
\begin{equation}\label{etalevel2}
  \wt\pD_{a'+}\wt\pD_{b'+}\bm\eta = -2\i\varepsilon_{a'b'}\partial_{++}\wt{\bm\eta}\ ,\quad   \wt\pD_{a'+}\wt\pD_{b'+}\w{\bm\eta} = -2\i\varepsilon_{a'b'}\partial_{++}\w{\wt{\bm\eta}}\ .
\end{equation} 
The above equations \eqref{etalevel2} can be explicitly checked by
using the expressions for $\wt\pD_{a'+}$ in \eqref{derdef}, the
complex conjugate derivatives in \eqref{derconj}, and the projective
constraints \eqref{etaprojconst}.

The $(0,4)$ supersymmetric action that describes the (free)
hypermultiplet is
\begin{align}\label{O1projact}
  \mc{S} &= -\frac{\i}{2}\int\ud^2x\oint_{\gamma}\frac{\ud\zeta}{2\pi \i}\wt\pD_{1'+}\wt\pD_{2'+}\, (\zeta^{-1}\w{\bm{\eta}}\partial_{--}\bm{\eta})\ .
\end{align}
Using the fact that the superspace Lagrangian is annihilated by
$\pD_{a'+}$, we can write it as an action in $(0,2)$ superspace as in
\eqref{04action}. We get
\begin{align}\label{O1projact02}
  \mc{S} &= \frac{\i}{2}\int\ud^2x\oint_{\gamma}\frac{\ud\zeta}{2\pi \i\zeta}\uD_+\w\uD_+\, (\zeta^{-1}\w{\bm{\eta}}\partial_{--}\bm{\eta})\ .
\end{align}
Next, we can obtain the component action by first performing the
$\zeta$-integral, pushing in the derivatives and using the definitions
\eqref{supfer02} and that $\w\eta{}^1$ and $\eta_2$ are $(0,2)$ chiral
superfields:
\begin{align}\label{04to02}
  \mc{S} &=  \frac{\i}{2}\uD_+\w\uD_+\int\ud^2x\oint_{\gamma}\frac{\ud\zeta}{2\pi \i\zeta} \big((\zeta^{-1}\w\eta{}^1 - \w\eta{}^2)\partial_{--}(\zeta\eta_1 + \eta_2)\big)\ ,\nonumber\\
         &=  \frac{\i}{2}\int\ud^2x\,\uD_+\w\uD_+\, (\w\eta{}^1\partial_{--}\eta_1 - \w\eta{}^2\partial_{--}\eta_2)\ ,\nonumber\\
         &= \int\ud^2x\, (-\partial_\mu\w\eta{}^{a}\partial^\mu \eta_{a} - \i\w\xi{}^{\,a'}_{+}\partial_{--}\xi_{a'+})\ .
\end{align}
(See Appendix \ref{02superspace} for a derivation of the component
action from the $(0,2)$ action in the second line in \eqref{04to02}.)
We can also obtain the same component action as above by pushing in
the derivatives $\wt\pD_{1'+}\wt\pD_{2'+}$ in \eqref{O1projact}, use
the definitions \eqref{supfer} and \eqref{etalevel2}, and finally
perform the $\zeta$ integral (see Appendix \ref{0400}).

The $\mc{O}(1)$ superfield can be described in ordinary $(0,4)$
superspace as well. Writing $\pD_{a'+} = u^a \uD_{aa'+}$ and
$\bm\eta = u^a\eta_a$, the projective constraints
$\pD_{a'+} \bm\eta = 0$ are equivalent to
\begin{equation}
  \uD_{aa'+} \eta_b + \uD_{ba'+} \eta_a = 0\ .
\end{equation}
As noted in \cite{Hull:2017hfa}, in contrast to an $\mc{O}(1)$
superfield in $(4,4)$ projective superspace, the above $(0,4)$
constraints do not put the $\mc{O}(1)$ superfield on-shell. Only the
antisymmetric part in $ab$ of $D_{aa'+}\eta_b$ is non-zero and it
gives the superpartner fermions defined in \eqref{supfer} (or
equivalently \eqref{supfer02}):
\begin{equation}
  \uD_{aa'+} \eta_b =: \sqrt{2} \varepsilon_{ab} \xi_{a'+}\ ,\quad \sqrt{2}  \varepsilon^{ab}  \w\xi{}^{a'}_{+} = -\varepsilon^{ac} \varepsilon^{a'c'} \uD_{cc'+} \w\eta{}^b\ .
\end{equation}
Note that the scalars $\eta_a$ are in an $F$-doublet whereas the
fermions $\xi_{a'+}$ are in an $F'$-doublet.

Recall from \eqref{etaprojconst} that $\w\eta{}^1$ and $\eta_2$ are
annihilated by $\w\uD_+$ and $\uQ_+$. Thus, we can write down a
manifestly $(0,4)$ supersymmetric action with the measure
$\uD_+\w\uQ_+$:
\begin{align}\label{O1chiralact}
  \mc{S} &= \frac{\i}{2}\int\ud^2x\,\uD_+\w\uQ_+\, (\w\eta{}^1\partial_{--}\eta_2)\ .
\end{align}
This is the projective superspace action \eqref{O1projact} after
plugging in $\wt\pD_{1'+} = \uD_+$, $\wt\pD_{2'+} = -\w\uQ_+$ and
performing the $\zeta$ integral; therefore, it also coincides with the
$(0,2)$ action \eqref{O1projact02}. The above action is not manifestly
R-symmetric, but a manifestly R-symmetric action also exists which
agrees with any of the above actions (up to total spacetime
derivatives):
\begin{align}
  \mc{S} &= \frac{\i}{2}\int\ud^2x\,\varepsilon^{a'b'}\uD_{aa'+}\uD_{bb'+}\,(\w\eta{}^a \partial_{--} \varepsilon^{bc}\eta_{c})\ .
\end{align}
However, the above action is not manifestly supersymmetric since the
measure does not involve all four superspace derivatives.

The $F$-projective superspace action \eqref{O1projact} does not seem
to be invariant under R-symmetry since the Lagrangian does not seem to
transform with $F$-weight $0$. To write a manifestly R-symmetric
action in projective superspace, we use the arctic realization of the
hypermultiplet, one that arises naturally from $(4,4)$ projective
superspace (see Appendix \ref{44stdhyp}).

%\paragraph{Arctic superfield}
\subsubsection{Arctic superfield}
Consider two arctic multiplets $\bUpsilon$ and $\bUpsilon_{--}$ with
$\zeta$-expansions
\begin{equation}
  \bUpsilon(\zeta) = \sum_{j=0}^\infty \iUpsilon_{j}\zeta^j\ ,\quad  \bUpsilon_{--}(\zeta) = \sum_{j=0}^\infty \iUpsilon_{j--}\zeta^j\ .
\end{equation}
The projective constraints $\pD_{a'+} \bUpsilon = 0$ give
\begin{align}\label{arcticprojconst}
  &\uQ_{+} \iUpsilon_0 = 0\ ,\quad \w{\uD}_{+} \iUpsilon_0 = 0\ ,\quad\uQ_{+} \iUpsilon_{j+1} = -\uD_{+} \iUpsilon_{j}\ ,\quad \w{\uQ}_+ \iUpsilon_{j} = \w{\uD}_+ \iUpsilon_{j+1}\quad\text{for}\quad j \geq 0\ ,
\end{align}
and similarly for $\bUpsilon_{--}$. The zeroth components
$\iUpsilon_0$ and $\iUpsilon_{0--}$ are $(0,2)$ chiral superfields
since $\w{\uD}_+ \iUpsilon_0 = \w\uD_+ \iUpsilon_{0--} = 0$ whereas
the $\iUpsilon_{j}$, $\iUpsilon_{j--}$, $j \geq 1$, are unconstrained
as $(0,2)$ superfields.

The $(0,4)$ supersymmetric action that describes the
(free) standard hypermultiplet is
\begin{align}\label{freesthyper}
  \mc{S} &= \int\ud^2x\oint_{\gamma}\frac{\ud\zeta}{2\pi \i}\wt\pD_{1'+}\wt\pD_{2'+}\, (\tfrac{\i}{2}\w{\bUpsilon}\partial_{--}\bUpsilon - \zeta \w{\bUpsilon} \bUpsilon_{--} + \zeta^{-1} \w{\bUpsilon}_{--} \bUpsilon)\ .
\end{align}
In fact, the above action is equivalent to that of an $\mc{O}(1)$
superfield when we go partially on-shell by performing the
$\zeta$-integral in the last two terms and integrating out the fields
$\iUpsilon_{j--}$ for $j \geq 1$:
\begin{align}\label{04lag}
&-\uD_+\w\uD_+\oint_\gamma\frac{\ud\zeta}{2\pi \i\zeta}\left(- \zeta\w{\bUpsilon}\bUpsilon_{--} + \zeta^{-1}\w{\bUpsilon}_{--}\bUpsilon\right)\ ,\nonumber\\
  &=-\uD_+\w\uD_+\left(\w{\iUpsilon}_1 \iUpsilon_{0--} + \w{\iUpsilon}_{0--} \iUpsilon_1 + \sum_{j=1}^\infty(-1)^{j+1}\left( - \w{\iUpsilon}_{j+1} \iUpsilon_{j--} - \w{\iUpsilon}_{j--} \iUpsilon_{j+1}\right)\right)\ .
\end{align}
Since the $\iUpsilon_{j--}$, $j \geq 1$, are unconstrained as $(0,2)$
superfields, we can integrate them out in the above superspace
action. This imposes $\w{\iUpsilon}_{j+1} = \iUpsilon_{j+1} = 0$
for $j \geq 1$ and retains only the $\zeta^0$ and $\zeta^1$ terms in
$\bUpsilon$. Integrating out $\iUpsilon_{j+1}$, we get
$\iUpsilon_{j--} = \frac{\i}{2}\partial_{--} \iUpsilon_{j+1}$ for
$j\geq 1$. We cannot integrate out $\iUpsilon_{0--}$ in the same way
and set $\iUpsilon_1 = 0$ since $\iUpsilon_{0--}$ is constrained as a
$(0,2)$ superfield, $\w{\uD}_+\iUpsilon_{0--} = 0$. Instead,
integrating out the constrained superfield $\iUpsilon_{0--}$
constrains $\iUpsilon_1$ to satisfy
$\w\uD_+ \w\iUpsilon_1 = 0$.\footnote{Here is the procedure to
  integrate out a constrained superfield: we first relax the
  constraint on $\iUpsilon_{0--}$ and introduce a Lagrange multiplier
  superfield $\Lambda_-$:
  $-\uD_+\w\uD_+\left( \w{\iUpsilon}_1 \iUpsilon_{0--} + \Lambda_-
    (\w\uD_+ \iUpsilon_{0--})\right)$. Integrating out $\Lambda_{-}$
  re-imposes the constraint $\w\uD_+ \iUpsilon_{0--} = 0$ whereas
  integrating out $\iUpsilon_{0--}$ gives
  $\w{\iUpsilon}_1 = - \w\uD_+\Lambda_-$, which indeed satisfies
  $\w\uD_+\w{\iUpsilon}_1 = 0$. We can conclude the same by going down
  to components, or at an intermediate stage by pushing in $\w\uD_+$
  in the first term in the Lagrangian \eqref{04lag} to get
  $-\uD_+ \left((\w\uD_+ \w{\iUpsilon}_1)
    \iUpsilon_{0--}\right)$. Since the remaining measure $\uD_+$ does
  not kill $\iUpsilon_{0--}$, we can integrate it out to conclude that
  $\w\uD_+ \w{\iUpsilon}_1 = 0$.\label{X0loose}}

Thus, we have two $(0,2)$ chiral superfields $\iUpsilon_0$ and
$\w{\iUpsilon}_1$ which we relabel as $\eta_2$ and $\w\eta{}^{1}$
respectively to make contact with the $\mc{O}(1)$ superfield
terminology \eqref{O1zeta}. Thus, $\bUpsilon$ becomes an $\mc{O}(1)$
superfield when we go partially on-shell by integrating out the
auxiliary superfield $\bUpsilon_{--}$:
\begin{equation}\label{partonshell}
  \bUpsilon = \iUpsilon_0 + \zeta \iUpsilon_1 =  \zeta \eta_1 + \eta_2\ ,\quad \w{\bUpsilon} = \w\eta{}^2 - \zeta^{-1} \w\eta{}^1 = -\zeta^{-1} (\w\eta{}^1 - \zeta \w\eta{}^2)\ ,
\end{equation}
and the action \eqref{freesthyper} becomes the $\mc{O}(1)$ action
\eqref{04to02}:
\begin{align}\label{stdhypcomp}
  \mc{S} &= \frac{\i}{2}\int\ud^2 x\,\uD_+\w\uD_+\left(\w\eta{}^1\partial_{--}\eta_1 - \w\eta{}^2\partial_{--}\eta_2\right) = \int\ud^2x\, (-\partial_\mu\w\eta{}^{a}\partial^\mu \eta_{a} - \i\w\xi{}^{\,a'}_{+}\partial_{--}\xi_{a'+})\ .
\end{align}

Since integrating out $\bUpsilon_{--}$ gives an $\mc{O}(1)$
superfield, it is consistent to give an $F$-weight of $+1$ to
$\bUpsilon$. However, the action does not seem to have $F$-weight $0$
and hence does not appear R-symmetric. But the action in ordinary
space \eqref{stdhypcomp} is certainly R-symmetric! Let us see how to
understand the R-symmetry of \eqref{freesthyper}.

The terms depending on $\bUpsilon_{--}$ can be made to have weight $0$
by declaring that $\bUpsilon_{--}$ is a weight $-1$
superfield. However, the kinetic term is still a problem. Since
$\bUpsilon_{--}$ is an auxiliary superfield, we can give it a
non-standard R-symmetry transformation so that it cancels that of the
kinetic term (this is motivated from the $(4,4) \to (0,4)$ reduction
in Appendix \ref{44stdhyp}):
\begin{equation}\label{ummmodtr}
  \bUpsilon_{--}(\zeta) \to j(g,\zeta)^{-1} \bUpsilon_{--}(g\cdot\zeta) - \tfrac{\i}{2}\w{b} \partial_{--}  \bUpsilon(g\cdot\zeta)\ ,\quad\text{where}\quad g = \pmat{a & b \\ -\w{b} & \w{a}} \in F\ .
\end{equation}
Recall that we must only perform infinitesimal $F$-transformations on
arctic superfields (see the discussion above eq.~\eqref{zetainf}). It
is easy to check that the Lagrangian \eqref{freesthyper} transforms
with weight zero when we transform $\bUpsilon_{--}$ according to the
above rule (see Appendix \ref{44stdhyp} for an explicit
demonstration).

We could write down the $(0,4)$ descendants directly by acting on
$\bUpsilon$ and $\bUpsilon_{--}$ with the derivatives
$\wt\pD_{a'+}$. We could then compute the component action
\eqref{stdhypcomp} by pushing the derivatives in the measure
$\wt\pD_{1'+}\wt\pD_{2'+}$ into the Lagrangian in the $(0,4)$ action
\eqref{freesthyper} and using the definition of the $(0,4)$
descendants. This procedure results in the same conclusions, namely
that $\bUpsilon$ is truncated to an $\mc{O}(1)$ superfield and
$\bUpsilon_{--}$ is auxiliary, and hence we do not describe it
here. However, see Appendix \ref{0400} for an illustration of this
method for an arctic fermi superfield.

\subsection{Twisted hypermultiplets}\label{twshyp}
A twisted hypermultiplet is described by a complex $\mc{O}(1)'$
superfield $\mH(\zeta')$ that is specified as
\begin{equation}
  \mH(\zeta') = \zeta' H_{1'} + H_{2'}\ ,\quad
 \w{\mH}(\zeta') = -\zeta' \w{H}{}^{2'} + \w{H}{}^{1'}\ .
\end{equation}
The $F'$-projective constraints $\pD_{a+} \mH = 0$ are given by
\begin{align}
  &\w\uQ_{+} H_{2'} = 0\ ,\ \, \w{\uD}_{+}H_{2'} = 0\ ,\ \, \uQ_{+} H_{1'} = 0\ ,\ \, \uD_{+}H_{1'} = 0\ ,\ \, \w\uQ_{+} H_{1'} = \uD_{+} H_{2'}\ ,\  \, \uQ_+ H_{2'} = -\w{\uD}_+ H_{1'}\ .
\end{align}
$\w{H}{}^{1'}$ and $H_{2'}$ are $(0,2)$ chiral superfields since
$\w{\uD}_+$ annihilates them. As for the standard hyper, the
superpartner fermions are defined by
\begin{equation}
  \uD_{aa'+} H_{b'} = \sqrt{2} \varepsilon_{a'b'} \xi_{a+}\ .
\end{equation}
The above definition makes it clear that the superpartner fermions
$\xi_{a+}$ of $H_{a'}$ are in the doublet of $F$. Explicitly, we have
\begin{equation}\label{twsuper}
\sqrt{2}  \xi_{1+} =  \uD_+ H_{2'}\ ,\quad \sqrt{2}\w\xi{}^1_+ = - \w\uD_+ \w{H}{}^{2'}\ ,\quad \sqrt{2}\xi_{2+} = - \w\uD_+ H_{1'}\ ,\quad \sqrt{2}\w\xi{}^{\,2}_+ = \uD_+ \w{H}{}^{1'}\ .
\end{equation}
The $(0,4)$ supersymmetric action that describes the twisted
hypermultiplet is
\begin{align}
\mc{S} &= -\int\ud^2x\oint_{\gamma'}\frac{\ud\zeta'}{2\pi \i}\wt\pD_{1+}\wt\pD_{2+}\, (\tfrac{\i}{2}\zeta'^{-1}\w{\mH}\partial_{--}\mH) = \int\ud^2x\oint_{\gamma'}\frac{\ud\zeta'}{2\pi \i\zeta'}\uD_+\w\uD_+\, (\tfrac{\i}{2}\zeta'^{-1}\w{\mH}\partial_{--}\mH)\ .
\end{align}
Performing the $\zeta'$ integral, we get
\begin{equation}
  \mc{S} = \frac{\i}{2}\int\ud^2x\, \uD_+\w\uD_+\left(\w{H}{}^{1'}\partial_{--}H_{1'} - \w{H}{}^{2'}\partial_{--}H_{2'}\right) = \int\ud^2x\, (-\partial_\mu\w{H}{}^{a'}\partial^\mu H_{a'} - \i \w\xi{}^{a}_{+}\partial_{--}\xi_{a+})\ ,
\end{equation}
which is the action for two $(0,2)$ chiral multiplets $\w{H}{}^{1'}$
and $H_{2'}$. The $H_{a'}$ form an $F'$-doublet and hence, the above
multiplet describes a twisted hyper. The description in terms of
$F'$-arctic superfields is analogous to that of the standard hyper.

\section{Fermi multiplets}\label{fermimult}

In this section, we describe matter fermi multiplets. We focus on
$F$-projective fermi superfields below; the $F'$-case follows
analogously. Like hypermultiplets, fermi multiplets can be realized
either as $\mc{O}(n)$ superfields or $F$-arctic superfields. We only
describe arctic superfields here since all our constructions use only
those and not the $\mc{O}(n)$ superfields.

%\paragraph{Arctic superfields}
%\subsection{Arctic multiplets}\label{arcticfermi}
Start with a weight $0$ $F$-arctic superfield
$\bm\Upsilon_- = \sum_{0}^\infty \iUpsilon_{j-}\zeta^j$ satisfying
\begin{equation}
  \pD_{a'+} \bm{\Upsilon}_- = 0\ .
\end{equation}
The constraints in terms of $\iUpsilon_{j-}$ are
\begin{equation}
  \uQ_+ \iUpsilon_{0-} = 0\ ,\quad \uQ_+ \iUpsilon_{j+1,-} + \uD_+ \iUpsilon_{j-} = 0\ ,\quad \w\uD_+ \iUpsilon_{0-} = 0\ ,\quad \w\uD_+ \iUpsilon_{j+1,-} - \w\uQ_+ \iUpsilon_{j-} = 0\ .
\end{equation}
The $\iUpsilon_{j-}$ for $j \geq 1$ are unconstrained $(0,2)$
superfields while $\iUpsilon_{0-}$ satisfies the chirality constraint
$\w\uD_+ \iUpsilon_{0-} = 0$. We relabel $\iUpsilon_{0-}$ as
$\psi_-$. The action is
\begin{align}\label{04feract}
  \mc{S} &= -\frac{1}{2}\int \ud^2x \oint \frac{\ud\zeta}{2\pi\i} \wt\pD_{1'+}\wt\pD_{2'+}\,(\w{\bm{\Upsilon}}_- \bm{\Upsilon}_-) = \frac{1}{2}\int \ud^2x \oint \frac{\ud\zeta}{2\pi\i\zeta} \uD_+\w\uD_+\,(\w{\bm{\Upsilon}}_- \bm{\Upsilon}_-)\ ,\nonumber\\
         &= \frac{1}{2}\int \ud^2x\,  \uD_+\w\uD_+(\w\psi_{-} \psi_{-}) + \frac{1}{2}\sum_{j=1}^\infty (-1)^{j} \int \ud^2x\,  \uD_+\w\uD_+\,(\w{\iUpsilon}_{j-}\iUpsilon_{j-})\ ,\nonumber\\
         &= \frac{1}{2}\int \ud^2x\, \uD_+\w\uD_+\,(\w\psi_{-} \psi_{-})\ .
\end{align}
In the last step, we have integrated out the (0,2) unconstrained
superfields $\iUpsilon_{j-}$ with $j \geq 1$. Note that this is
consistent with the $F$-transformations discussed in Section
\ref{Rsym} only for weight $k=0$. In more detail, the $F$
transformation rules for the fields $\iUpsilon_{j-}$ in
\eqref{arctictr} preserve the auxiliary field equations
$\iUpsilon_{j-} = 0$, $j \geq 1$, only for weight $0$.

To get the component action, we push the measure derivatives
$\uD_+\w\uD_+$ into the Lagrangian:
\begin{equation}
  \mc{S} =  \int \ud^2x\, (\w{G} G + \i \partial_{++} \w\psi_- \psi_-)\ ,
\end{equation}
where the auxiliary field $G$ is defined as
$-\sqrt{2} G = \uD_+\psi_-$. In Appendix \ref{0400}, we define the
ordinary space components of $\bUpsilon_-$ directly without going to
$(0,2)$ superspace by acting on $\bUpsilon_-$ with $\wt\pD_{a'+}$
successively. We also compute the above component action by directly
pushing in the $(0,4)$ measure $\wt\pD_{1'+}\wt\pD_{2'+}$ in
\eqref{04feract} and using the definitions of the components that were
just alluded to, and finally perform the $\zeta$-integral.

\section{Interactions}\label{interaction}

The criteria for $(0,4)$ supersymmetry are closure of the algebra
$\pD_+^2 = 0$ on all the superfields and the invariance of the action
(see the comments at the end of Section \ref{algsupact}). In this
section, we use these criteria to discover possible $(0,4)$
supersymmetric interactions between twisted hypers, standard hypers
and fermis.

As indicated in the Introduction (Section \ref{intro}), interactions
could be $E$-terms, gauge interactions, or of the nonlinear sigma
model type. Nonlinear sigma models have been discussed for $\mc{O}(1)$
standard hypers in $(0,4)$ projective superspace \cite{Hull:2016khc}
and arctic standard hypers in $(4,4)$ projective superspace
\cite{Lindstrom:1987ks}. We have not explored all the possibilities
for $E$-term interactions. In this paper, we consider the combination
of $F$-arctic standard hypers, $F$-arctic fermis and $\mc{O}(1)'$
twisted hypers with the R-charge assignments given previously (of
course, everything we say can be used for the mirror combination where
we swap the two R-symmetry groups).

Consider $F$-arctic fermi multiplets $\bUpsilon_-$, arctic standard
hypermultiplets $(\bUpsilon, \bUpsilon_{--})$ and $\mc{O}(1)'$ twisted
hypermultiplets $\mH$ with the following projective constraints:
\begin{alignat}{2}\label{condconst}
\pD_+ \bUpsilon &= 0\ ,\quad  \pD_{+} \bUpsilon_- = -\sqrt{2}\wh{\mC}\bUpsilon\ ,\quad &&\pD_{+} \bUpsilon_{--} = \frac{1}{\sqrt{2}}\mC\bUpsilon_-\ ,\quad \pD_+ \mH = 0\ ,\nonumber\\
\pD_+ \w\bUpsilon &= 0\ ,\quad  \pD_{+} \w\bUpsilon_- = \sqrt{2}\zeta\w\bUpsilon\w{\wh{\mC}}\ ,\quad
  &&\pD_{+} \w\bUpsilon_{--} =
  \frac{1}{\sqrt{2}}\zeta\w\bUpsilon_-\w{\mC}\ ,\quad \pD_+ \w{\mH} = 0\ .
\end{alignat}
where $\pD_{+} = u^a v^{a'}\uD_{aa'+}$ is the fully contracted
derivative (see \eqref{derfullcont} in Section \ref{sspace}),
${\mC} = v^{a'} C_{a'}$ and $\wh{\mC} = v^{a'}\wh{C}_{a'}$ are
$\mc{O}(1)'$ superfields which are functions of the various
superfields in the model. The second line in \eqref{condconst} is
obtained by applying extended complex conjugation on the first line
and using the appropriate definitions of extended complex conjugates
from Section \ref{extconjdef}.

The closure of the supersymmetry algebra $\pD_+^2 = 0$ on
$\bUpsilon_-$ and $\bUpsilon_{--}$ give
\begin{equation}\label{04EJcond}
  \pD_{+} \mC = 0\ ,\quad  \pD_{+} \wh{\mC} = 0\ ,\quad \mC\wh{\mC}  = 0\ .
\end{equation}

The action $\mc{S}$ for the above superfields splits into an action
$\mc{S}_F$ in $F$-projective superspace for the standard hypers and
the fermis, and an action $\mc{S}_{F'}$ in $F'$-projective superspace
for the twisted hypers, i.e., $\mc{S} = \mc{S}_F + \mc{S}_{F'}$ with
\begin{align}\label{indiact}
  \mc{S}_F &= \int\ud^2x \oint \frac{\ud\zeta}{2\pi\i} \wt\pD_{1'+}\wt\pD_{2'+} \left(\tfrac{\i}{2}\w\bUpsilon \partial_{--} \bUpsilon - \zeta\w\bUpsilon\bUpsilon_{--} + \zeta^{-1}\w\bUpsilon_{--}\bUpsilon - \tfrac{1}{2}\w\bUpsilon_-\bUpsilon_-\right)\ ,\nonumber\\
  \mc{S}_{F'} &= \int\ud^2x \oint \frac{\ud\zeta'}{2\pi\i} \wt\pD_{1+}\wt\pD_{2+} \left(-\tfrac{\i}{2}\zeta'^{-1}\w{\mH} \partial_{--} \mH\right)\ .
\end{align}

The action $\mc{S}_F$ and $\mc{S}_{F'}$ in \eqref{indiact} are $(0,4)$
invariant if the Lagrangians are annihilated by $\pD_{+}$. This is
obvious for $\mc{S}_{F'}$. The action of $\pD_+$ on the Lagrangian in
$\mc{S}_F$ is
\begin{align}\label{actinvcond}
  &\pD_{+} \left(\tfrac{\i}{2} \w\bUpsilon \partial_{--} \bUpsilon - \zeta \w\bUpsilon \bUpsilon_{--} + \zeta^{-1} \w\bUpsilon_{--} \bUpsilon - \tfrac{1}{2}\w\bUpsilon_- \bUpsilon_-\right) =  - \tfrac{1}{\sqrt{2}}\zeta \w\bUpsilon (\mC + \w{\wh{\mC}}) \bUpsilon_- +  \tfrac{1}{\sqrt{2}}\w\bUpsilon_- (\w{\mC} - \wh{\mC})\bUpsilon\ .
\end{align}
For the right hand side to be zero, the following conditions then have
to be satisfied:
\begin{equation}\label{EJrel}
  \ol{\mC} = \wh{\mC}\ ,\quad \mC = -\w{\wh{\mC}}\ ,\quad\text{i.e.,}\quad \w{C}{}^{a'} = \varepsilon^{a'b'} \wh{C}_{b'}\ .% \w{C}{}^{1'} - \zeta' \w{C}{}^{2'} = \zeta' \wh{C}_{1'} + \wh{C}_{2'}\ .
\end{equation}
(the two conditions are consistent with each other since we have
$\w{\w{\bm{\Phi}}} = - \bm{\Phi}$ for an $\mc{O}(1)'$ superfield
$\bm\Phi$.)

Upon using \eqref{EJrel}, the constraints $\mC \wh\mC = 0$ in
\eqref{04EJcond} become
\begin{equation}
  \mC\w\mC = 0\quad \Leftrightarrow\quad C_{a'} \w{C}{}^{b'} = \frac{1}{2}C_{c'} \w{C}{}^{c'}\, \delta_{a'}^{b'}\ .
\end{equation}
$\mC$ and $\wh{\mC}$ are \emph{a priori} functions of both standard
and twisted hypers. We restrict ourselves to the case where $\mC$ and
$\wh{\mC}$ are polynomials in the standard and twisted hypers. Recall
that the $F$-weights of $\pD_+$, $\bUpsilon$, $\bUpsilon_-$ and
$\bUpsilon_{--}$ are $+1$, $+1$, $0$ and $-1$ respectively. Since the
$F$-weight has to be preserved in the constraint equations
\eqref{condconst} above, $\mC$ and $\wh{\mC}$ should have $F$-weight
$0$. Further, since we restrict $\mC$ and $\w\mC$ to be polynomials in
the superfields, they must simply be independent of the standard
hypers $\bUpsilon$.

The reality constraints \eqref{EJrel} are also consistent with $\mC$
and $\wh{\mC}$ being independent of standard hypers. However, note
that $\mC$ and $\wh{\mC}$ can be chosen to be more general $F$-weight
0 functions of the standard hypers (e.g.~rational functions) and these
may have good Taylor expansions around both $\zeta = 0$ and
$\zeta = \infty$. Then it is possible to satisfy the reality
constraint \eqref{EJrel} even when $\mC$ and $\wh{\mC}$ depend on
arctic standard hypers non-trivially.

Since $\mC$ is an $\mc{O}(1)'$ superfield which is assumed to be a
polynomial in the twisted hypers and is annihilated by $\pD_{a+}$, it
must be linear in the $\mc{O}(1)'$ twisted hypers $\mH$. Thus, $\mC$
must take the form
\begin{equation}\label{Jsol}
  \mC = \mK + L \mH\ ,
\end{equation}
where $\mK$ is $\mc{O}(1)'$ and constant, and $L$ is constant.

Recall from Sections \ref{stdhyp}, \ref{twshyp} and \ref{fermimult}
that the dynamical components of the arctic standard hyper are
$(\eta_a, \xi_{a'+})$, those of the twisted hyper are
$(H_{a'}, \xi_{a+})$ and that of the fermi is $(\psi_-)$. The full
component action for these fields that follows from the projective
superspace action \eqref{indiact} is worked out in Appendix
\ref{components}. We give the result here:
\begin{align}\label{Stotfin}
  \mc{S} &= \int\ud^2x\,\left( - \w{{\partial}_\mu H}{}^{a'} \partial_\mu H_{a'}  - {\i}\w\xi{}^a_+ \partial_{--} \xi_{a+}  - \w{{\partial}_\mu \eta}{}^{a} \partial_\mu \eta_{a}  - {\i}\w\xi{}^{a'}_+ \partial_{--} \xi_{a'+} - \i \w\psi_-\partial_{++} \psi_-\right) \nonumber\\
         &+ \int\ud^2x\,\left(- \tfrac{1}{2} \w\eta{}^a C_{a'} \w{C}{}^{a'} \eta_a +  \left(- \w\xi{}^{a'}_+ C_{a'} \psi_{-} -\w\eta{}^a L \xi_{a+} \psi_{-}   + \text{c.c.} \right)\right)\ .
\end{align}

\section{Example: ADHM sigma model}\label{ADHM}
In this section we consider an interacting model with standard hypers,
fermis and twisted hypers. This is a particular $(0,4)$ linear sigma
model which flows to a nonlinear sigma model with target space a
$k$-instanton solution in Yang-Mills theory in four dimensions. This
model was written in $(0,1)$ superspace in
\cite{Douglas:1996u,Witten:1994tz} and in harmonic superspace in
\cite{Galperin:1994qn,Galperin:1995pq}.

This linear sigma model for $\text{U}(n)$ instantons is realised by
the following nested D-brane configuration in Type IIB theory
\cite{Douglas:1996u}: 1 D1-brane $\subset$ $k$ D5-branes $\subset$ $n$
D9-branes. The $k$ D5-branes appear as $k$-instanton configurations in
the D9-brane $\text{U}(n)$ gauge theory and the D1-brane probes this
configuration. The $1+1$ dimensional linear sigma model is the theory
on the D1-brane worldsheet.

The D1-brane worldsheet theory includes a $\text{U}(1)$ gauge
multiplet arising from the D1-D1 open string spectrum. However, the
$\text{U}(1)$ multiplet does not have an effect on the computation of
the instanton connection on target space in the classical theory on
the D1-brane \cite{Douglas:1996u}. We describe the classical
$\text{U}(n)$ instanton model without the $\text{U}(1)$ gauge
multiplet in Section \ref{uninst} and show that it reproduces the
calculation in \cite{Douglas:1996u}, and redo the analysis more
carefully in the companion paper \cite{PRcomp} with the gauge
multiplet included. The novelty of the projective superspace approach
is that twistor space and the relevant holomorphic bundles on twistor
space required for describing instantons
\cite{Ward:1977ta,Atiyah:1977pw,Atiyah:1978ri} appear explicitly in
the description of the model which we describe below.

For $\text{SO}(n)$ instantons, we add an O9$^-$-plane to the above
D-brane configuration. The orientifold projection requires an even
number of D5-branes which we take to be $2k$, and after the projection
pairs of D5-branes are stuck and cannot be separated. The projection
reduces the D9-brane gauge group to $\text{SO}(n)$, that of the
D5-branes to $\text{Sp}(k)$ and projects out the vector multiplet on
the D1-brane. For $\text{Sp}(n)$ instantons, we start with $2n$
D9-branes, $k$ D5-branes and $2$ D1-branes and add an O9$^+$-plane
which results in an $\text{Sp}(n)$ gauge group on the D9-branes, an
$\text{SO}(k)$ gauge group on the D5-branes and an $\text{Sp}(1)$
gauge group on the D1-branes (again, the two D1-branes cannot be
separated). These facts may be found in, e.g., \cite{Gimon:1996rq}.
Since the $\text{Sp}(n)$ instanton sigma model requires a gauge
multiplet, and both $\text{SO}(n)$ and $\text{Sp}(n)$ models require
orientifolds, we describe both sigma models together in the companion
paper \cite{PRcomp}.

\subsection{\tp{$\text{U}(n)$}{U(n)} instantons}\label{uninst}

The superfield content consists of
\begin{enumerate}
\item $2k'$ twisted hypers $\mH_{Y'}$, $Y'=1',\ldots,2k'$ (we consider
  $2k' = 2$ for most of the discussion),

\item $k$ standard hypers $(\bUpsilon_Y, \bUpsilon_{Y--})$,
  $Y = 1,\ldots,k$,

\item $2k + n$ fermis $\bUpsilon_{A-}$, $A = 1,\ldots,2k+n$.
\end{enumerate}
The above superfields (for $2k'=2$) are a subset of the low-energy
spectrum of the various D$p$-D$q$ open strings in the D-brane
configuration described above. Since we are interested in the
low-energy theory on the D1-brane, we retain only those fields that
appear from the D1-D$p$ open string sectors for $p=1,5,9$. The two
twisted hypers $\mH_{Y'}$ arise from the D1-D1 strings in the
directions transverse to the D1-brane and D5-branes. The $k$ standard
hypers $\bUpsilon_{Y}$ arise from D1-D5 strings and the $2k+n$ fermis
$\bUpsilon_{A-}$ arise from the D1-D5 strings ($2k$ fermis) and the
D1-D9 strings ($n$ fermis). Part of the couplings $\mC$ described
below arise from the D5-D9 open string degrees of freedom which are
frozen from the point of view of the D1-brane, and they contain the
instanton moduli.

We suppress the flavour indices $Y'$, $Y$ and $A$ on the twisted
hypers, standard hypers and fermis respectively unless we wish to
explicitly exhibit the flavour properties of the superfields. We work
with a given symplectic structure $\omega^{Y'Z'}$ on the space of
twisted hypers. This allows for a reality condition:
 \begin{equation}\label{realtw}
   \w{\mH}{}^{Y'} = \omega^{Y'Z'} \mH_{ Z'}\ ,\quad\text{i.e.,}\quad \w{H}{}^{a'Y'} = \varepsilon^{a'b'} \omega^{Y'Z'} H_{ b'Z'}\ .
\end{equation}
(note that according to the above condition
$\w{\w{\mH}}_{Y'} =- \mH_{Y'}$ since
$\omega^{Y'Z'} \omega_{Z'X'} = -\delta_{X'}^{Y'}$. This is consistent
with the result $\w{\w{\mH}} = -\mH$ for an $\mc{O}(1)'$
multiplet). 
% \begin{figure}[htbp]
%   \centering \input{r4quiver.pdf_t}
%   \caption{The quiver for instantons on $\mbb{R}^4$. The big node
%     stands for the worldsheet gauge group $G$ (and its corresponding
%     vector multiplet), the small nodes are the $\text{U}(k)$ the
%     $\text{U}(n)$ flavours. The solid blue lines are a pair of twisted
%     hypers with a reality condition between them and there are a total
%     of $k'$ such lines. The solid red lines are standard hypers
%     and the dashed lines are fermis. Every line is in the
%     bifundamental of the two nodes it connects.}\label{r4quiver}
% \end{figure}
% \begin{table}\centering
%   \begin{tabular}{c|c}%@{\extracolsep{\fill}}}
%     \toprule%\hline
%     Multiplets & Role in target space \\ \midrule
%     Twisted hypers & Target spacetime \\ 
%     Fermis & Fibres of instanton bundle \\ 
%     Standard hypers & $?$ \\ \bottomrule
%   \end{tabular}
% \caption{Various multiplets and their role in the ADHM sigma model.}
% \end{table}
The most general $(0,4)$ constraints are those given in
\eqref{condconst}:
\begin{align}\label{projconst}
  &\pD_{+} \bUpsilon = 0\ ,\quad \pD_{+} \bUpsilon_{-} = -\sqrt{2}\wh{\mC}\bUpsilon\ ,\quad \pD_{+} \bUpsilon_{--} = \frac{1}{\sqrt{2}}\mC\,\bUpsilon_{-}\ ,\quad \pD_{+} \mH = 0\ ,
\end{align}
where recall from Section \ref{interaction} that $\mC = v^{a'}C_{a'}$,
$\wh\mC = v^{a'} \wh{C}_{a'}$ are $\mc{O}(1)'$ superfields. As
discussed in Section \ref{interaction}, $\mC$ and $\wh\mC$ are
independent of the standard hypers $\bUpsilon$ and are linear in the
$\mc{O}(1)'$ twisted hypers $\mH$. The constraints on the couplings
$\mC$ and $\wh\mC$ that follow from the closure of the $(0,4)$
superalgebra are \eqref{04EJcond} which we reproduce here for convenience:
\begin{equation}\label{algconst}
  \pD_+ \mC = 0\ ,\quad \pD_+ \wh\mC = 0\ ,\quad \mC \wh\mC = 0\ .
\end{equation}
$\mC$ and ${\wh{\mC}}$ and are $k \times (2k+n)$ and $(2k+n) \times k$
matrices respectively; with the flavour indices explicitly displayed,
the matrices are resp.~written as $\mC_Y^A$ and $\wh\mC_A^Y$ . Recall from
the discussion around \eqref{Jsol} that $\mC$ has to be of the form
\begin{equation}\label{ADHMcoupling}
  \mC = \mK  + L^{Y'} \mH_{Y'}\ ,%\quad\text{i.e.}\quad C^A_{a'Y} = K^A_{a' Y} + L^{A Y'}_Y \Phi_{a'Y'}\ ,
\end{equation}
where $\mH_{Y'}$ are the twisted hypermultiplets. The coupling $\mK$
is a constant $k \times (2k+n)$ matrix $\mc{O}(1)'$ superfield and the
$L^{Y'}$ are constant $k \times (2k+n)$ matrices (one matrix for each
$Y' \in \{1,\ldots,2k'\}$).

\paragraph{Twistor space} Let us consider two twisted hypers, i.e.,
$2k' = 2$ (everything we say for two twisted hypers can be extended to
general $k'$). The twisted hyper superfields $H_{a'Y'}$ are
coordinates on the target space $\mbb{R}^4$. The $\text{SU}(2)'$
doublet $v^{a'}$ together with the projective superfields $\mH_{Y'}$
can be interpreted as homogeneous coordinates
$\mZ = (v^{1'}, v^{2'}, \mH_{1'}, \mH_{2'})$ for a $\mbb{CP}^3$ which
is in fact the twistor space of $\mbb{S}^4$ (the one-point
compactification of the target $\mbb{R}^4$). The symplectic structure
$\omega^{Y'Z'}$ on the space of twisted hypers and the symplectic
structure $\varepsilon^{a'b'}$ on the space of $F'$-doublets together
give an antiholomorphic involution
$v^{a'} \to \varepsilon^{a'b'} \w{v}_{b'}$,
$\mH_{Y'} \to \omega_{Y'Z'} \w{\mH}{}^{Z'}$, on the $\mbb{CP}^3$ which
squares to $-1$. The $(v^{a'}, H_{Y'b'})$ serve as coordinates on the
correspondence space and the \emph{incidence relations}
$\mH_{Y'} = H_{Y'a'} v^{a'}$ are simply the definition of the
$\mH_{Y'}$ as projective superfields.

\paragraph{Monads on twistor space} Next, we show that the couplings
$\mC$ and $\wh\mC$ encode the data of a \emph{monad} on
$\mbb{CP}^3$. Let $V_S$ and $\wh{V}_{S}$ be the vector spaces of
$\bUpsilon$ and $\bUpsilon_{--}$ respectively with
$\dim V_S = \dim \wh{V}_{S} = k$ and $V_F$ be the vector space of
fermis with $\dim V_F = 2k + n$. Then, the couplings $\mC$ and
$\wh\mC$ can be interpreted as elements of $\Hom(V_F,\wh{V}_S)$ and
$\Hom(V_S,V_F)$ respectively, as is clear from the constraints
\eqref{projconst}. Recall that these maps are linear in the
homogeneous coordinates $\mZ = \{v^{a'}, \mH_{Y'}\}$ since
$\mC = K_{a'} v^{a'} + L^{Y'} \mH_{Y'}$. We thus have
\begin{equation}\label{monadcomp}
  V_S \overset{\wh\mC}{\longrightarrow} V_F \overset{\mC}{\longrightarrow} \wh{V}_S\ .
\end{equation}
The constraint $\mC\wh\mC = 0$ that follows from the closure of the
algebra \eqref{algconst} makes \eqref{monadcomp} a complex. We further
require that $\wh\mC$ is injective and $\mC$ is surjective: this
imposes non-degeneracy conditions on the couplings $K_{a'}$ and
$L^{Y'}$. Then the above complex is precisely a monad and the
cohomology at $V_F$, i.e., $\ker \mC / \im \wh\mC$ is a holomorphic
rank $n$ vector bundle $\mc{E}$ on $\mbb{CP}^3$ which is trivial when
restricted to lines in $\mbb{CP}^3$, and has $c_2(\mc{E}) = k$. Thus,
the data that goes into choosing the off-shell superfield content of
our linear sigma model is precisely the same data that goes into
defining a holomorphic bundle on twistor space $\mbb{CP}^3$ that is
trivial on lines.

We get a symplectic structure on the bundle $\mc{E}$ also from the
requirement that the action is $(0,4)$ supersymmetric. Some reality
conditions (which were implicit in the previous sections) are
necessary on the vector spaces $V_F$, $V_S$ and $\wh{V}_S$ to write
down an action for the projective superfields $\bUpsilon_{-}$,
$\bUpsilon$ and $\bUpsilon_{--}$. They are (1) a hermitian structure
on $V_F$ that identifies ${V}^*_F \simeq {V}^\vee_F$, and (2) the
identification ${\wh{V}}^*_S \simeq {V}^\vee_S$, where ${V}^*$ and
$V^\vee$ stand for the complex conjugate and dual of a vector space
$V$ respectively. With these at hand, the $(0,4)$ invariance of the
action gives the following constraint \eqref{EJrel} on the couplings
$\mC$ and $\wh\mC$:
\begin{equation}\label{actconst}
  \w\mC =  \wh\mC\ ,
\end{equation}
where the bar on $\w{\mC}$ acts the hermitian conjugate on the matrix
components and extended conjugate on the $\mc{O}(1)'$ superfield. This
imposes a symplectic structure on the bundle $\mc{E}$ obtained from
the monad \eqref{monadcomp}. By the Penrose-Ward-Atiyah correspondence
\cite{Ward:1977ta,Atiyah:1977pw}, the bundle $\mc{E}$ on twistor space
with the symplectic structure described above corresponds to a
self-dual $\text{SU}(n)$ connection on $\mbb{R}^4$ (more precisely, on
the one-point compactification $S^4$ of $\mbb{R}^4$). The ADHM
construction \cite{Atiyah:1978ri} gives an explicit expression for the
instanton gauge field in terms of the data described above. The
constraints $\mC \w{\mC} = 0$ are precisely the ADHM equations that
describe the instanton moduli space \cite{Atiyah:1978ri}.

Next, we show that the model flows to an $\text{SU}(n)$ instanton
solution in the infrared by explicitly obtaining the expression for
the instanton gauge field given by the ADHM construction
\cite{Atiyah:1978ri}. The material in the rest of this section is not
new and follows the calculations in
\cite{Witten:1994tz,Douglas:1996u}. In Section \ref{symminst} below,
we choose particular bases for the vector spaces of superfields to
give the usual standard characterization of the ADHM instanton moduli
space in terms of finite dimensional matrices. Again, most of the
material is standard except for a formula of the virtual dimension of
the instanton moduli space on $\mbb{R}^{4k'}$ for $k' \geq 2$.

\paragraph{Instantons on $\mbb{R}^4$} The potential energy density of
the model described above can be read off from the general expression
in \eqref{Stotfin} and is positive-definite:
\begin{align}\label{ADHMpot}
  V &=  \tfrac{1}{2}\w\eta{}^{a} C_{a'} \w{C}{}^{a'}\eta_{a} = \tfrac{1}{2} \w\eta{}^{aY} C_{a'Y}^A \w{C}{}^{a'Z}_A \eta_{aZ}\ .
\end{align}
Recall that $C_{a'} = K_{a'} + L^{Y'} H_{a'Y'}$ and the $\eta_{aY}$
are components of the arctic standard hyper
$\bUpsilon_Y = \zeta \eta_{1Y} + \eta_{2Y}$ once we eliminate the
auxiliary superfields accompanying higher powers of $\zeta$ (see
\eqref{partonshell} and the discussion around it). Suppose the
constant matrices $K_{a'}$ and $L^{Y'}$ are sufficiently generic so
that $\tfrac{1}{2} C_{a'}\w{C}{}^{a'} \equiv f^{-1}$ is an invertible
$k \times k$ matrix, i.e., all its eigenvalues are non-zero, for any
value of $H_{a'Y'}$. Then, the vacuum corresponds to setting the
$\eta_{aY} = 0$ for every flavour $Y = 1,\ldots,k$.

About this vacuum, the potential $V$ vanishes and in particular does
not give a mass for the twisted hyper scalars: there is a classical
moduli space of vacua $\mbb{R}^{4}$ parametrized by the four twisted
hyper scalars with the reality condition \eqref{realtw}. Under the
genericity assumption on $K_{a'}$ and $L^{Y'}$, the eigenvalues of the
standard hyper mass matrix $f^{-1}$ are all (1) positive since
$f^{-1}$ is a positive-definite matrix, and (2) strictly positive
since $f^{-1}$ is invertible. We list them as
$(m^2_1, m^2_2, \ldots, m_k^2)$. Then, the mass of the standard hyper
scalars $\eta_{aY}$ for a given $Y$ is $m_Y$. The Yukawa couplings can
also be read off from \eqref{Stotfin}:
\begin{equation}\label{ADHMyuk}
  -\w\xi{}^{a'}_+ C_{a'} \psi_- - \w\eta{}^a L\xi_{a+} \psi_- - \w\psi_-\w{C}{}^{a'} \xi_{a'+} - \w\psi_- \w\xi{}^a_+\w{L} \eta_a \ .
\end{equation}
On the classical vacuum moduli space characterised by
$\eta_{aY} = \w\eta{}^{aY} = 0$ and arbitrary $H_{Y'a'}$, the twisted
hyper fermions $\xi_{Y'a+}$ again have no mass terms. Let us look at
the mass terms for the standard hyper fermions $\xi_{Ya'+}$:
\begin{equation}
  -\w\xi{}^{a'}_+ C_{a'} \psi_- - \w\psi_- \w{C}{}^{a'} \xi_{a'+} = - \w\xi{}^{a'Y}_+ C_{a'Y}^A \psi_{A-} - \w\psi{}^{A}_- \w{C}{}^{a'Y}_A \xi_{a'Y+}\ ,
\end{equation}
where we have displayed the flavour indices explicitly. Recall that we
have diagonalized $f^{-1} = C_{1'}\w{C}{}^{1'} =
C_{2'}\w{C}{}^{2'}$. By using an appropriate $\text{U}(2k + n)$
transformation, we can further cast the $2k \times (2k+n)$ matrix
$\pmat{C_{1'Y}^A \\ C_{2'Y}^A}$ into a block form with a non-trivial
$2k \times 2k$ block and a zero $2k \times n$ block:
\begin{equation}
\pmat{C_{1'Y}^A \\ C_{2'Y}^A} = \pmat{\bigstar_{2k \times 2k} & \mbb{0}_{2k \times n}}\ ,
\end{equation}
where the non-trivial $2k \times 2k$ block is
$\text{diag}(m_1, \ldots, m_k, m_1,\ldots, m_k)$. For a fixed flavour
$Y$ of the standard hyper, the two fermions $\xi_{1'Y+}$, $\xi_{2'Y+}$
and the two fermis $\psi_{Y,-}$, $\psi_{k+Y,-}$ interact through the
$2\times 2$ mass matrix $\pmat{m_Y & 0 \\ 0 & m_Y}$ and become massive
with mass $m_Y$. Recall that the standard hyper scalars $\eta_{aY}$
also have the same mass $m_Y$. The zero block of size $2k \times n$
implies that the $n$ fermis $\psi_{A-}$, $A=2k+1,\ldots, 2k+n$ are
massless. Thus, for generic values of the couplings $K_{a'}$ and
$L^{Y'}$, we have $k$ massive standard hypers, $2k$ massive fermis and
$n$ massless fermis about any point of the classical vacuum moduli
space that is parametrized by the massless twisted hypers.

The $n$ massless fermis can be characterised more generally as the
solutions of the equation
\begin{equation}\label{ferzeromodes}
\sum_{A=1}^{2k+n}  \mC_{Y}^A \psi_{A-} = 0\ .
\end{equation}
Let the $n$ massless solutions be arranged into the $(2k+n)\times n$
matrix $\mc{V}_A{}^i$ with the normalisation
$(\mc{V}^\dag)_j{}^A \mc{V}_A{}^i = \delta_{j}^i$. The most general
massless solution is then
\begin{equation}
  \psi_{A-} = \sum_{i = 1}^n \mc{V}_{A}{}^i \lambda_{i-}\ .
\end{equation}
Plugging in the above expression for $\psi_{A-}$ in its kinetic term,
we get the kinetic term for the massless modes $\lambda_{i-}$:
\begin{equation}\label{lambdakin}
  \w\psi{}^A_- \partial_{++} \psi_{A-} = \w\lambda{}^i_- (\mc{V}^\dag)_i{}^A \partial_{++}(\mc{V}_A{}^j \lambda_{j-}) = \w\lambda{}^i_- \left[\delta_i{}^j\partial_{++} + \partial_{++} \w{H}{}^{Y'a'} (\mc{V}^\dag)_i{}^A \frac{\partial \mc{V}_A{}^j}{\partial \w{H}{}^{Y'a'}}\right]\lambda_{j-}\ .
\end{equation}
We see that the massless fermis have now acquired an additional
connection which is the pullback of a connection $\mc{A}$ on target
space $\mbb{R}^4$:
\begin{equation}\label{gaugefield}
  (\mc{A}_{Y'a'})_i{}^j := \i (\mc{V}^\dag)_i{}^A \frac{\partial \mc{V}_A{}^j}{\partial \w{H}{}^{Y'a'}}\ .
\end{equation}
This is the connection for a $k$-instanton solution with $\text{U}(n)$
gauge group, a fact that follows from standard results in the ADHM
construction. Since we have assumed the instanton to be
non-degenerate, the $\text{U}(1)$ part of the connection is trivial
and $\mc{A}_{Y'a'}$ is in fact an $\text{SU}(n)$ instanton
connection. We study the degenerate cases carefully in \cite{PRcomp}
where we shall find that the $\text{U}(1)$ gauge multiplet on the
D1-brane worldsheet plays an important role.

\subsection{The instanton moduli space and symmetries}\label{symminst}
The constraints $\mC\w\mC = 0$ and the fermi zero modes
\eqref{ferzeromodes} (and in turn, the formula for the instanton gauge
field) are unaffected by $\text{GL}(k,\mbb{C})$ transformations on the
space of standard hypermultiplets and $\text{U}(2k+n)$ transformations
of the space of fermis:
\begin{equation}\label{glkcu2kn}
  \mC \to S \cdot \mC\cdot U^\dag\ ,\quad S \in \text{GL}(k,\mbb{C})\ ,\quad U \in \text{U}(2k+n)\ .
\end{equation}
Thus, two different solutions of $\mC\w\mC = 0$ that are related by a
$\text{GL}(k,\mbb{C}) \times \text{U}(2k+n)$ transformation as in
\eqref{glkcu2kn} correspond to the same instanton solution. This
redundancy allows us to choose a simple form for the coupling $\mC$
and the equations $\mC \w\mC = 0$.

Plugging in the explicit form $\mC = \mK + L^{Y'} \mH_{Y'}$, we get
\begin{align}\label{twistorADHM}
0 = \mC \w{\mC} &=  \mK  \w{\mK} + \mK \w{L}_{Z'}\, \w\mH{}^{Z'}  +  \mH_{Y'} L^{Y'} \w{\mK} + L^{Y'} \w{L}_{Z'} \mH_{Y'} \w\mH{}^{Z'}\ ,\nonumber\\
              &=  \mK  \w{\mK} + \mK \w{L}_{Z'}\, \omega^{Z'X'} \mH_{X'}  + \mH_{Y'} L^{Y'}  \w{\mK} + L^{Y'} \w{L}_{Z'} \mH_{Y'} \omega^{Z'X'} \mH_{X'}\ .
\end{align}
We have used the reality condition \eqref{realtw} on the twisted
hypers in going to the second line above.
% \begin{equation}
%  C_{b'} \w{C}{}^{c'} =  K_{b'}  \w{K}{}^{c'} + K_{b'} \w{L}_{Z'}\, \w\Phi{}^{c'Z'}  + L^{Y'} \Phi_{b'Y'} \w{K}{}^{c'} + L^{Y'} \w{L}_{Z'} \Phi_{b'Y'} \w\Phi{}^{c'Z'}\ ,
% \end{equation}
% and, consequently, using the reality condition \eqref{realtw} on the
% twisted hypers, we get
% \begin{equation}\label{halfADHM}
%   C_{b'} \w{C}{}^{c'} \varepsilon_{c'a'} = K_{b'}
%   \w{K}{}^{c'}\varepsilon_{c'a'} + K_{b'} \w{L}_{Z'}\,
%   \omega^{Z'Y'} \Phi_{a'Y'} + L^{Y'} \Phi_{b'Y'}
%   \w{K}{}^{c'} \varepsilon_{c'a'} + L^{Y'} \w{L}_{Z'}
%   \Phi_{b'Y'} \omega^{Z'X'} \Phi_{a'X'}\ .
% \end{equation}
Terms with different numbers of twisted hypers must vanish
separately. Let us study each of them in turn:
\begin{enumerate}
\item The constant part $\mK \w\mK$ of \eqref{twistorADHM} satisfies
  $\mK \w\mK = 0$. Displaying the $\text{SU}(2)'$ indices explicitly,
  we have
  \begin{equation}\label{Kconst}
    K_{b'}\w{K}{}^{c'} \varepsilon_{c'a'} +   K_{a'}\w{K}{}^{c'} \varepsilon_{c'b'} = 0\ ,\quad\text{i.e.,}\quad     K_{b'}\w{K}{}^{c'} = \mu\, \delta_{b'}{}^{c'}\ ,
  \end{equation}
  where $\mu$ is a positive-definite $k \times k$ matrix.
  
\item The vanishing of the terms linear in $\mH_{Y'}$ in
  \eqref{twistorADHM} requires
  \begin{equation}\label{philinconst}
     \mK \w{L}_{Z'} \omega^{Z'Y'} + L^{Y'} \w{\mK} = 0\ ,\quad\text{or, with $\text{SU}(2)'$ indices,}\quad  K_{b'} \w{L}_{Z'} \omega^{Z'Y'} = -L^{Y'} \w{K}{}^{c'}\varepsilon_{c'b'}\ .
   \end{equation}
   
 \item The term quadratic in the twisted hypers $L^{Y'} \w{L}_{Z'} \omega^{Z'X'} \mH_{Y'} \mH_{X'}$ vanishes when
   \begin{equation}\label{Lconst}
    L^{Y'} \w{L}_{Z'} \omega^{Z'X'} + L^{X'} \w{L}_{Z'} \omega^{Z'Y'} = 0\ ,\quad\text{that is}\quad     L^{Y'} \w{L}_{Z'} \omega^{Z'X'} =  \nu^{Y'X'}\ ,
  \end{equation}
  where $\nu^{Y' X'}$ is antisymmetric in $Y' X'$ and is an arbitrary
  hermitian $k \times k$ matrix for each $X',Y' \in
  \{1,\ldots,2k'\}$. For the special case $k' = 1$, i.e., when there are
  two twisted hypers, the antisymmetric matrix $\nu^{Y'X'}$ is
  proportional to the symplectic form $\omega^{Y'X'}$:
  \begin{equation}\label{Lconst2}
    L^{Y'} \w{L}_{Z'} \omega^{Z'X'} =  \nu\ \omega^{Y'X'}\ ,\quad\text{that is}\quad L^{Y'} \w{L}_{Z'}  =  \nu\ \delta^{Y'}{}_{Z'}\ ,
  \end{equation}
  where $\nu$ is now a single positive-definite $k \times k$ matrix.
\end{enumerate}
The couplings $\mK$ and $L^{Y'}$ transform under the
$\text{GL}(k,\mbb{C}) \times \text{U}(2k+n)$ \eqref{glkcu2kn} as
\begin{equation}\label{kltr}
  \mK \to S\cdot \mK\cdot U^\dag\ ,\quad L^{Y'} \to S\cdot L^{Y'} \cdot U^\dag\ ,
\end{equation}
with the same $\text{GL}(k,\mbb{C})$ matrix $S$ and $\text{U}(2k+n)$
matrix $U$ for all $Y'$. This freedom can be used to choose a
convenient form for $L^{Y'}$ and $\mK$ as follows.

First, the $L^{Y'}$ satisfy the constraints \eqref{Lconst}
$L^{Y'} \w{L}_{Z'} \omega^{Z'X'} + L^{X'} \w{L}_{Z'} \omega^{Z'Y'} =
0$. Suppose we choose the symplectic form canonically to be
\begin{equation}
\omega_{Y'}{}^{Z'} = \diag_{k'/2}\{\omega_2,\omega_2,\ldots,\omega_2\}\ ,\quad\text{with}\quad  \omega_2 = \pmat{0 & 1 \\ -1 & 0}\ ,
\end{equation}
where $\diag_\ell$ indicates that length of the diagonal matrix is
$\ell$. Let us look at the pair of matrices $L^{1'}$, $L^{2'}$. They
satisfy
\begin{equation}
  L^{1'} \w{L}_{1'} = L^{2'} \w{L}_{2'} = \nu^{1'2'}\ ,\quad L^{1'}\w{L}_{2'} = 0\ .
\end{equation}
By an appropriate $\text{GL}(k,\mbb{C})$ transformation $S$
\eqref{kltr}, we can transform $\nu^{1'2'}$ into the $k \times k$
identity matrix. Then, the $2k \times (2k+n)$ matrix
$\pmat{L^{1'} \\ L^{2'}}$ satisfies
\begin{equation}
  \pmat{L^{1'} \\ L^{2'}} \pmat{\w{L}_{1'} & \w{L}_{2'}} = \pmat{\mbbm{1}_k & 0_k \\ 0_k & \mbbm{1}_k}\ ,
\end{equation}
where $\mbbm{1}_k$ and $0_k$ are the $k \times k$ identity and zero
matrices respectively. Using an appropriate $\text{U}(2k+n)$
transformation $U$ \eqref{kltr}, we can cast the above
$2k \times (2k+n)$ matrix into the form
\begin{equation}\label{Lform}
  \pmat{L^{1'} \\ L^{2'}} = \pmat{\mbbm{1}_k & {0}_k & {0}_{k\times n} \\ {0}_k & \mbbm{1}_k & {0}_{k\times n}}\ .
\end{equation}
There is a residual $\text{U}(k) \times \text{U}(n)$ subgroup of
$\text{GL}(k,\mbb{C}) \times \text{U}(2k+n)$ which preserves the above
configuration \eqref{Lform} which corresponds to
\begin{equation}\label{resid}
  S = \mc{U}\ ,\quad U = \pmat{\mc{U} & {0}_{k} & {0}_{k \times n} \\ {0}_{k} & \mc{U} & {0}_{k \times n} \\ {0}_{n \times k} & {0}_{n \times k} & \wt{\mc{U}}}\ ,\quad\text{where}\quad \mc{U} \in \text{U}(k)\ ,\quad \wt{\mc{U}} \in \text{U}(n)\ .
\end{equation}
The reality constraint \eqref{philinconst} for $Y' = 1',2'$, i.e., 
\begin{equation}
-  \mK \w{L}_{2'} + L^{1'} \w\mK = 0\ ,\quad \mK \w{L}_{1'} + L^{2'} \w\mK = 0\ ,
\end{equation}
is solved by the following expression for $\mK$:
\begin{equation}\label{Ksol}
  \mK = \pmat{ \zeta' B^{(1')}_1  + B^{(1')}_2{}^\dag & -\zeta' B^{(1')}_2 + B^{(1')}_1{}^\dag & \zeta' I^{(1')} + J^{(1')}{}^\dag}\ .%  K_{a'Y}{}^A = \pmat{B_1 & -B_2 & I \\  B_2^\dag & B_1^\dag & J^\dag}_{2k \times (2k+n)}\ ,\quad 
\end{equation}
where $I^{(1')}$, $J^{(1')}{}^\dag$ are $k \times n$ matrices and
$B^{(1')}_1$, $B^{(1')}_2$ are $k \times k$ matrices. The remaining
matrices $L^{Y'}$, $Y'=3',4',\ldots,2k'$ can also be simplified to a
form similar to \eqref{Ksol} using the constraints
\begin{equation}
  L^{1'} \w{L}_{2y'-1} - L^{2y'} \w{L}_{2'} = 0\ ,\quad L^{2'} \w{L}_{2y'-1} + L^{2y'} \w{L}_{1'} = 0\ ,\quad y' = 2',\ldots,k'\ ,
\end{equation}
where we have introduced the index $y' = 2',\ldots,k'$, such that the
pairs $\{2y'-1,2y'\}$ cover the index $Y' \in \{3',4',\ldots,2k'\}$
(later, we will append the value $y' = 1'$ as well). We then get the
simplified form
\begin{equation}\label{Lsol}
  \pmat{L^{2y'-1} \\ L^{2y'}} = \pmat{ B_1^{(y')} & -B_2^{(y')} & I^{(y')} \\ B_2^{(y')}{}^\dag & B_1^{(y')}{}^\dag & J^{(y')}{}^\dag}\ .
\end{equation}
Thus, the degrees of freedom that remain after fixing the
$\text{GL}(k,\mbb{C}) \times \text{U}(2k+n)$ symmetries are
\begin{equation}
  \{B^{(y')}_1\ ,\ B^{(y')}_2\ ,\ I^{(y')}\ ,\ J^{(y')}\}\ ,\quad\text{for}\quad y'=1',\ldots,k'\ .
\end{equation}
There $k'(2k^2 + 2k^2 + 2kn + 2kn) = k'(4k^2 + 4kn)$ real degrees of
freedom.

The remaining constraints on the $K_{a'}$ and $L^{Y'}$,
$Y'=3',4',\ldots$, are
\begin{equation}
  K_{a'} \w{K}{}^{c'} \varepsilon_{c'b'} +   K_{a'} \w{K}{}^{c'} \varepsilon_{c'b'} = 0\ ,\quad K_{a'} \w{L}_{Z'} \omega^{Z'Y'} + L^{Y'} \w{K}{}^{c'}\varepsilon_{c'a'} = 0\ ,\quad L^{Y'} \w{L}_{Z'} \omega^{Z'X'} + L^{X'} \w{L}_{Z'} \omega^{Z'Y'} = 0\ . 
\end{equation}
In terms of the matrices $B^{(y')}_1$, $B^{(y')}_2$, $I^{(y')}$ and
$J^{(y')}$, $y'=1',\ldots,k'$, we have the equations
\begin{align}\label{insteq}
  [B_1^{(y')}, B_2^{(z')}] +   [B_1^{(z')}, B_2^{(y')}] + I^{(y')} J^{(z')} + I^{(z')} J^{(y')} &= 0\ ,\nonumber\\
  [B_1^{(y')}, B_1^{(z')}{}^\dag] +   [B_2^{(y')}, B_2^{(z')}{}^\dag] + I^{(y')} I^{(z')}{}^\dag - J^{(z')}{}^\dag J^{(y')} &= 0\ ,\quad\text{for all}\quad y',z' = 1',\ldots,k'\ .
\end{align}
Let us get a count of the number of such equations. The above
equations are symmetric in $y'$, $z'$. For $y'=z'$, the last equation
in \eqref{insteq} is manifestly real whereas the first equation is
complex. Thus, for $y'=z'$, we have $k' \times 3k^2$ real
equations. For $y'\neq z'$, it is sufficient to restrict $y' < z'$,
and both equations in \eqref{insteq} are complex. This gives a count
of $\tfrac{1}{2}k'(k'-1)\times 4k^2$. In total, the number of
equations is $k^2k'(2k'+1)$. For $k'=1$, the target space is
$\mbb{R}^4$ and the above equations are precisely the ADHM equations.

We must also remember that the instanton connection \eqref{gaugefield}
is invariant under the residual $\text{U}(k)$ transformations
\eqref{resid}. We treat the residual $\text{U}(n)$ in \eqref{resid} as
a symmetry of framings at $\infty$ of the instanton solution. The
$B^{(y')}_1$, $B^{(y')}_2$ are inert under framing whereas the
$I^{(y')}$ and $J^{(y')}$ transform as
\begin{equation}\label{framing}
  I^{(y')} \to  I^{(y')} \wt{\mc{U}}\ ,\quad  J^{(y')} \to  \wt{\mc{U}}{}^\dag J{}^{(y')}\ .
\end{equation}
Thus, the moduli space of framed instantons is described by
\begin{equation}
  \left\{\textsc{Fields}\ \Big|\ \textsc{Equations} \right\} \Big/ \textsc{Symmetries}\ ,
\end{equation}
with
\begin{enumerate}
\item \textsc{Fields}:
  $B^{(y')}_1, B^{(y')}_2, I^{(y')}, J^{(y')}$,
  $y'=1',\ldots,k'$,
\item \textsc{Equations}: the equations \eqref{insteq}, and
\item \textsc{Symmetries}: the residual $\text{U}(k)$ symmetry in \eqref{resid} which acts on the various fields as
  \begin{equation}
    B^{(y')}_1 \to \mc{U} B^{(y')}_1 \mc{U}^\dag\ ,\quad  B^{(y')}_2 \to \mc{U} B^{(y')}_2 \mc{U}^\dag\ ,\quad     I^{(y')} \to \mc{U} I^{(y')} \ ,\quad J^{(y')} \to J{}^{(y')} \mc{U}^\dag\ .
  \end{equation}
\end{enumerate}
The virtual dimension of the moduli space of framed instantons is then
\begin{multline}
\dim_{\mbb{R}} \{\textsc{Fields}\} - \dim_{\mbb{R}}
\{\textsc{Equations}\} - \dim_{\mbb{R}} \{\textsc{Symmetries}\} \\ 
=  k'(4k^2 + 4kn) - k^2k'(2k' + 1) - k^2 = 4k'kn - k^2(2k' - 1)(k'-1)\ .
\end{multline}
When $k'= 1$, this becomes $4kn$ which is the virtual dimension (in
fact, the dimension itself) of the $\text{SU}(n)$ $k$-instanton moduli
space on $\mbb{R}^4$.

{\hfil\textbf{Acknowledgements}\hfil}\\ The authors thank N.~Nekrasov,
S.~Shatashvili and J.~P.~Ang for useful discussions during early
stages of the work. N.~P.~would like to acknowledge the support of the
Jawaharlal Nehru Postdoctoral Fellowship for part of the duration of
the work. The work of M.~R.~was supported in part by NSF grants
PHY-19-15093 and PHY-22-15093.

\appendix

\section{\tp{$(0,1)$}{(0,1)} and \tp{$(0,2)$}{(0,2)} superspace}\label{0102app}
  
\subsection{\tp{$(0,1)$}{(0,1)} superspace}\label{01superspace}
$(0,1)$ superspace has coordinates $(x^{\pm\pm}, \theta^+)$ where
$\theta^+$ is a real Grassmann variable. The corresponding
supercovariant derivatives are $(\partial_{\pm\pm}, \mc{D}_+)$ which
satisfy the algebra
\begin{equation}
  \mc{D}_+^2 = \i \partial_{++}\ .
\end{equation}
with all other commutators being zero.

Multiplets of the $(0,1)$ supersymmetry algebra are not
constrained. The most common ones are the scalar multiplet (spin $0$),
the fermi multiplet (spin $\tfrac{1}{2}$, left-handed) and the gauge
multiplet (spin $1$). The multiplets are irreducible representations
of the algebra when they are real (or hermitian).

A real scalar superfield $\phi$ has components
\begin{equation}\label{01realscalar}
  \phi_\bvert\ ,\quad  \i \xi_+ = (\mc{D}_+ \phi)_\bvert\ ,
\end{equation}
where $\phi_\bvert$ is a real scalar field and $\xi_+$ is a real
right-handed fermion. We follow the usual convention of denoting the
lowest component of a superfield by the same symbol and drop the
`slash' $\bvert$ from here on. A supersymmetric action with the lowest
number of derivatives is
\begin{align}
  \mc{S}_{\rm scalar} &= \frac{\i}{2}\int\ud^2x\, \mc{D}_+\left(-(\mc{D}_+\phi)\,\partial_{--}\phi\right) = \frac{1}{2}\int\ud^2x \left(-\partial^{\mu}{\phi}\,\partial_{\mu}\phi - \i\xi_{+} \partial_{--}\xi_{+}\right)\ .
\end{align}
A real fermi superfield $\psi_-$ has the components
\begin{equation}\label{01realfermi}
  \psi_- \ ,\quad  F = \mc{D}_+ \psi_-\ ,
\end{equation}
where $\psi_-$ is a real left-handed fermion and $F$ is a real
auxiliary field, with the action
\begin{align}
  \mc{S}_{\rm fermi}
  &=  \frac{1}{2}\int\ud^2x\, \mc{D}_+\left(\psi_- \mc{D}_+ \psi_-\right) =  \frac{1}{2} \int \ud^2x \left(-\i\psi_- \partial_{++} \psi_- + F^2 \right)\ .
\end{align}
One can add a potential term in the action via a term that is linear
in the fermi superfields $\psi_{\alpha-}$ in the theory:
\begin{align}
  \mc{S}_{\rm potential}
  &= \int \ud^2x\, \mc{D}_+ \left(\psi_{\alpha-} M^\alpha\right) = \int \ud^2x\, \left( F_\alpha M^\alpha   - \psi_{\alpha-} \frac{\partial M^\alpha}{\partial \phi_i} \xi_{i+} \right)\ ,
\end{align}
where $M^\alpha := M^\alpha(\phi)$ are functions of the scalar
superfields in the theory.

\subsection{\tp{$(0,2)$}{(0,2)} superspace}\label{02superspace}
$(0,2)$ superspace has coordinates
$(x^{\pm\pm}, \theta^+, \w\theta{}^+)$ where $\theta^+$ and
$\w\theta{}^+$ are left-handed spinors. We denote the corresponding
supercovariant derivatives by $(\partial_{++}, \uD_{+},
\w{\uD}_+)$. They satisfy the algebra
\begin{equation}
  \uD_+^2 = \w{\uD}{}^2_+ = 0\ ,\quad \{\uD_+\, ,\w{\uD}_+\} = 2\i\partial_{++}\ .
\end{equation}
We review various constrained superfields that are required to write
down supersymmetric actions in superspace.

\paragraph{Chiral} A scalar chiral superfield (or, simply a chiral
superfield) $\phi$ is a Lorentz scalar and satisfies
$\w\uD_+ \phi = 0$ and has components
\begin{equation}
  \phi\ ,\quad \w\phi \ ,\quad \sqrt{2}\,\xi_{+} := \uD_+\phi\ ,\quad -\sqrt{2}\,\w\xi_{+} := \w\uD_+\w\phi\ ,
\end{equation}
and consequently, $\w\uD_+\uD_+\phi = 2\i\partial_{++}
\phi$. The action for a free chiral superfield is
\begin{align}
  \mc S_{\text{chiral}}
  &= -\frac{\i}{2}\int\ud^2x\,\uD_+\w\uD_+\ (\w\phi\, \partial_{--} \phi) = \int\ud^2x \left(-\w{\partial^{\mu}\phi}\,\partial_{\mu}\phi - \i\w\xi_{+} \partial_{--}\xi_{+}\right)\ .
\end{align}

\paragraph{Fermi} A Fermi superfield $\psi_-$ is a left-handed spinor
and satisfies the constraint $\w\uD_+ \psi_- = 0$. It has components
\begin{equation}
\psi_-\ ,\quad \w\psi_-\ ,\quad -\sqrt{2}\,G := \uD_+\psi_-\ ,\quad -\sqrt{2}\,\w{G} := \w\uD_+\w\psi_-\ .
\end{equation}
The action for a free Fermi multiplet $\psi_-$ is
\begin{align}
  \mc S_{\text{Fermi}} &= \frac{1}{2}\int \ud^2x\, \uD_+ \w\uD_+\ (\w\psi_-\psi_-) = \int \ud^2x \left(-\i\w\psi_- \partial_{++} \psi_- + \w{G} G\right)\ .
\end{align} 
We see that the left-handed fermion $\psi_-$ satisfies the equation of
motion $\partial_{++}\psi_-=0$ and hence is right-moving on-shell. The
field $G$ is auxiliary with equation of motion $G=0$.

\paragraph{Potential terms}
Let $\phi_i$ collectively denote all the $(0,2)$ chiral superfields in
the theory and $\psi_{\alpha-}$ the $(0,2)$ Fermi superfields. We can
modify the constraint $\w\uD_+ \psi_{\alpha-} = 0$ to
\begin{equation}
  \w\uD_+ \psi_{\alpha-} = \sqrt{2} E_\alpha(\phi)\ ,
\end{equation}
where the $E_\alpha(\phi)$ are holomorphic functions of the chiral
multiplets $\phi_i$. This modification results in additional
interaction terms in the action for the fermi superfields:
\begin{align}\label{02fermiact}
  \mc S_{\text{Fermi}} &= \frac{1}{2}\int \ud^2x\, \uD_+ \w\uD_+\ (\w\psi{}^\alpha_-\psi_{\alpha-})\ ,\nonumber\\
                       &= \int \ud^2x \left(-\i\w\psi{}^\alpha_- \partial_{++} \psi_{\alpha-} + \w{G}{}^\alpha G_\alpha  - \w{E}{}^\alpha(\w\phi) E_\alpha(\phi) + \w\psi{}^\alpha_-\,\frac{\partial E_\alpha}{\partial \phi_i}\ \xi_{ i+} + \frac{\partial \w E{}^\alpha}{\partial \w\phi{}^i}\,\w\xi{}^i_{+}\,\psi_{\alpha-}\right)\ .
\end{align} 
We can also write a superpotential term, known as a ``$J$-term'' in
$(0,2)$ literature:
\begin{align}
  \mc S_{J} &=  -\frac{1}{\sqrt{2}}\int \ud^2x\, \uD_+ \left(J^{\alpha}(\phi) \psi_{\alpha-}\right)  + \text{h.c.}\ ,\nonumber\\
            &= \int \ud^2x \left(J^\alpha(\phi) G_\alpha  + \w{G}{}^\alpha \w{J}_\alpha(\w{\phi}) - \frac{\partial J^\alpha}{\partial \phi_j}\xi_{j+}\psi_{\alpha-} - \w\psi{}^\alpha_{-}\frac{\partial \w{J}_\alpha}{\partial \w{\phi}{}^j}\w\xi{}^j_{+}\right)\ .
\end{align}
Since the superspace measure in the $J$-term involves only half the
supercovariant derivatives, its invariance under $(0,2)$
supersymmetry requires the integrand to be chiral, i.e.,
$\w\uD_+ (\psi_{\alpha-} J^{\alpha}) = 0$. This implies
\begin{equation}\label{EJconstr}
  E\cdot J :=  \sum_\alpha E_\alpha J^\alpha = 0\ .
\end{equation}
If the above constraint is not satisfied, supersymmetry is softly
broken down from $(0,2)$ to $(0,1)$, even though the $J$-term is
written in $(0,2)$ superspace.

\paragraph{Reduction to $\mc{N} = (0,1)$ superspace}
Define the derivatives
\begin{equation}\label{02der01}
  \mc{D}_+ = \frac{\uD_+ + \w\uD_+}{\sqrt{2}}\ ,\ \mc{Q}_+ = \frac{\uD_+ - \w{\uD}_+}{\sqrt{2}}\quad \text{with}\quad \mc{D}_+^2 = \i\partial_{++}\ ,\quad \mc{Q}^2_+ = -\i\partial_{++}\ ,\quad \{\mc{D}_+, \mc{Q}_+\} = 0\ .
\end{equation}
$\mc{D}_+$ is the real $(0,1)$ super derivative and $\mc{Q}_+$ is
the generator of the extra (non-manifest) supersymmetry.

The $(0,2)$ chiral and fermi multiplets (and their antichiral
counterparts) become complex $(0,1)$ scalar and fermi multiplets with
components
\begin{align}
  \text{Chiral}:&\quad  \phi\ ,\quad \mc{D}_+\phi =  \xi_{+} \ , \quad \mc{D}_+\w\phi = -\w\xi_+ \ ,\nonumber\\
  \text{Fermi}:&\quad \psi_{-}\ ,\quad \mc{D}_+\psi_{-} = G + E =: F\ ,\quad \mc{D}_+\w\psi_- = \w{G} + \w{E} =: \w{F}\ .
\end{align}
% The above $(0,1)$ superfields transform as follows under the extra
% supersymmetry:
% \begin{align}
%   \wt\uQ_+ \phi_j = -i\zeta_{j+}\ ,\quad \wt\uQ_+\psi_{a-} = -i(G_a - E_a)\ , \quad \wt\uQ_+\lambda^D_- = F\ ,\quad \wt\uQ_+\lambda^F_- = -D\ .
% \end{align}
We have $\uD_+\w\uD_+ = -\i\mc{D}_+\mc{Q}_+ + \i\partial_{++}$. We can
discard the second term since it gives rise to a total derivative term
in the action. Using that $\mc{Q}_+$ acts as $-\i\mc{D}_+$ on
superfields satisfying $\w\uD_+(\cdot) = 0$, we can write the
$(0,2)$ actions in $(0,1)$ superspace:
\begin{align}\label{02in01act}
  \mc{S}_{\text{chiral}} &=\frac{\i}{2}\int\ud^2x\, \mc{D}_+\left(-\mc{D}_+\w\phi{}^i\,\partial_{--}\phi_i - \partial_{--}\w\phi{}^i\mc{D}_+\phi_{i}\right)\ ,\nonumber\\
\mc{S}_{\text{fermi}} &= \int\ud^2x\, \mc{D}_+\left(\w\psi{}^\alpha_- \left(\tfrac{1}{2}\mc{D}_+\psi_{\alpha-} - \mu_\alpha\right)  + \left(\tfrac{1}{2}\mc{D}_+\w\psi{}^\alpha - \w\mu{}^\alpha\right) \psi_{\alpha-}\right)\ ,
\end{align}
where $\mu_\alpha = E_\alpha + \w{J}_\alpha$.

\section{\tp{$(4,4)$}{(0,4)} projective superspace and  \tp{$(4,4) \to (0,4)$}{(4,4) to (0,4)}}\label{44app}

\subsection{Definitions}
We start with the $(4,4)$ real supercharges
$(\mc{Q}_{m+}, \mc{Q}_{\tilde{m}-})$ with $m$, $\tilde{m} =
1,2,3,4$. The R-symmetry group is
$$\text{Spin}(4)_L \times \text{Spin}(4)_R \simeq \text{SU}(2)_L \times
\text{SU}(2)_L' \times \text{SU}(2)_R \times \text{SU}(2)_R''\ .$$ We
restrict our attention to the subgroup $F \times F' \times F''$ where
$F = \text{SU}(2)_\Delta$, the diagonal subgroup of
$\text{SU}(2)_L \times \text{SU}(2)_R$, $F' = \text{SU}(2)_L'$ and
$F''= \text{SU}(2)_R''$. The supercharges can then be written as
$(\mc{Q}_{aa'+}, \mc{Q}_{aa''-})$ where $a$, $a'$ and $a''$ are
doublet indices of $F$, $F'$ and $F''$ respectively. This restriction
of the R-symmetry group to a subgroup seems to be required to obtain
the vector multiplet via gauged supercovariant derivatives and the
relevant superspace constraints \cite{Siegel:1984pw}.

The algebra of $(4,4)$ supercovariant derivatives $\uD_{aa'+}$ and
$\uD_{aa''-}$ is
\begin{align}
&\{\uD_{aa'+}\, ,\uD_{bb'+}\} = 2\i\varepsilon_{ab}\varepsilon_{a'b'} \partial_{++}\ ,\quad \{\uD_{aa''-}\, ,\uD_{bb''-}\} = 2\i\varepsilon_{ab}\varepsilon_{a''b''} \partial_{--}\ ,\quad \{\uD_{aa'+}\, ,\uD_{bb''-}\} = 0\ .
\end{align}
The reality conditions on the derivatives are
\begin{equation}\label{44conj}
  \uD_{aa'\pm} = \w{\uD}{}^{bb'}_\pm\varepsilon_{ba}\varepsilon_{b'a'}\ .
\end{equation}
It will be useful to define the $(2,2)$ subalgebra spanned by the
derivatives
\begin{equation}
\uD_+ := \uD_{11'+}\ ,\quad \uD_- := \uD_{11''-}\ ,\quad \w\uD_+ := \uD_{22'+}\ ,\quad \w\uD_- := \uD_{22''-}\ ,
\end{equation}
which satisfy
\begin{equation}
  \{\uD_\pm, \w\uD_\pm\} =  2\i\partial_{\pm\pm}\ .
\end{equation}
The non-manifest $(4,4)$ supersymmetry generators are then
$\uQ_+ := \uD_{21'+}$, $\w\uQ_+ := -\uD_{12'+}$ and
$\uQ_- := \uD_{21''-}$, $\w\uQ_- := -\uD_{12''-}$.

The general projective superspace corresponding to
$F \times F' \times F''$ is described by introducing a doublet for
each of the $\text{SU}(2)$s in the R-symmetry group:
$u^a = (\zeta, 1)$, $v^{a'} = (\zeta', 1)$ and
$w^{a''} = (\zeta'', 1)$ for the subgroups $F = \text{SU}(2)_\Delta$,
$F'=\text{SU}(2)'_L$ and $F''=\text{SU}(2)'_R$ respectively.

We then define the following projective supercovariant derivatives:
\begin{align}\label{44projder}
  &\pD_{a'+} := u^a \uD_{aa'+}\ ,\quad\text{i.e.,}\quad \pD_{1'+} = \zeta \uD_+ + \uQ_+\ ,\quad \pD_{2'+} = -\zeta \w\uQ_+ + \w\uD_+\ ,\nonumber\\
  &\pD_{a+} := v^{a'} \uD_{aa'+}\ ,\quad\text{i.e.,}\quad \pD_{1+} = \zeta' \uD_+ - \w\uQ_+\ ,\quad \pD_{2+} = \zeta' \uQ_+ + \w\uD_+\ ,\nonumber\\
  &\pD_{a''-} := u^a \uD_{aa''-}\ ,\quad\text{i.e.,}\quad \pD_{1''-} = \zeta \uD_- + \uQ_-\ ,\quad \pD_{2''-} = -\zeta \w\uQ_- + \w\uD_-\ ,\nonumber\\
  &\pD_{a-} := w^{a''} \uD_{aa''-}\ ,\quad\text{i.e.,}\quad \pD_{1-} = \zeta'' \uD_- - \w\uQ_-\ ,\quad \pD_{2-} = \zeta'' \uQ_- + \w\uD_-\ .
\end{align}
We also introduce the doublets $\wt{u}{}^a$, $\wt{v}{}^{a'}$ and
$\wt{w}{}^{a''}$ as was done in the main text above
eq.~\eqref{projder}. We again choose $\wt{u}{}^a = ( 1,0)$,
$\wt{v}{}^{a'} = (1,0)$ and $\wt{w}{}^{a''} = (1,0)$ and define
the linearly independent derivatives:
\begin{align}\label{44projdertl}
  &\wt\pD_{a'+} := \wt{u}^a \uD_{aa'+}\ ,\quad\text{i.e.,}\quad \wt\pD_{1'+} =  \uD_+\ ,\quad \wt\pD_{2'+} =  -\w\uQ_+\ ,\nonumber\\
  &\wt\pD_{a+} :=  \wt{v}^{a'} \uD_{aa'+}\ ,\quad\text{i.e.,}\quad \wt\pD_{1+} =  \uD_+\ ,\quad \wt\pD_{2+} =  \uQ_+\ ,\nonumber\\
  &\wt\pD_{a''-} := \wt{u}^a \uD_{aa''-}\ ,\quad\text{i.e.,}\quad \wt\pD_{1''-} = \uD_-\ ,\quad \wt\pD_{2''-} = - \w\uQ_-\ ,\nonumber\\
  &\wt\pD_{a-} :=  \wt{w}^{a''} \uD_{aa''-}\ ,\quad\text{i.e.,}\quad \wt\pD_{1-} = \uD_- \ ,\quad \wt\pD_{2-} = \uQ_-\ .
\end{align}
We consider projective superfields which are functions of one
projective coordinate from the left moving sector ($\zeta$ or
$\zeta'$) and one projective coordinate from the right-moving sector
($\zeta$ or $\zeta''$) and are annihilated by the corresponding set of
projective derivatives. For example, an $(F, F'')$ projective
superfield $\Phi$ is a function of $\zeta$ and $\zeta''$ and is
annihilated by $\pD_{a'+}(\zeta)$ and $\pD_{a-}(\zeta'')$. The
$(4,4)$ supersymmetric action is
\begin{align}
  \mc{S}[\bm\Phi] &= \int\ud^2x \oint_\gamma\frac{\ud\zeta}{2\pi\i} \oint_{\gamma''}\frac{\ud\zeta''}{2\pi\i}\,
                        \wt\pD_{1'+}\wt\pD_{2'+}\wt\pD_{1-}\wt\pD_{2-}\,\mK(\bm\Phi)\ .
\end{align}
Using that $\mK(\bm\Phi)$ is annihilated by $\pD_{a'+}$ and $\pD_{a-}$
and $\wt\pD_{2'+} = \zeta^{-1}\pD_{2'+} - \zeta^{-1}\w\uD_+$,
$\wt\pD_{2-} = \zeta''^{-1} \pD_{2-} - \zeta''^{-1} \w\uD_-$, we can
replace the measure by the $(2,2)$ measure and do the $\zeta$, $\zeta''$ integrals to get an action in $(2,2)$ superspace:
\begin{align}
  \mc{S}[\bm\Phi] &= \int\ud^2x\, \uD_+\w\uD_+\uD_-\w\uD_- \oint_\gamma\frac{\ud\zeta}{2\pi\i\zeta} \oint_{\gamma''}\frac{\ud\zeta''}{2\pi\i\zeta''}\,
                        \mK(\bm\Phi)\ .
\end{align}
There are many choices for projective superfields: they can be a
polynomial or a power series in each of the projective coordinates
that they depend on. A polynomial $\mc{O}(n)$ superfield with respect
to $F$, $F'$ and $F''$ will be respectively denoted as $\mc{O}(n)$,
$\mc{O}(n')$ and $\mc{O}(n'')$. Power series superfields are typically
denoted as $F$-arctic, $F$-antarctic, $F'$-arctic and so on. Below, we
discuss the $(F,F)$ arctic superfield, i.e., an arctic superfield
which is a power series only in $\zeta$ and is annihilated by
$\pD_{a'+}(\zeta)$ and $\pD_{a''-}(\zeta)$.

\subsection{\tp{$(4,4)$}{N=(4,4)} standard hypermultiplet}\label{44stdhyp}
Consider an $(F,F)$ arctic superfield
$\bUpsilon(\zeta) = \sum_{i = 0}^\infty \iUpsilon_i \zeta^i$ with
alternate notation $\Phi$ and $\Sigma$ for $\iUpsilon_0$ and
$\iUpsilon_1$ respectively. The constraints
$\pD_{a'+}\bUpsilon = \pD_{a''-}\bUpsilon = 0$ give the $(2,2)$
constraints
\begin{align}
  &\uQ_{\pm} \Phi = 0\ ,\quad \w{\uD}_{\pm}\Phi = 0\ ,\quad \w{\uQ}_\pm\Phi = \w{\uD}_\pm \Sigma \Rightarrow \w{\uD}_+\w{\uD}_- \Sigma = 0\ ,\nonumber\\
  &\uQ_{\pm} \iUpsilon_{j+1} = -\uD_{\pm} \iUpsilon_{j}\ \text{for}\ j\geq 0\
    ,\quad\text{and}\quad \w{\uQ}_\pm \iUpsilon_j = \w{\uD}_\pm \iUpsilon_{j+1}\ \text{for}\ j \geq 1\ .
\end{align}
$\Phi$ is chiral as an $(2,2)$ superfield since
$\w{\uD}_\pm \Phi = 0$, $\Sigma$ is complex linear since
$\w\uD_+\w\uD_- \Sigma = 0$, whereas the $\iUpsilon_{j\geq 2}$ are
unconstrained as $(2,2)$ superfields.

The action for the arctic superfield is
\begin{align}\label{44hyperact1}
  \mc{S}[\bUpsilon] = \frac{1}{4}\int\ud^2x\oint_{\gamma}\frac{\ud\zeta}{2\pi \i}\wt\pD_{1'+}\wt\pD_{2'+}\wt\pD_{1''-}\wt\pD_{2''-}\, (\zeta\w\bUpsilon\bUpsilon)\ .
\end{align}
This action is R-symmetric since the measure has $F$-weight $-2$ (
$+2$ from $\ud\zeta$, $-2$ from $\wt\pD_{1'+}\wt\pD_{2'+}$ and $-2$
from $\wt\pD_{1''-}\wt\pD_{2''-}$) and the Lagrangian has $F$-weight
$+2$ ($+1$ each from $\bUpsilon$ and $\zeta\w\bUpsilon$, see the
paragraph after equation \eqref{Rarctic} in Section \ref{Rsym}).

Next, we obtain the $(0,4)$ content by applying $\wt\pD_{a''-}$ to
$\bUpsilon$:
\begin{align}\label{44hypercomp}
  \bUpsilon_{a''-} &\equiv \frac{1}{\sqrt{2}}\wt\pD_{a''-}\bUpsilon\ ,\quad \bUpsilon_{--} \equiv -\frac{1}{4}\wt\pD_{1''-}\wt\pD_{2''-}\bUpsilon\ .
\end{align}
Recall from \eqref{derconjarctic} that the conjugate of
$\wt\pD_{a''-}$ when acting on arctic superfields is
\begin{equation}
  \breve{\wt\pD}{}^{a''}_- = \varepsilon^{a''b''}(-\zeta\wt\pD_{b''-} +
\pD_{b''-})\ .
\end{equation}
Using this, we get
\begin{equation}
  \w\bUpsilon{}^{a''}_- = \frac{1}{\sqrt{2}}\varepsilon^{a''b''} (\zeta\wt\pD_{b''-} - \pD_{b''-})\w\bUpsilon\ ,\quad \w{\bUpsilon}_{--} = \frac{1}{4}(-\zeta \wt\pD_{2''-} + \pD_{2''-}) (\zeta\wt\pD_{1''-} - \pD_{1''-})\w\bUpsilon\ .
\end{equation}
Using $\pD_{a''-}\w\bUpsilon = 0$ and
$\{\pD_{2''-}\,, \wt\pD_{1''-}\} = 2\i \partial_{--}$, we get
\begin{equation}\label{bardef}
  \w\bUpsilon{}^{a''}_- = \frac{1}{\sqrt{2}} \zeta\varepsilon^{a''b''}\wt\pD_{b''-}\w\bUpsilon\ ,\quad \w{\bUpsilon}_{--} = \frac{1}{4} \zeta^2 \wt\pD_{1''-} \wt\pD_{2''-} \w\bUpsilon + \frac{\i}{2} \zeta\partial_{--} \w\bUpsilon\ .
\end{equation}
The $(0,4)$ supersymmetric action is obtained by pushing in the $\wt\pD_{a''-}$ derivatives in the measure:
\begin{align}\label{44hyperact}
  \mc{S}[\bUpsilon]
  &= \frac{1}{4}\int\ud^2x\oint_{\gamma}\frac{\ud\zeta}{2\pi \i}\wt\pD_{1'+}\wt\pD_{2'+}\wt\pD_{1''-}\wt\pD_{2''-}\, (\zeta\w\bUpsilon\bUpsilon)\ ,\nonumber\\
  % &= \frac{1}{4}\int\ud^2x\oint_{\gamma}\frac{\ud\zeta}{2\pi \i}\wt\pD_{1'+}\wt\pD_{2'+}\wt\pD_{1''-}\, (\zeta(\wt\pD_{2''-}\w\bUpsilon)\bUpsilon + \zeta \w\bUpsilon \wt\pD_{2''-}\bUpsilon)\ ,\nonumber\\
  % &= \frac{1}{4}\int\ud^2x\oint_{\gamma}\frac{\ud\zeta}{2\pi \i}\wt\pD_{1'+}\wt\pD_{2'+}\, \zeta\Big((\wt\pD_{1''-}\wt\pD_{2''-}\w\bUpsilon)\bUpsilon - (\wt\pD_{2''-}\w\bUpsilon)\wt\pD_{1''-}\bUpsilon +  \wt\pD_{1''-}\w\bUpsilon \wt\pD_{2''-}\bUpsilon +  \w\bUpsilon \wt\pD_{1''-}\wt\pD_{2''-}\bUpsilon\Big)\ ,\nonumber\\
  &= \int\ud^2x\oint_{\gamma}\frac{\ud\zeta}{2\pi \i}\wt\pD_{1'+}\wt\pD_{2'+}\, (\zeta^{-1}\w{\bUpsilon}_{--} \bUpsilon - \tfrac{\i}{2} (\partial_{--} \w\bUpsilon) \bUpsilon - \tfrac{1}{2}\w\bUpsilon{}^{1''}_-\bUpsilon_{1''-} - \tfrac{1}{2}\w\bUpsilon{}^{2''}_-\bUpsilon_{2''-}  - \zeta \w\bUpsilon\bUpsilon_{--})\ .
\end{align}
Let us study the R-symmetry invariance of the above action in more
detail. Recall that $\bUpsilon$ and $\wt\pD_{a''-}$
(see \eqref{derconjarctic}) transform under $F$ as
\begin{equation}\label{upsdtltr}
\bUpsilon(\zeta) \to \bUpsilon'(\zeta) = j(g,\zeta) \bUpsilon(g\cdot\zeta)\ ,\quad  \wt\pD_{a''-} \to j(g,\zeta)^{-1}\wt\pD_{a''-}(g\cdot \zeta) - \w{b} \pD_{a''-}(g\cdot\zeta)\ ,
\end{equation}
where $\bUpsilon'$ is a new superfield which is evaluated at $\zeta$
whose expression is given by expanding the right hand side
$j(g,\zeta) \bUpsilon(g\cdot\zeta)$ around $\zeta = 0$. The
transformations of all the other superfields can be obtained by using
the above. We first summarize the results and then detailed
calculations. The hypers $\bUpsilon$, $\zeta\w\bUpsilon$ transform as
weight $1$ objects:
\begin{equation}
  \bUpsilon(\zeta) \to j(g,\zeta) \bUpsilon(g\cdot\zeta)\ ,\quad   \zeta\w\bUpsilon(-\zeta^{-1}) \to j(g,\zeta)\ (g\cdot\zeta)\ \w\bUpsilon(-(g\cdot\zeta)^{-1})\ ,
\end{equation}
the fermis $\bUpsilon_{a''-}$, $\w\bUpsilon{}^{a''}_-$ transform as
weight $0$ objects:
\begin{equation}
  \bUpsilon_{a''-}(\zeta) \to \bUpsilon_{a''-}(g\cdot\zeta)\ ,\quad    \w\bUpsilon{}^{a''}_-(-\zeta^{-1}) \to \w\bUpsilon{}^{a''}_-(-(g\cdot\zeta)^{-1})\ ,
\end{equation}
and $\bUpsilon_{--}$, $\zeta^{-1}\w\bUpsilon_{--}$ transform as weight $-1$
objects, along with an additional shift:
\begin{align}\label{upmmtrfin}
  \bUpsilon_{--}(\zeta) &\to j(g,\zeta)^{-1}\bUpsilon_{--}(g\cdot\zeta) - \frac{\i}{2} \w{b} \partial_{--}\bUpsilon(g\cdot\zeta)\ ,\nonumber\\
  \zeta^{-1}\w\bUpsilon'_{--}(-\zeta^{-1}) &\to j(g,\zeta)^{-1} (g\cdot\zeta)^{-1}  \w\bUpsilon_{--}(-(g\cdot\zeta)^{-1}) + \zeta^{-1}b\frac{\i}{2}\partial_{--} \w\bUpsilon(-(g\cdot\zeta)^{-1})\ .
\end{align}
Using these, the $(0,4)$ supersymmetric action \eqref{44hyperact} can
be checked to be R-symmetric, a fact which was already demonstrated
for the $(4,4)$ action \eqref{44hyperact1}.

\subsubsection*{The derivation of R-symmetry
  transformations}

\textbf{Note:} In the following calculations, a $'$ on superfields
denotes the transformed superfield and must not be confused with the
$'$ on the R-symmetry indices.

Given the transformation of $\bUpsilon$ in \eqref{upsdtltr}, the
transformation of $\w\bUpsilon$ is
\begin{equation}\label{upbartr}
\w\bUpsilon(-\zeta^{-1}) \to \w\bUpsilon'(-\zeta^{-1}) = (a + b\zeta^{-1}) \w\bUpsilon\left(\frac{-\w{a}\zeta^{-1} + \w{b}}{a + b \zeta^{-1}}\right) = \frac{1}{\zeta} \times (\w{a} - \w{b}\zeta) \times \frac{a \zeta + b}{\w{a} - \w{b}\zeta}  \times \w\bUpsilon\left(\frac{-\w{a}\zeta^{-1} + \w{b}}{a + b \zeta^{-1}}\right)\ ,
\end{equation}
which implies that
\begin{equation}\label{upbartr1}
\zeta \w\bUpsilon(-\zeta^{-1}) \to  \zeta\w\bUpsilon'(-\zeta^{-1}) = j(g,\zeta)\  (g\cdot\zeta) \ \w\bUpsilon(-(g\cdot\zeta)^{-1})\ .
\end{equation}
This tells us that $\zeta\w\bUpsilon$ transforms as a weight $1$ field
as well. $\bUpsilon_{--}$ transforms as
\begin{align}\label{upmmtr}
  &\bUpsilon_{--}(\zeta)\nonumber\\
%  &\to -\frac{1}{4}\Big(j(g,\zeta)^{-1}\wt\pD_{1''-}(g\cdot \zeta) - \w{b} \pD_{1''-}(g\cdot\zeta)\Big)\Big(j(g,\zeta)^{-1}\wt\pD_{2''-}(g\cdot \zeta) - \w{b} \pD_{2''-}(g\cdot\zeta)\Big) \Big( j(g,\zeta) \bUpsilon(g\cdot\zeta)\Big)\ ,\nonumber\\
  &\to -\frac{1}{4}\Big(j(g,\zeta)^{-1}\wt\pD_{1''-}(g\cdot \zeta) - \w{b} \pD_{1''-}(g\cdot\zeta)\Big) \Big(j(g,\zeta)^{-1}\wt\pD_{2''-}(g\cdot \zeta)\Big) \Big( j(g,\zeta) \bUpsilon(g\cdot\zeta)\Big)\ ,\nonumber\\
  &=  j(g,\zeta)^{-1} \bUpsilon_{--}(g\cdot\zeta) - \w{b} \frac{\i}{2} \partial_{--}\bUpsilon(g\cdot\zeta)\ ,
\end{align}
that is,
\begin{equation}
  \bUpsilon_{--}(\zeta) \to \bUpsilon'_{--}(\zeta) = j(g,\zeta)^{-1} \bUpsilon_{--}(g\cdot\zeta) - \w{b}\frac{\i}{2}\partial_{--}\bUpsilon(g\cdot\zeta)\ .
\end{equation}
Thus, $\bUpsilon_{--}$ transforms as a weight $-1$ superfield but with
an additional shift term proportional to
$\partial_{--}\bUpsilon$. Finally, we need the transformation of
$\w\bUpsilon_{--}$. Analogous to \eqref{upbartr}, we have
\begin{align}
  \w\bUpsilon'_{--}(-\zeta^{-1})
  &= \frac{1}{a + b \zeta^{-1}} \w\bUpsilon_{--}\left(\frac{-\w{a}\zeta^{-1} + \w{b}}{a + b \zeta^{-1}}\right) + b \frac{\i}{2}\partial_{--}\w\bUpsilon\left(\frac{-\w{a}\zeta^{-1} + \w{b}}{a + b \zeta^{-1}}\right)\ ,\nonumber\\
  &= \zeta \frac{\w{a} - \w{b}\zeta}{a\zeta + b }\frac{1}{\w{a} - \w{b}\zeta} \w\bUpsilon_{--}\left(\frac{-\w{a}\zeta^{-1} + \w{b}}{a + b \zeta^{-1}}\right) + b \frac{\i}{2}\partial_{--}\w\bUpsilon\left(\frac{-\w{a}\zeta^{-1} + \w{b}}{a + b \zeta^{-1}}\right)\ ,
\end{align}
which gives
\begin{equation}\label{barupmmtr}
  \zeta^{-1}\w\bUpsilon'_{--}(-\zeta^{-1}) = j(g,\zeta)^{-1} (g\cdot\zeta)^{-1}  \w\bUpsilon_{--}(-(g\cdot\zeta)^{-1}) + \zeta^{-1}b\frac{\i}{2}\partial_{--} \w\bUpsilon(-(g\cdot\zeta)^{-1})\ .
\end{equation}
Again, we see that $\zeta^{-1}\w\bUpsilon_{--}$ transforms as a weight
$-1$ superfield, along with an additional shift term proportional to
$\partial_{--} \w\bUpsilon$. We can also start with the definition of
$\w\bUpsilon_{--}$ in \eqref{bardef} and arrive at the above
result. In detail, we have
\begin{align}
  &\zeta\wt\pD_{1''-}\wt\pD_{2''-}\w\bUpsilon(-\zeta^{-1})\nonumber\\
  &\qquad\to \zeta\wt\pD_{1''-}\wt\pD_{2''-}\w\bUpsilon'(-\zeta^{-1})\nonumber\\
  &\qquad= \zeta\Big(j(g,\zeta)^{-1}\wt\pD_{1''-}(g\cdot \zeta) - \w{b} \pD_{1''-}(g\cdot\zeta)\Big) \Big(j(g,\zeta)^{-1}\wt\pD_{2''-}(g\cdot \zeta)\Big) \frac{a\zeta+b}{\zeta} \w\bUpsilon(-(g\cdot\zeta)^{-1})\nonumber\\
  &\qquad=  j(g,\zeta)^{-1}\ g\cdot\zeta\  \wt\pD_{1''-}\wt\pD_{2''-}\w\bUpsilon(-(g\cdot\zeta)^{-1}) + 2\i \w{b}\ g\cdot\zeta\ \partial_{--}\w\bUpsilon(-(g\cdot\zeta)^{-1})\ .
\end{align}
Also using the transformation of $\w\bUpsilon$ from \eqref{upbartr1},
we get
\begin{align}\label{barupmmsimplify}
  &\zeta \wt\pD_{1''-}\wt\pD_{2''-}\w\bUpsilon'(-\zeta^{-1}) + 2\i\partial_{--}\w\bUpsilon'(-\zeta^{-1})\nonumber\\
  &=  j(g,\zeta)^{-1} \ g\cdot\zeta \  \wt\pD_{1''-}\wt\pD_{2''-}\w\bUpsilon(-(g\cdot\zeta)^{-1}) + 2\i \left(\w{b}\ g\cdot\zeta + \frac{a\zeta+b}{\zeta}\right)  \partial_{--}\w\bUpsilon(-(g\cdot\zeta)^{-1}) \nonumber\\
  &=  j(g,\zeta)^{-1}\left(g\cdot\zeta \  \wt\pD_{1''-}\wt\pD_{2''-}\w\bUpsilon(-(g\cdot\zeta)^{-1}) + 2\i \partial_{--}\w\bUpsilon_{--}(-g\cdot\zeta)^{-1})\right)\nonumber\\
  &\qquad\qquad\qquad\qquad\qquad\qquad\qquad\qquad+ 2\i \left(\w{b}\ g\cdot\zeta + \frac{a\zeta+b}{\zeta} - j(g,\zeta)^{-1}\right)  \partial_{--}\w\bUpsilon(-(g\cdot\zeta)^{-1})\ .
\end{align}
The quantity in the parentheses in the last line simplifies to give
$b/\zeta$. Plugging this into \eqref{barupmmsimplify} and dividing by
$4$, we get \eqref{barupmmtr}.

The fermis $\bUpsilon_{a''-}$ transform as weight $0$ objects:
\begin{multline}\label{upfertr}
  \bUpsilon_{a''-}(\zeta) \to \Big(j(g,\zeta)^{-1}\wt\pD_{a''-}(g\cdot \zeta) - \w{b} \pD_{a''-}(g\cdot\zeta)\Big) \Big( j(g,\zeta) \bUpsilon(g\cdot\zeta)\Big) \\ = \wt\pD_{a''-}(g\cdot \zeta)\bUpsilon(g\cdot\zeta) = \bUpsilon_{a''-}(g\cdot\zeta)\ .
\end{multline}
The conjugates $\w\bUpsilon{}^{a''}_{-}$ also transform with weight
$0$, a fact which can be seen either by complex conjugating the
expressions in \eqref{upfertr} or by direct calculation using the
expression for $\w\bUpsilon{}^{a''}_-$ in \eqref{bardef}:
\begin{align}
  \zeta\wt\pD_{a''-}\w\bUpsilon(-\zeta^{-1}) \to \zeta\wt\pD_{a''-}\w\bUpsilon'(-\zeta^{-1})
  &= \zeta\Big(j(g,\zeta)^{-1}\wt\pD_{a''-}(g\cdot \zeta) - \w{b} \pD_{a''-}(g\cdot\zeta)\Big)\frac{a\zeta+b}{\zeta}\w\bUpsilon(-(g\cdot\zeta)^{-1})\nonumber\\
  &= g\cdot\zeta\wt\pD_{a''-}(g\cdot \zeta)\w\bUpsilon(-(g\cdot\zeta)^{-1})\ ,
\end{align}
which gives
\begin{equation}
  \w\bUpsilon{}^{a''}_-(-\zeta^{-1}) \to   \w\bUpsilon{}^{a''}_-(-(g\cdot\zeta)^{-1})\ .
\end{equation}

\section{Component actions}\label{components}
In this appendix, we derive the action for the ordinary space
components of the various superfields in two ways: (1) by reducing to
$(0,2)$ superspace and using standard results from Appendix
\ref{02superspace}, and (2) by reducing directly to ordinary space by
pushing in the $\wt\pD_{a'+}$ in the superspace measure.

\subsection{\tp{$(0,4) \to (0,2) \to (0,0)$}{(0,4) to (0,2) to (0,0)}}
  Recall from Sections \ref{hypermult} and \ref{fermimult} the $\zeta$
and $\zeta'$ expansions of the various superfields:
\begin{equation}
  \bUpsilon = \sum_{j=0}^\infty \iUpsilon_{j}\ ,\quad  \bUpsilon_{--} = \sum_{j=0}^\infty \iUpsilon_{j--}\ ,\quad  \bUpsilon_- = \sum_{j=0}^\infty \iUpsilon_{j-}\ ,\quad \mH = \zeta' H_{1'} + H_{2'}\ .
\end{equation}
Also recall that we relabelled some low-lying components of the above
superfields since they were constrained as $(0,2)$ superfields:
\begin{equation}\label{apprelabel}
 \iUpsilon_{0} \to \eta_2\ ,\quad \iUpsilon_1 \to \eta_1\ ,\quad \text{and}\quad \iUpsilon_{0-} \to \psi_-\ .
\end{equation}
We reproduce here the projective superspace constraints on the various
superfields given in \eqref{condconst}:
\begin{alignat}{2}\label{appcondconst}
\pD_+\bUpsilon &= 0\ ,\quad  \pD_{+} \bUpsilon_{-} = -\sqrt{2}\wh{\mC}\bUpsilon\ ,\quad &&\pD_{+} \bUpsilon_{--} = \frac{1}{\sqrt{2}}\mC\bUpsilon_-\ ,\quad \pD_+ \mH = 0\ ,\nonumber\\
\pD_+\w\bUpsilon &= 0\ ,\quad  \pD_{+} \w\bUpsilon_- = \sqrt{2}\zeta\w\bUpsilon\w{\wh{\mC}}\ ,\quad &&\pD_{+} \w\bUpsilon_{--} = \frac{1}{\sqrt{2}}\zeta\w\bUpsilon_-\w{\mC}\ ,\quad \pD_+ \w{\mH} = 0\ ,
\end{alignat}
The actions are given by
\begin{align}\label{appactions}
  \mc{S}_F &= \int\ud^2x \oint \frac{\ud\zeta}{2\pi\i} \wt\pD_{1'+}\wt\pD_{2'+} \left(\tfrac{\i}{2}\w\bUpsilon \partial_{--} \bUpsilon - \zeta\w\bUpsilon\bUpsilon_{--} + \zeta^{-1}\w\bUpsilon_{--}\bUpsilon - \tfrac{1}{2}\w\bUpsilon_{-}\bUpsilon_{-}\right)\ ,\nonumber\\
  \mc{S}_{F'} &= \int\ud^2x \oint \frac{\ud\zeta'}{2\pi\i} \wt\pD_{1+}\wt\pD_{2+} \left(-\tfrac{\i}{2}\zeta'^{-1}\w{\mH} \partial_{--} \mH\right)\ ,
\end{align}
The closure of the projective superspace algebra $\pD_+^2 = 0$ on
$\bUpsilon_{--}$ gives the constraints
\begin{equation}\label{appcoupcons}
  \pD_+ \mC = \pD_+ \wh{\mC} = 0\ ,\quad \mC\wh{\mC} = 0\ ,\quad\text{i.e.,}\quad \pD_{a+}\mC = \pD_{a+} \wh\mC = 0\ ,\quad C_{(a'} \wh{C}_{b')} = 0\ ,
\end{equation}
and the $(0,4)$ invariance of the above actions gives
\begin{equation}\label{reality}
  \w\mC = \wh{\mC}\ ,\quad\text{that is}\ ,\quad \wh{C}_{a'} = \w{C}{}^{b'} \varepsilon_{b'a'}\ .
\end{equation}
The assumption that the $\mC$ are polynomials in the various
superfields constrains $\mC$ to take the form $\mC = \mK + L \mH$. The
constraints \eqref{appcondconst} lead to the following $E$-terms for
the $(0,2)$ superfield $\psi_- = \iUpsilon_{0-}$ and
$\iUpsilon_{0--}$:
\begin{equation}
  \w\uD_+ \psi_- = -\sqrt{2} E = - \sqrt{2} \wh{C}_{2'} \eta_2\\ ,\quad   \w\uD_+ \iUpsilon_{0--} = \frac{1}{\sqrt{2}} C_{2'} \psi_{-}\ .
\end{equation}
Integrating out the auxiliary superfield $\bUpsilon_{--}$ proceeds in
the same way as in the free case, with one important difference due to
the $E$-term for $\iUpsilon_{0--}$ above. Unconstraining
$\iUpsilon_{0--}$ in the standard way (see Footnote \ref{X0loose}), we
get
\begin{equation}
  -\uD_+\w\uD_+ (\w\iUpsilon_1 \iUpsilon_{0--} + \Lambda_- (\w\uD_+ \iUpsilon_{0--} - \tfrac{1}{\sqrt{2}} C_{2'} \psi_{-}))\ .
\end{equation}
Integrating out $\iUpsilon_{0--}$ gives
$\w\iUpsilon_1 = -\w\uD_+\Lambda_-$ which implies that $\w\iUpsilon_1$
is a $(0,2)$ chiral superfield which we labelled as $\w\eta{}^1$. In
addition, there is now a $(0,2)$ $J$-term:
\begin{equation}
  -\tfrac{1}{\sqrt{2}}\uD_+\w\uD_+ (- \Lambda_- C_{2'}\psi_{-}) = -\tfrac{1}{\sqrt{2}}\uD_+ (\w\eta{}^1 C_{2'} \psi_{-} -  \sqrt{2}  \Lambda_- C_{2'}\wh{C}_{2'}\eta_2) = -\tfrac{1}{\sqrt{2}}\uD_+ (\w\eta{}^1 C_{2'} \psi_{-})\ ,
\end{equation}
where, in the last equality, we have used the constraint
$C_{2'} \wh{C}_{2'} = 0$ that follows from \eqref{appcoupcons}. 

Rewriting the projective superspace measure in \eqref{appactions} as
$-\zeta^{-1} \uD_+\w\uD_+$ and performing the $\zeta$- and
$\zeta'$-integrals, we get the following $(0,2)$ superspace actions:
\begin{align}
  \mc{S}_{F'} &= \int\ud^2x\, \uD_+\w\uD_+\left(\tfrac{\i}{2}\w{H}{}^{1'}\partial_{--}H_{1'} - \tfrac{\i}{2}\w{H}{}^{2'}\partial_{--}H_{2'}\right)\ ,\nonumber\\
  \mc{S}_F &=  \int \ud^2x\, \uD_+\w\uD_+ \Big(\tfrac{\i}{2}\w\eta{}^1 \partial_{--} \eta_1 - \tfrac{\i}{2}\w\eta{}^2 \partial_{--} \eta_2 + \tfrac{1}{2}\w\psi_{-}\psi_{-}\Big) + \int \ud^2x\,\uD_+ ( - \tfrac{1}{\sqrt{2}}\w\eta{}^1 C_{2'}\psi_{-}) + \text{c.c.}\ ,
\end{align}
with $\w\uD_+ \psi_- = \sqrt{2}E = -\sqrt{2} \wh{C}_{2'} \eta_2$.

Now we further push in the derivatives in the $(0,2)$ actions above
and compute the component actions according to Appendix
\ref{02superspace}.  Recall that the superspace components of $H_{a'}$
are
\begin{equation}
\sqrt{2}  \xi_{1+} =  \uD_+ H_{2'}\ ,\quad \sqrt{2}\w\xi{}^1_+ = - \w\uD_+ \w{H}{}^{2'}\ ,\quad \sqrt{2}\xi_{2+} = - \w\uD_+ H_{1'}\ ,\quad \sqrt{2}\w\xi{}^{\,2}_+ = \uD_+ \w{H}{}^{1'}\ ,
\end{equation}
and the superspace components of $\eta_a$ are
\begin{equation}
 \sqrt{2} \xi_{1'+} = \uD_+ \eta_2 \ ,\quad \sqrt{2} \w\xi{}^{1'}_+ = -\w\uD_+\w\eta{}^2\ ,\quad  -\sqrt{2} \xi_{2'+} = \w\uD_+ \eta_1\ ,\quad \sqrt{2} \w\xi{}^{2'}_+ = \uD_+ \w\eta{}^1\ .
\end{equation}
The components of the fermi $\psi_{-}$ are
\begin{equation}
  \uD_+\psi_{-} = -\sqrt{2}G\ ,\quad \w\uD_+ \w\psi_{-} = -\sqrt{2} \w{G}\ .
\end{equation}
Let us work out the twisted hyper part of $\mc{S}_{F'}$ first. We have
\begin{align}\label{twcomp}
  \mc{S}_{F'}[H_{a'}] &= \int\ud^2x\,\uD_+ (-\tfrac{\i}{\sqrt{2}} \w{H}{}^{1'} \partial_{--} \xi_{2+} + \frac{\i}{\sqrt{2}}\w\xi{}^{1}_+ \nabla_{--} H_{2'}) = \int\ud^2x\,( - \w{{\partial}_\mu H}{}^{a'} \partial_\mu H_{a'}  - {\i}\w\xi{}^a_+ \partial_{--} \xi_{a+})\ .
\end{align}
The standard hyper part of $\mc{S}_{F}$ is given by
\begin{align}\label{stcomp}
  \mc{S}_{F}[\eta_{a}] &= \int\ud^2x\,\uD_+ ( - \tfrac{\i}{\sqrt{2}} \w{\eta}{}^{1} \partial_{--} \xi_{2'+} + \tfrac{\i}{\sqrt{2}}\w\xi{}^{1'}_+ \partial_{--} \eta_{2}) = \int\ud^2x\,( - \w{{\partial}_\mu \eta}{}^{a} \partial_\mu \eta_{a}  - {\i}\w\xi{}^{a'}_+ \partial_{--} \xi_{a'+})\ ,
\end{align}
whereas the fermi part of $\mc{S}_{F}$ is given by
\begin{align}\label{fercomp02}
  \mc{S}_F[\psi_{-}]
  &= \frac{1}{\sqrt{2}}\int \ud^2x\,\uD_+ (- \w{G} \psi_{-} - \w\psi_{-} E) + \int\ud^2x\,\left(\uD_+(- \tfrac{1}{\sqrt{2}}\w\eta{}^1 C_{2'} \psi_{-}) + \text{c.c.}\right)\ ,\nonumber\\
  &=\int \ud^2x\,\Big(\i(\partial_{++}\w\psi_{-})  \psi_{-} + \w{G} G - \w\eta{}^2 \w{\wh{C}}{}^{2'}\wh{C}_{2'} \eta_2 \nonumber\\
  &\qquad\qquad\qquad + \tfrac{1}{\sqrt{2}}\w\eta{}^2 (\w\uD_+ \w{\wh{C}}{}^{2'}) \psi_{-} - \tfrac{1}{\sqrt{2}} \w\psi_{-} (\uD_+\wh{C}_{2'})\eta_2 - \w\xi{}^{1'}_+ \w{\wh{C}}{}^{2'} \psi_{-} - \w\psi_{-} \wh{C}_{2'}\xi_{1'+}\Big) \nonumber\\
  &+ \int\ud^2x\,\Big( - \w\xi{}^{2'}_+ C_{2'} \psi_{-} - \tfrac{1}{\sqrt{2}} \w\eta{}^1 (\uD_+ C_{2'}) \psi_{-} + \w\eta{}^1 C_{2'} G + \text{c.c.}\Big)\ ,
\end{align}
Integrating out the auxiliary fields $G$, $\w{G}$, we get
\begin{equation}
 \w{G} = -\w\eta{}^1 C_{2'}\ ,\quad G = -\w{C}{}^{2'} \eta_1\ ,
\end{equation}
and the fermi action becomes
\begin{align}\label{fercomp}
  \mc{S}_F[\psi_{-}]
  &= \int \ud^2x\,\Big(\i(\uD_{++}\w\psi_{-})  \psi_{-}  - \w\eta{}^1 C_{2'} \w{C}{}^{2'}  \eta_1 - \w\eta{}^2 \w{\wh{C}}{}^{2'}\wh{C}_{2'} \eta_2 \nonumber\\
  &\qquad\qquad\qquad + \tfrac{1}{\sqrt{2}}\w\eta{}^2 (\w\uD_+ \w{\wh{C}}{}^{2'}) \psi_{-} - \tfrac{1}{\sqrt{2}} \w\psi_{-} (\uD_+\wh{C}_{2'})\eta_2 - \w\xi{}^{1'}_+ \w{\wh{C}}{}^{2'} \psi_{-} - \w\psi_{-} \wh{C}_{2'}\xi_{1'+} \Big) \nonumber\\
  &+ \int\ud^2x\,\Big( - \w\xi{}^{2'}_+ C_{2'} \psi_{-} - \tfrac{1}{\sqrt{2}} \w\eta{}^1 (\uD_+ C_{2'}) \psi_{-} + \text{c.c.}\Big)\ .
\end{align}
Let us look at the potential terms:
\begin{align}
  -V &= -\w\eta{}^1 C_{2'} \w{C}{}^{2'} \eta_1 -  \w\eta{}^2 \w{\wh{C}}{}^{2'}  \wh{C}_{2'} \eta_2 = -\w\eta{}^1 C_{2'} \w{C}{}^{2'} \eta_1-  \w\eta{}^2 C_{1'} \w{C}{}^{1'}\eta_2\ ,
\end{align}
where, in the second step, we have used the reality constraint
\eqref{reality}. The above form does not seem invariant under
R-symmetry. However, it follows from $\mC\w{\mC} = 0$ that $C_{1'}
\w{C}{}^{1'} = C_{2'} \w{C}{}^{2'}$ which allows us to write the
potential in manifest R-symmetry form:
\begin{equation}
  -V = - \tfrac{1}{2} \w\eta{}^a C_{a'} \w{C}{}^{a'} \eta_a\ .
\end{equation}
Next, let us collect all the Yukawa couplings from the fermi action
\eqref{fercomp}:
\begin{equation}
\left(\tfrac{1}{\sqrt{2}}\w\eta{}^2 (\w\uD_+C_{1'}) \psi_{-} - \w\xi{}^{1'}_+ C_{1'} \psi_{-} - \w\xi{}^{2'} C_{2'} \psi_{-} - \tfrac{1}{\sqrt{2}} \w\eta{}^1 \uD_+ C_{2'}\psi_{-}\right) + \text{c.c.}
\end{equation}
We have $C_{a'} = K_{a'} + L H_{a'}$. Then, we get
\begin{equation}
  \left(-\w\eta{}^2 L \xi_{2+} \psi_{-} - \w\xi{}^{1'}_+ (K_{1'} + L H_{1'}) \psi_{-} - \w\xi{}^{2'} (K_{2'} + L H_{2'}) \psi_{-} - \w\eta{}^1 L \xi_{1+} \psi_{-}\right) + \text{c.c.}
\end{equation}
We thus get the manifest R-symmetric form of the Yukawa couplings
\begin{equation}
 \left(- \w\xi{}^{a'}_+ K_{a'} \psi_{-} -\w\eta{}^a L \xi_{a+} \psi_{-}  - \w\xi{}^{a'}_+ L H_{a'} \psi_- \right) + \text{c.c.}\ ,
\end{equation}
where the first term and its complex conjugate together are mass terms
which contain the fermis and the superpartners of the standard
hypers. The other terms are Yukawa couplings which involve the standard
hypers, the twisted hypers and the fermis.

\subsection{\tp{$(0,4) \to (0,0)$}{(0,4) to (0,0)}}\label{0400}
In this subsection, we directly go from $(0,4)$ superspace to ordinary
space. We give two illustrative examples, an $\mc{O}(1)$ standard
hyper and an arctic fermi.

\paragraph{$\mc{O}(1)$ standard hyper} Recall from \eqref{supfer} and
\eqref{etalevel2} that the $(0,4)$ descendants of $\bm\eta$ are, at
the first level,
\begin{equation}\label{appsupfer}
  \sqrt{2}\xi_{a'+} := \wt\pD_{a'+}\bm\eta\ ,\quad \sqrt{2}\w\xi{}^{a'}_+ := - \varepsilon^{a'b'} \wt\pD_{b'+} \w{\bm\eta}\ ,
\end{equation}
and at the second level,
\begin{equation}\label{appetalevel2}
  \wt\pD_{a'+}\wt\pD_{b'+}\bm\eta = -2\i\varepsilon_{a'b'}\partial_{++}\wt{\bm\eta}\ ,\quad   \wt\pD_{a'+}\wt\pD_{b'+}\w{\bm\eta} = -2\i\varepsilon_{a'b'}\partial_{++}\w{\wt{\bm\eta}}\ .
\end{equation} 
The action is 
\begin{align}\label{appO1projact}
  \mc{S} &= -\frac{\i}{2}\int\ud^2x\oint_{\gamma}\frac{\ud\zeta}{2\pi \i}\wt\pD_{1'+}\wt\pD_{2'+}\, (\zeta^{-1}\w{\bm{\eta}}\partial_{--}\bm{\eta})\ .
\end{align}
Pushing in the derivatives in the measure and using \eqref{appsupfer} and
\eqref{appetalevel2}, we get
\begin{align}\label{04actprocess}
  \mc{S}% &= -\frac{\i}{2}\int\ud^2x\oint_{\gamma}\frac{\ud\zeta}{2\pi \i\zeta}\, (\wt\pD_{1'+}\wt\pD_{2'+}\w{\bm{\eta}}\partial_{--}\bm{\eta} -\wt\pD_{2'+}\w{\bm{\eta}}\partial_{--} \wt\pD_{1'+}\bm{\eta} +  \wt\pD_{1'+}\w{\bm{\eta}}\partial_{--}\wt\pD_{2'+}\bm{\eta} + \w{\bm{\eta}}\partial_{--} \wt\pD_{1'+}\wt\pD_{2'+}\bm{\eta})\ ,\nonumber\\
         &= -{\i}\int\ud^2x\oint_{\gamma}\frac{\ud\zeta}{2\pi \i\zeta}\, (-\i\partial_{++}\w{\wt{\bm{\eta}}}\partial_{--}\bm{\eta} + \w\xi{}^{1'}_+\partial_{--} \xi_{1'+} + \w\xi{}^{2'}_+\partial_{--}\xi_{2'+} - \i \w{\bm{\eta}}\partial_{--} \partial_{++}\wt{\bm{\eta}})\ ,\nonumber\\
          &= \int\ud^2x\oint_{\gamma}\frac{\ud\zeta}{2\pi \i\zeta}\, (\partial_{++}\w\eta{}^2\partial_{--}\eta_2 -\i \w\xi{}^{a'}_+\partial_{--} \xi_{a'+} -\w\eta{}^1\partial_{--} \partial_{++}\eta_1)\ ,\nonumber\\
            &= \int\ud^2x\, (-\partial_\mu\w\eta{}^{a}\partial^\mu \eta_{a} - \i\w\xi{}^{\,a'}_{+}\partial_{--}\xi_{a'+})\ .
\end{align}
where, in going to the second line, we have used the explicit
expressions $\wt{\bm\eta} = \eta_1$ and $\w{\wt{\bm\eta}} = -\w\eta{}^2$
(to compute $\w{\wt{\bm\eta}}$, we follow the same steps as for
$\wt\pD_{a'+}$ in Section \ref{extconjdef}).

\paragraph{The arctic fermi superfield} We look at the weight $0$
arctic fermi superfield. The descendants are
\begin{alignat}{2}
  \sqrt{2}  \mF_{a'} &= \wt\pD_{a'+}\bUpsilon_-\ ,\quad && \sqrt{2} \w\mF{}^{a'} = -\varepsilon^{a'b'} \zeta \wt\pD_{b'+}\w\bUpsilon_-\ ,\nonumber\\
  \mX_+ &= \wt\pD_{1'+}\wt\pD_{2'+}\bUpsilon_-\ ,\quad &&\w\mX_+ = -\zeta^2 \wt\pD_{1'+}\wt\pD_{2'+}\w\bUpsilon_- - 2\i\zeta \partial_{++}\w\bUpsilon_-\ .
\end{alignat}
The action is
\begin{align}\label{appfermiact}
  \mc{S} &= -\frac{1}{2}\int \ud^2x \oint \frac{\ud\zeta}{2\pi\i} \wt\pD_{1'+}\wt\pD_{2'+}\,(\w{\bm{\Upsilon}}_- \bm{\Upsilon}_-)\ .
\end{align}
Pushing in the derivatives in the measure, we get
\begin{align}\label{appfermiact1}
  \mc{S} &= -\frac{1}{2} \int\ud^2x \oint\frac{\ud\zeta}{2\pi\i} \left( - 2\i\zeta^{-1}\partial_{++} \w\bUpsilon_- \bUpsilon_- - \zeta^{-2} \w\mX_+\bUpsilon_- + \w\bUpsilon_-\mX_+ - 2 \zeta^{-1} \w\mF{}^{a'} \mF_{a'}\right)\ .
\end{align}
The superfields $\mX_+$ and $\mF_{a'}$ are auxiliary and can be
integrated out. The terms involving $\mX_+$ are
\begin{equation}
  \oint \frac{\ud\zeta}{2\pi\i} (\w\bUpsilon_-\mX_+- \zeta^{-2} \w\mX_+\bUpsilon_-) = -\sum_{j\geq 0} (-1)^j\w\iUpsilon_{j+1,-} X_{j+} + \text{c.c.}\ .
\end{equation}
Integrating out $X_{j+}$ gives
\begin{equation}
  \iUpsilon_{j,-} = 0\quad\text{for}\quad j \geq 1\ .
\end{equation}
Thus, the weight $0$ superfield $\bUpsilon_{-}$ which was locally
defined on $\mbb{CP}^1$ becomes a constant on $\mbb{CP}^1$, which is
nothing but a globally defined weight $0$ superfield. Integrating out
$\mF_{a'}$ just sets them to zero. Relabelling
$\iUpsilon_{0-} \to \psi_-$, the action becomes
\begin{equation}
  \mc{S} = \int \ud^2 x\,(-\i)\, \w\psi_- \partial_{++}  \psi_-\ .
\end{equation}

 \printbibliography
%\bibliography{refs}

%\begin{thebibliography}{ACGH}
%\addtolength{\leftmargin}{0.2in}
%\setlength{\itemindent}{-0.2in}

\end{document}